\pdfoutput=1
\documentclass[twocolumn]{aastex631}
\usepackage[utf8]{inputenc}
\usepackage{amsmath,amssymb}

\usepackage{graphicx}
\usepackage[caption=false]{subfig}
\usepackage{hyperref}
\usepackage{color}
\usepackage{nameref}
\usepackage{wrapfig}

\definecolor{asparagus}{rgb}{0.53, 0.66, 0.42}

\submitjournal{ApJ}

\shorttitle{2D/3D PAIR-PLASMA MRI WITH PARTICLE-IN-CELL}
\shortauthors{F. Bacchini et al.}

\newcommand{\D}[2]{\frac{{\rm d} #2}{{\rm d} #1}}

\newcommand\bb[1]{\mbox{\boldmath{$#1$}}}
\newcommand\grad{\bb{\nabla}}
\newcommand\bcdot{\,\bb{\cdot}\,}
\newcommand\bdbldot{\,\bb{:}\,}
\newcommand\btimes{\,\bb{\times}\,}

\newcommand{\rmd}{\textrm{d}}

\newcommand{\vecom}{\bb{\Omega}_0} 
\newcommand{\om}{\Omega_0}
\newcommand{\omp}{\omega_\mathrm{p}}
\newcommand{\omrat}{\omega_{\rm C}/\Omega_0}
\newcommand{\lmri}{\lambda_\mathrm{MRI}}
\newcommand{\hatvece}{\hat{\bb{e}}} 
\newcommand{\vecv}{\bb{v}} 
\newcommand{\vecE}{\bb{E}} 
\newcommand{\vecB}{\bb{B}} 
\newcommand{\veceps}{\bb{\epsilon}} 
\newcommand{\vectau}{\bb{\tau}} 
\newcommand{\tilveceps}{\widetilde{\bb{\epsilon}}} 
\newcommand{\tileps}{\widetilde{\epsilon}}
\newcommand{\tilvectau}{\widetilde{\bb{\tau}}} 
\newcommand{\tiltau}{\widetilde{\tau}}
\newcommand{\bgam}{\bar{\gamma}}
\newcommand{\vecJ}{\bb{J}} 
\newcommand{\vecu}{\bb{u}} 
\newcommand{\vecx}{\bb{x}} 
\newcommand{\vecF}{\bb{F}} 
\newcommand{\pd}{\partial}
\newcommand{\ppar}{p_{\|}}
\newcommand{\pperp}{p_\perp}
\newcommand{\KSB}{\hyperref[eq:rotframeB]{KSB}}
\newcommand{\KSBSC}{\hyperref[eq:riquelmeB]{KSB-SC}}
\newcommand{\KSBOA}{\hyperref[eq:OABfinal]{KSB-OA}}
\DeclareMathAlphabet\mathbfcal{OMS}{cmsy}{b}{n}
\begin{document}

\title{Fully kinetic shearing-box simulations of magnetorotational turbulence in 2D and 3D. I. Pair plasmas}

\correspondingauthor{F. Bacchini}\email{fabio.bacchini@colorado.edu}

\author[0000-0002-7526-8154]{Fabio Bacchini}
\affiliation{Center for Integrated Plasma Studies, Department of Physics, University of Colorado, 390 UCB, Boulder, CO 80309-0390, USA}
\author[0000-0002-5263-9274]{Lev Arzamasskiy}
\affiliation{School of Natural Sciences, Institute for Advanced Study, Princeton, NJ 08544, USA}
\author[0000-0001-8363-839X]{Vladimir Zhdankin}
\affiliation{Department of Astrophysical Sciences, Princeton University, 4 Ivy Lane, Princeton, NJ 08544, USA}
\affiliation{Center for Computational Astrophysics, Flatiron Institute, 162 Fifth Avenue, New York, NY 10010, USA}
\author[0000-0001-9039-9032]{Gregory R.\ Werner}
\affiliation{Center for Integrated Plasma Studies, Department of Physics, University of Colorado, 390 UCB, Boulder, CO 80309-0390, USA}
\author[0000-0003-0936-8488]{Mitchell C.\ Begelman}
\affiliation{JILA, University of Colorado and National Institute of Standards and Technology, 440 UCB, Boulder, CO 80309-0440, USA}
\affiliation{Department of Astrophysical and Planetary Sciences, University of Colorado, 391 UCB, Boulder, CO 80309-0391, USA}
\author[0000-0001-8792-6698]{Dmitri A.\ Uzdensky}
\affiliation{Center for Integrated Plasma Studies, Department of Physics, University of Colorado, 390 UCB, Boulder, CO 80309-0390, USA}

\begin{abstract}
The magnetorotational instability (MRI) is a fundamental mechanism determining the macroscopic dynamics of astrophysical accretion disks. In collisionless accretion flows around supermassive black holes, MRI-driven plasma turbulence cascading to microscopic (i.e.\ kinetic) scales can result in enhanced angular-momentum transport and redistribution, nonthermal particle acceleration, and a two-temperature state where electrons and ions are heated unequally. However, this microscopic physics cannot be captured with standard magnetohydrodynamic (MHD) approaches typically employed to study the MRI. In this work, we explore the nonlinear development of MRI turbulence in a pair plasma, employing fully kinetic Particle-in-Cell (PIC) simulations in two and three dimensions. First, we thoroughly study the axisymmetric MRI with 2D simulations, explaining how and why the 2D geometry produces results that differ substantially from MHD expectations. We then perform the largest (to date) 3D simulations, for which we employ a novel shearing-box approach, demonstrating that 3D PIC models can reproduce the mesoscale (i.e.\ MHD) MRI dynamics in sufficiently large runs. With our fully kinetic simulations, we are able to describe the nonthermal particle acceleration and angular-momentum transport driven by the collisionless MRI. Since these microscopic processes ultimately lead to the emission of potentially measurable radiation in accreting plasmas, our work is of prime importance to understand current and future observations from first principles, beyond the limitations imposed by fluid (MHD) models. While in this first study we focus on pair plasmas for simplicity, our results represent an essential step toward designing more realistic electron-ion simulations, on which we will focus in future work.
\end{abstract}



\section{Introduction}
\label{sec:intro}

Astrophysical accretion disks are ubiquitous in the Universe, often surrounding massive compact objects such as black holes and neutron stars. Plasmas in accretion disks rotate around the central object with a differential-rotation profile determined by the object's gravity; gravity also drives accretion of gas onto the central object, thus requiring a mechanism of angular-momentum transport and redistribution throughout the disk. Such a mechanism has not been completely understood so far. The magnetorotational instability (MRI, \citealt{velikhov1959,chandrasekhar1960,balbushawley1991,balbushawley1998}) has been proposed as a possible pathway to promote angular-momentum transport in accretion disks: the MRI causes a dramatic amplification of any pre-existing seed magnetic field, however small, driving and sustaining turbulence at macroscopic scales. The development of the MRI thus creates the conditions to connect the plasma dynamics at the smallest and largest length scales in accretion disks. Although the development of strong turbulence is an inevitable consequence of the MRI, it remains to be understood whether, and under which conditions, this turbulence can also be responsible for efficient angular-momentum transport.

Several studies have been conducted on the linear MRI theory (\citealt{balbushawley1991,goodmanxu1994,latter2009,pessahgoodman2009}), but the full nonlinear development of this instability cannot be treated analytically. Decades of research efforts have employed magnetohydrodynamic (MHD) simulations to study the complex, turbulent state that arises during the nonlinear stage of the MRI. In particular, local models of accretion disks --- where only a small sector of the disk is simulated --- have attracted widespread recognition, owing to the possibility of studying the MRI development separately from its embedding astrophysical environment. From the very beginning, such simulations have been conducted with the so-called \emph{shearing-box} approach (\citealt{Hill1878,hawley1995}), where a small simulation box with locally Cartesian coordinates is taken as representative of the large-scale behavior of the whole disk. In the shearing-box paradigm without density stratification, the vertical ($z$, parallel to the rotation axis) and toroidal ($\varphi$, the direction of rotation; locally $y$) coordinates are periodic, and the radial ($r$, locally $x$) coordinate employs \emph{shearing-periodic} conditions that model the background differential rotation of the disk. 
Several MHD codes implementing the shearing box have been developed  (e.g.\ \citealt{gresselziegler2007,stonegardiner2010}) and applied (e.g.\ \citealt{matsumototajima1995,stone1996,sanoinutsuka2001,baistone2013,hirai2018}) to study the MRI. These works have found that, when the initial conditions involve a weak, purely vertical magnetic field, the system initially evolves to a state of so-called ``channel flows'' where strong radial and toroidal magnetic fields develop. The magnetic-field polarity and flow bulk velocity change sign at the channel interfaces, which become susceptible to the development of parasitic instabilities (e.g.\ tearing, Kelvin-Helmholtz, or drift-kink modes among others; \citealt{goodmanxu1994,pessahgoodman2009}). These secondary modes feed off the primary instability, eventually destroying the macroscopic channels and driving a turbulent state.

While MHD studies have provided fundamental insight into the physics of the MRI, plasmas in astrophysical accretion disks should in many cases be treated with kinetic approaches. Particularly for radiatively inefficient accretion flows (RIAFs) around supermassive black holes (SMBHs) accreting at low rates, it is expected that plasma particles (electrons and ions) can travel long distances between binary collisions, i.e.\ their mean-free-path is comparable to the macroscopic scales in these environments.
As a consequence, electrons and ions in RIAFs can be thermally decoupled at short distances from the central object; both species can experience strong nonthermal acceleration, but energy partition can be unequal. In particular, electron acceleration to high energies can result in the emission of most of the outgoing observable disk radiation.
Such a collisionless, two-temperature, nonthermal, radiative state cannot be fully captured with the standard MHD simulations typically employed for accretion-disk modeling; it is however imperative to investigate plasmas in these conditions with appropriate methods to correctly interpret current and future observations, such as the first direct images of the accretion flow around M87* and SgrA* (\citealt{EHT2019e,EHT2022e}). So far, no self-consistent (i.e.\ first-principles) model of collisionless MRI-driven turbulence, including reconnection, angular-momentum transport, plasma heating, and particle acceleration has been produced.

This lack of accurate models has profound impact on our understanding of accretion physics. For example, recent observational campaigns are affected by heavy uncertainties due to a lack of a reliable theory for the electron/ion energy partition in accretion disks. One of the most widely employed models of plasma heating in general-relativistic magnetohydrodynamic (GRMHD) simulations used to interpret Event Horizon Telescope (EHT) images \citep[e.g.,][]{ressler2015,chael2018,EHT2019e} is based on the non-relativistic gyrokinetic (GK) approximation \citep{schekochihin2009}, in which the magnetic moment of particles is assumed to be conserved.
This assumption reduces the dimensionality of the problem and makes it much less computationally challenging, allowing accurate calculations of energy partition between different species in turbulence \citep{Howes2010,Kawazura2018}, including MRI turbulence \citep[e.g.,][in the reduced-MHD limit]{kawazura2021}.
However, the GK assumptions prevent one from studying non-adiabatic heating mechanisms (in which the magnetic moment is not conserved). Additionally, GK heating models do not take into account deviations of the particle distribution function from thermodynamic equilibrium (e.g. pressure anisotropy). Such deviations can significantly modify the rate of Landau damping \citep{Kunz2015,Kunz2018}, which is the main dissipation mechanism in~GK. In addition, in realistic astrophysical plasmas, the magnetic moment of particles is not expected to be conserved on global dynamical timescales, allowing for a number of heating mechanisms not included in~GK. For example, if high-frequency fluctuations are present, turbulence can be dissipated due to cyclotron resonance \citep{KennelEngelmann1966,Isenberg2004,IsenbergVasquez2007,arzamasskiy2019,Squire2022}; alternatively, if electric-field fluctuations at scales comparable to particle gyro-radii have sufficiently large amplitudes, particles can start experiencing a random walk in velocity space and extract energy from these fluctuations \citep[``stochastic heating'',][]{McChesney1987,Chandran2010,Hoppock2018,Cerri2021}. It is unclear whether these two mechanisms are viable in the context of accretion flows because both are expected to become less and less important at large scale separation (i.e. the ratio between global scales and the scales of particle gyromotion). Additionally, if plasmas significantly deviate from local thermodynamic equilibrium, various kinetic micro-instabilities can be excited, such as firehose \citep{Rosenbluth1956,Parker1958,Chandrasekhar1958,VedenovSagdeev1958,kunz2014,Riquelme2015}, mirror \citep{ShapiroShevchenko1964,kunz2014,Riquelme2015}, ion-cyclotron \citep{SagdeevShafranov1960,ley2019} and whistler \citep{SagdeevShafranov1960,Riquelme2016}. These instabilities create small-scale fluctuations which scatter particles, and introduce an effective viscosity in the plasma, which becomes the dominant dissipation mechanism of the cascade (Arzamasskiy et al., in prep). The partition of energy due to this gyro-viscosity has been studied by \citet{sharma2006} (with ``kinetic MHD'' methods, similar to the Braginskii-MHD approach) who concluded that the ion-to-electron heating ratio depends on which instabilities are excited and how pressure anisotropy is mediated by them, making the fully kinetic modeling of these processes crucial for understanding plasma heating.

Importantly, plasma energization can also occur in the presence of current sheets due to the tearing instability \citep[leading to magnetic reconnection; see, e.g.][]{zweibelyamada2009} and the drift-kink instability \citep[DKI, ][]{pritchettetal1996,zenitanihoshino2007,hoshino2020}. 
Both these instabilities can operate in turbulent plasmas, where myriad small-scale current sheets can form \citep[e.g.][]{zhdankin2013,zhdankin2015,mallet2017,zhdankin2018,loureiroboldyrev2020}. These mechanisms can rapidly convert built-up magnetic energy to plasma energy, accelerating a significant fraction of particles to very high (nonthermal) energies \citep[as shown by kinetic simulations, e.g.][]{zenitanihoshino2008,guo2014,sironispitkovsky2014,werner2016,zhdankin2017,comissosironi2018,werner2018,comissosironi2019,zhdankin2019,wong2020}.
In particular, the tearing instability tends to disrupt thin current sheets; concurrently, the DKI can cause a current sheet to kink (ripple) and become distorted, dissipating magnetic energy as the sheet turbulently folds over on itself \citep{werneruzdensky2021}. These processes play an important role in dissipating magnetic fields generated by the MRI, and require a kinetic treatment to be captured self-consistently. Recently, prescriptions for electron/ion heating in magnetic reconnection obtained from kinetic simulations (e.g.\ \citealt{werner2018,ball2018pic}) have been employed in GRMHD as a way to include the missing microphysics leading to plasma heating and subsequent emission (\citealt{dexter2021,scepi2022}; Hankla et al.\ 2022, in prep.).

Studying the MRI with kinetic approaches is important beyond the electron/ion thermal- and nonthermal-energization problem. A primary question is how the aforementioned angular-momentum transport is realized in collisionless disks; turbulence could in principle supply an effective ``collisionless'' viscosity and resistivity determined by electromagnetic, bulk, and anisotropic-pressure stresses. However, the latter is not included in standard MHD models, which typically assume an isotropic pressure. Based on kinetic theory (\citealt{quataert2002}), \cite{sharma2006} conducted simulations with a more sophisticated ``kinetic'' (i.e.\ Braginskii) MHD model which evolves the parallel and perpendicular pressure independently, limiting the allowed pressure anisotropy with an ad-hoc prescription. These simulations have shown that pressure anisotropy can enhance the effective viscosity, but a first-principles explanation for the observed dynamics is still lacking.

Turbulent angular-momentum transport and particle acceleration in collisionless plasmas can be self-consistently modeled from first principles with Particle-in-Cell (PIC) simulations. \cite{riquelme2012} and \cite{hoshino2013} presented the first attempts to conduct shearing-box PIC simulations in~2D. These pioneering works focused on pair plasmas (for simplicity), and found that reconnection and turbulence during the nonlinear MRI stage can produce substantial nonthermal particle acceleration and momentum transport. \cite{hoshino2015} carried out the first-ever 3D shearing-box PIC simulation for a pair plasma and found similar results during the nonlinear MRI stage. These findings were corroborated by additional large-scale 2D pair-plasma simulations by \cite{inchingolo2018}, who pointed out the role of drift-kink instabilities in driving the turbulent state. All these results emphasized that magnetic reconnection and turbulence driven by the MRI could produce efficient angular-momentum transport and nonthermal particle acceleration. This aligns well with 3D simulations conducted by \cite{kunz2016} with a nonrelativistic hybrid approach (although these employed a different initial magnetic-field geometry), where ions are fully kinetic and electrons are treated as a charge-neutralizing massless fluid. 

\begin{figure}
\centering
\includegraphics[width=1\columnwidth, trim={0mm 60mm 52.5mm 0mm}, clip]{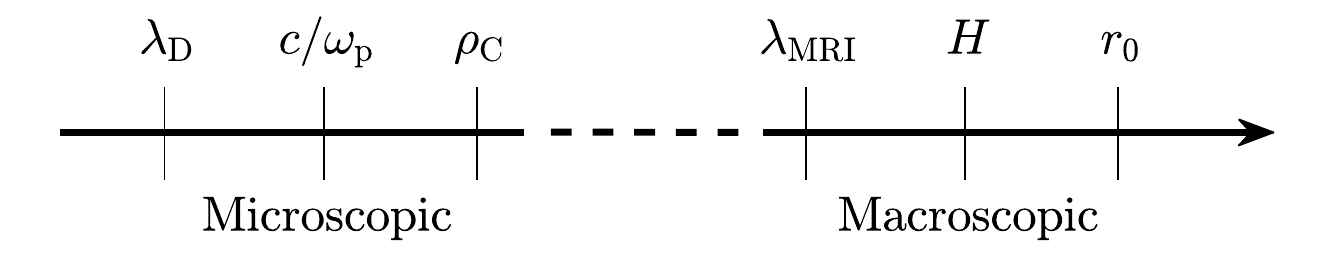}
\caption{Representation of the hierarchy from microscopic (kinetic) to macroscopic (fluid) length scales that are relevant for kinetic shearing-box simulations of the MRI.}
\label{fig:scales}
\end{figure}

Although they have provided new, fundamental insight, the aforementioned PIC studies of the MRI so far presented in literature have been rather inconclusive. The main issue impeding a comprehensive understanding of the collisionless MRI is the presence of very different length and time scales that have to be included in PIC simulations for these to achieve the converged, mesoscale dynamics that is expected from MHD numerical experiments. In nonrelativistic astrophysical plasmas, the kinetic(microscopic) length scales for each particle species $j$ subjected to a magnetic field with strength $B$ are the Debye length $\lambda_{\mathrm{D},j} = \sqrt{kT_j/(4\pi n_j q_j^2)}$, the skin depth $c/\omega_{\mathrm{p},j}$ (with the plasma frequency $\omega_{\mathrm{p},j} = \sqrt{4\pi n_j q_j^2/m_j}$), and the Larmor radius $\rho_{\mathrm{C},j}=m_j c u_{\perp,j}/(q_j B)$ (where $u_\perp$ represents the particle momentum perpendicular to~$\vecB$); here, $T$ and $n$ are the plasma temperature and number density, and $q$ and $m$ are the particle charge and mass. For an accretion disk subjected to the MRI, the macroscopic length scales at a specific distance $r_0$ from the central object are the most-unstable MRI wavelength $\lmri=2\pi v_\mathrm{A}/\om$ and the disk scale-height $H=v_\mathrm{th}/\om$, where $v_\mathrm{A} = B/\sqrt{4\pi \sum_j n_j m_j}$
is the (nonrelativistic) Alfv\'en speed, $v_\mathrm{th}$ is the plasma thermal speed, and $\om$ is the disk rotational frequency at~$r_0$. The relative ordering of microscopic (kinetic) and macroscopic (fluid) spatial scales for low-$\beta$ plasmas is depicted in Figure~\ref{fig:scales}. The separation between macroscopic and microscopic scales in real astrophysical systems is typically enormous, and including a realistic separation in PIC simulations translates into prohibitive computational costs. To face the problem, PIC methods typically employ a reduced (i.e.\ unrealistic) separation between fluid and kinetic scales. Unrealistic physical parameters, however, can have a substantial effect on simulation results; studying how a certain choice of parameters translates into a specific simulated result is a key element in interpreting many kinetic studies presented in the literature, which has so far been underexplored.  In addition, the reduced dimensionality of 2D simulations can dramatically alter the MRI evolution  (\citealt{sanostone2002,masadasano2008}), adding to the uncertainty of the PIC results presented in literature.

In this work, we aim at filling the existing knowledge gap on the physics of the collisionless MRI by carrying out a thorough exploration of PIC simulations in two and three dimensions. We assess the effect of all relevant physical parameters on kinetic simulations, and most importantly, we conduct large-scale, fully kinetic 3D simulations of unprecedented size. With these runs, we can analyze the phenomenology of MRI-driven turbulence, angular-momentum transport, and particle acceleration self-consistently, exploring a wide parameter space. In this first study, we focus on pair plasmas for simplicity, with the aim of approaching the more relevant electron-ion case in a follow-up work.

This paper is organized as follows:
in section \ref{sec:equations}, we present the relevant equations for shearing-box PIC simulations and the methods employed to carry out our numerical study. In section~\ref{sec:parameters} we discuss the choice of simulation parameters and the implications for the numerical cost of the simulations we run. In section \ref{sec:2D} we present a large parameter-space exploration of the 2D axisymmetric pair-plasma MRI, discussing its phenomenology, the numerical convergence of simulations, and the mechanisms of angular-momentum transport and particle acceleration. In section \ref{sec:3D} we present the first large-scale, 3D, fully kinetic simulations of the MRI conducted with a newly developed ``orbital-advection'' shearing-box method; finally, in section \ref{sec:conclusions} we discuss our results and present our main conclusions.

\section{Kinetic shearing-box equations}
\label{sec:equations}

In this section we present the theoretical models employed for the numerical simulations presented in section~\ref{sec:2D} and onward. First, we briefly describe the particle and field equations in a rotating frame (section~\ref{sec:KSB}) and the transformation to shearing coordinates (section~\ref{sec:KSBSC}). Then, we present a new formulation of the governing equations that simplifies the existing approaches and is appropriate for 3D simulations (section~\ref{sec:equationsKSBOA}). We will call this new paradigm \emph{kinetic orbital advection}.

\subsection{Equations in a rotating frame}
\label{sec:KSB}

We consider a small Cartesian sector of an accretion disk with coordinates $(r,\varphi,z)$, i.e.\ the radial, toroidal, and vertical (parallel to the central object's axis of rotation) directions. We apply a boost to a frame of reference which moves with rotational velocity $\vecv_0=\vecom \btimes \bb{r}$, where $\vecom=(0,0,\om)$ is the rotational frequency. The local box assumes a coordinate system $(x,y,z)$, where $r_0+x$ gives the radial position by means of a (small) increment $x$ with respect to a reference radius $r_0$; $y$ is the azimuthal (toroidal) direction, ignoring effects of curvature; and $z$ is still the vertical direction. The rotational velocity is assumed to be  nonrelativistic, i.e.\ $v_0^2 \ll c^2$. The disk is assumed to be differentially rotating, implying that an observer in the corotating frame, at some radius $r_0$ with velocity $\vecv_0(r_0)$, measures a background shear in the toroidal direction. Across short distances, the corresponding background velocity profile can be linearized as $\vecv_\mathrm{s}(x) = -s\om x\hatvece_y$, with $s=3/2$ for a Keplerian disk (note that the equations we write below remain applicable for arbitrary~$s$).

Maxwell's equations for the electromagnetic fields in this frame become (e.g.\ \citealt{schiff1939,hoshino2013}\footnote{Note that there is a sign typo in \cite{hoshino2013} in the first $\vecv_0$ term on the right-hand side of equation \eqref{eq:rotframeE}.})
\begin{equation}
 \pd_t\bb{B} = -c\grad\btimes\bb{E},
 \label{eq:rotframeB}
\end{equation}

\begin{equation}
\begin{aligned}
 \pd_t\left(\bb{E}-\frac{\bb{v}_0}{c}\btimes\bb{B}\right) = & c\grad\btimes\bb{B} - 4\pi\bb{J} \\
 & - \grad\btimes\left(\bb{v}_0\btimes\left(\bb{E}-\frac{\bb{v}_0}{c}\btimes\bb{B}\right)\right),
 \end{aligned}
 \label{eq:rotframeE}
\end{equation}
\begin{equation}
 \grad\bcdot\bb{B}=0,
 \label{eq:rotframedivB}
\end{equation}
\begin{equation}
 \grad\bcdot\left(\bb{E}-\frac{\bb{v}_0}{c}\btimes\bb{B}\right)=4\pi\rho,
 \label{eq:rotframedivE}
\end{equation}
for the electric and magnetic fields $\vecE$ and $\vecB$, evolving in time according to the source terms $\vecJ$ and $\rho$, i.e.\ current and charge density.
We note that, in the corotating frame, there exist four vector fields $\vecE,\vecE^\textbf{*},\vecB,\vecB^\textbf{*}$ obeying the relations (\citealt{arendtjr1998}),
\begin{equation}
 \vecE=\vecE^\textbf{*}+\frac{\vecv_0}{c}\btimes\vecB,
\end{equation}
\begin{equation}
 \vecB^\textbf{*}=\vecB-\frac{\vecv_0}{c}\btimes\vecE^\textbf{*},
\end{equation}
and one could write different (but equally valid) versions of Maxwell's equations in terms of any set of such fields. Equations \eqref{eq:rotframeB}--\eqref{eq:rotframedivE} employ $\vecE,\vecB$ as this choice makes the resulting equations more suitable for numerical integration.

The relativistic particle equations of motion in the rotating frame are
\begin{equation}
 \frac{{\rm d} \vecx}{{\rm d} t} = \frac{\vecu}{\gamma},
 \label{eq:rotframex}
\end{equation}
\begin{equation}
 \frac{{\rm d} \vecu}{{\rm d} t} = \frac{q}{m}\left(\vecE+\frac{\vecu}{c\gamma}\btimes\vecB\right) + 2\vecu\btimes\vecom +2\gamma s\om^2 x \hatvece_x,
 \label{eq:rotframeu}
\end{equation}
for a particle with position $\vecx$ and spatial part of the 4-velocity $\vecu=(u_x,u_y,u_z)$ (with the Lorentz factor $\gamma=\sqrt{1+u^2/c^2}$). On the right-hand side of equation \eqref{eq:rotframeu} the Coriolis, centrifugal, and radial gravitational forces (simplified to the small-box approximation) add to the usual electromagnetic Lorentz force acting on particles. Equations \eqref{eq:rotframeB}--\eqref{eq:rotframeu} will be referred to as the \emph{kinetic shearing-box} (\KSB) system from now on.
  
Because of the background shearing profile, peculiar boundary conditions are necessary in shearing-box simulations. Maxwell's equations \eqref{eq:rotframeB}--\eqref{eq:rotframedivE} require a special treatment at the $x$-boundaries, reflecting the fact that the shear advects and distorts field lines and fluid bulk motions. These {\it shearing-periodic boundary conditions} are
\begin{equation}
\begin{aligned}
 \vecE(x_\mathrm{i},y,z) = & \vecE(x_\mathrm{o},y-\Delta y_\mathrm{s},z) \\ & +\vecv_\mathrm{s}(L_x)\btimes\vecB(x_\mathrm{o},y-\Delta y_\mathrm{s},z)/c,
\end{aligned}
 \label{eq:shearingBCE}
\end{equation}
\begin{equation}
 \vecB(x_\mathrm{i},y,z) = \vecB(x_\mathrm{o},y-\Delta y_\mathrm{s},z),
 \label{eq:shearingBCB}
\end{equation}
\begin{equation}
\begin{aligned}
 \vecJ(x_\mathrm{i},y,z) = & \vecJ(x_\mathrm{o},y-\Delta y_\mathrm{s},z) \\
 & + \vecv_\mathrm{s}(L_x)\rho(x_\mathrm{o},y-\Delta y_\mathrm{s},z),
\end{aligned}
 \label{eq:shearingBCJ}
\end{equation}
\begin{equation}
 \rho(x_\mathrm{i},y,z) = \rho(x_\mathrm{o},y-\Delta y_\mathrm{s},z),
\end{equation}
where $x_\mathrm{i}$ and $x_\mathrm{o}$ represent the position of the inner and outer $x$-boundaries and $L_x=x_\mathrm{o}-x_\mathrm{i}$ is the box size along $x$. The difference in $y$-coordinate $\Delta y_\mathrm{s}=s\om L_x t$ increases with time, and at the \emph{periodic times} $t^* = N L_y/(s\om L_x)$ with $N\in \mathbb{N}$ the simulation box is strictly periodic again along $x$. The $y$- and $z$-directions always remain strictly periodic.

Finally, note that in particle simulations special boundary conditions must be applied to particles crossing $x$-boundaries. First, the velocity of each particle must be modified by adding or subtracting (upon crossing the outer or inner boundary, respectively) the global velocity offset $s\om L_x$. Second, the position of each crossing particle must be shifted by a distance $\Delta y_\mathrm{s}$ along the positive or negative $y$-direction (for particles crossing the outer or inner boundary, respectively).

The numerical implementation of the \KSB~system presents significant complications due to the shearing-periodic boundary conditions. These difficulties mainly arise in computer codes employing MPI parallelization: because corresponding points along the inner and outer $x$-boundaries shift in position as time passes, the communication between processors across these boundaries is substantially more complex than in the case of standard periodic boundary conditions. Additionally, the values of $\vecE$ and $\vecJ$ must be corrected (with $\vecB$ and $\rho$, respectively) prior to communication, which requires additional interpolation since these quantities are typically not defined at the same spatial locations. 

\subsection{Equations in shearing coordinates}
\label{sec:KSBSC}

A simpler alternative approach for PIC simulations has been proposed by \cite{riquelme2012} in the form of a transformation to the so-called \emph{shearing coordinates} (originally developed and employed for MHD calculations, e.g.\ \citealt{goldreichlyndenbell1965,goldreichtremaine1978,narayan1987,kaisig1989}). The transformation consists of a boost along $y$ to a reference frame moving with velocity $\vecv_\mathrm{s}(x)$, thus obtaining the cancellation of velocity offsets at the $x$-boundaries\footnote{Note that all terms containing $\vecv_0$ in the \KSB~system are neglected prior to boosting the reference frame, under the assumption that $v_0^2\ll c^2$.}. Assuming a 2D ($x$--$z$) geometry, Maxwell's equations can then be rewritten and simplified to obtain
\begin{equation}
 \pd_t\vecB = -c\grad\btimes\vecE - s\om B_x \hatvece_y,
 \label{eq:riquelmeB}
\end{equation}
\begin{equation}
 \pd_t\vecE = c\grad\btimes\vecB - s\om E_x\hatvece_y - 4\pi\vecJ,
 \label{eq:riquelmeE}
\end{equation}
\begin{equation}
 \grad\bcdot\vecB=0,
\end{equation}
\begin{equation}
 \grad\bcdot\vecE=4\pi\rho,
\end{equation}
which only differ from the standard Maxwell's equations by two source terms that drive the shearing of electric- and magnetic-field lines along $y$ (see \citealt{riquelme2012} for further details).

The particle equations of motion are rewritten by subtracting the background shearing velocity, such that
\begin{equation}
 \frac{\rmd\vecx}{\rmd t} = \frac{\vecu}{\gamma},
\end{equation}
\begin{equation}
 \frac{\rmd\vecu}{\rmd t} = \frac{q}{m}\left(\vecE+\frac{\vecu}{c\gamma}\btimes\vecB\right) + 2\vecu\btimes\vecom + s\om u_x \hatvece_y.
 \label{eq:riquelmeu}
\end{equation}
We will refer to equations \eqref{eq:riquelmeB}--\eqref{eq:riquelmeu} as the \emph{kinetic shearing box in shearing coordinates} (\KSBSC) system. Note that, because of the boost, all quantities (electromagnetic fields, currents, velocities, etc.) measured in this reference frame are not equivalent to those measured in the standard corotating frame employed for the \KSB~system. In particular, particle velocities here do not include the background shearing-velocity part, implying that the current density measured in this frame lacks the corresponding contribution from the background shearing motion.

The \KSBSC~equations above are adequate for 2D axisymmetric simulations such as those conducted by \cite{riquelme2012} and \cite{inchingolo2018} (as well as in section \ref{sec:2D} of this work). These equations are advantageous for such simulations also because they are far less complicated to solve numerically than those of the \KSB~system, since the $x$-boundaries are here strictly periodic. However, the full 3D version of the \KSBSC~equations includes several additional terms that explicitly depend on the $y$-coordinate (\citealt{riquelme2012}); the numerical implementation of these terms is substantially more complicated than that of the original \KSB~equations. For this reason, the \KSBSC~system has so far only been employed for 2D simulations.

\subsection{A new approach: kinetic orbital advection}
\label{sec:equationsKSBOA}
Here, we propose a new approach to kinetic shearing-box simulations in 3D. This is based on a combination of existing techniques that have been employed in previous MHD and PIC studies of the MRI. Our method is similar to the implementation of the \KSB~system as presented by \cite{hoshino2013,hoshino2015}, but simplifies the required boundary conditions and employs equations similar to those of the \KSBSC~system, while remaining fully applicable in 3D. The construction of our method follows the so-called ``orbital advection'' approach that is widely applied in MHD shearing-box implementations (e.g.\ \citealt{gresselziegler2007,stonegardiner2010} and references therein). As we will show in the following, the basic idea is that electric fields and particle momenta are replaced with corresponding quantities boosted to a frame moving with $\vecv_\mathrm{s}$, but the coordinates of the simulation box are kept in the lab frame. We note that a similar strategy was employed in nonrelativistic hybrid simulations by \cite{kunz2014,kunz2014code,kunz2016}.

To construct our set of equations, we start from the \KSB~system. First, as done in all previous works, we neglect the terms involving $\vecv_0$ in Maxwell's equations (\citealt{riquelme2012,hoshino2013,hoshino2015,inchingolo2018}). Then, assuming $v_\mathrm{s}^2 \ll c^2$, we apply a Galilean transformation,
\begin{equation}
 \vecE'\simeq\vecE+\frac{\vecv_\mathrm{s}}{c}\btimes\vecB,
\end{equation}
where $\vecE'$ is the electric field in the comoving frame. Substituting for $\vecE'$ in Maxwell's equations we get
\begin{equation}
 \pd_t\vecB = -c\grad\btimes\left(\vecE'-\frac{\vecv_\mathrm{s}}{c}\btimes\vecB\right),
 \label{eq:OAB}
\end{equation}
\begin{equation}
 \pd_t\left(\vecE'-\frac{\vecv_\mathrm{s}}{c}\btimes\vecB\right) = c\grad\btimes\vecB - 4\pi\vecJ,
 \label{eq:OAE}
\end{equation}
\begin{equation}
 \grad\bcdot\vecB=0,
 \label{eq:OAdivB}
\end{equation}
\begin{equation}
 \grad\bcdot\left(\vecE'-\frac{\vecv_\mathrm{s}}{c}\btimes\vecB\right)=4\pi\rho.
 \label{eq:OAdivE}
\end{equation}
Further, substituting equation \eqref{eq:OAB} into \eqref{eq:OAE} we get
\begin{equation}
\begin{aligned}
 \pd_t\vecE' = & c\grad\btimes\vecB - 4\pi\vecJ \\
 & - \vecv_\mathrm{s}\btimes\left(\grad\btimes\vecE'\right) + \frac{\vecv_\mathrm{s}}{c}\btimes\left(\grad\btimes\left(\frac{\vecv_\mathrm{s}}{c}\btimes\vecB\right)\right),
\end{aligned}
 \label{eq:OAE_1}
\end{equation}
where, for consistency with the assumption $v_\mathrm{s}^2\ll c^2$, we will ignore the last term on the right-hand side. Note that having removed the bulk-motion electric field $\vecv_\mathrm{s}\btimes\vecB/c$ from $\vecE$, one would expect $E'\ll E$; however, $E'$ can still be large due to nonideal effects --- e.g.\ at reconnection sites --- hence we retain the term $\vecv_\mathrm{s}\btimes(\grad\btimes\vecE')$ in equation \eqref{eq:OAE_1}. A clear advantage of a formulation involving $\vecE'$ as the electric field is that the boundary condition along radial boundaries becomes
\begin{equation}
 \vecE'(x_\mathrm{i},y,z) = \vecE'(x_\mathrm{o},y-\Delta y_\mathrm{s},z),
 \label{eq:OABCE}
\end{equation}
which, in contrast with equation \eqref{eq:shearingBCE} for the \KSB~system, does not require to correct the electric field at the radial boundaries with the corresponding magnetic field. This simplifies the numerical implementation of boundary conditions significantly.

Before rewriting Maxwell's equations into a form that is suitable for a numerical treatment, it is convenient (for reasons that will become apparent later) to substitute the current, $\vecJ$, with the comoving current $\vecJ'$. By applying another Galilean transformation we get
\begin{equation}
 \vecJ' \simeq \vecJ - \rho\vecv_\mathrm{s}.
\end{equation}
Note that using $\vecJ'$ as a source term for Maxwell's equations modifies the boundary condition \eqref{eq:shearingBCJ}, which becomes
\begin{equation}
 \vecJ'(x_\textrm{i},y,z) = \vecJ'(x_\textrm{o},y-\Delta y_\mathrm{s},z)
\end{equation}
and is manifestly simpler to implement numerically since it requires no correction of the current at the boundaries. We can now employ equations \eqref{eq:OAdivB} and \eqref{eq:OAdivE} to obtain the following evolution equations for $\vecE'$ and $\vecB$,
\begin{equation}
 \begin{cases}
  \pd_t B_x & = -c\pd_y E_z' +c\pd_z E_y' - v_{\mathrm{s},y}\pd_y B_x \\
  \pd_t B_y & = c\pd_x E_z' -c\pd_z E_x' - v_{\mathrm{s},y}\pd_y B_y - s\om B_x \\
  \pd_t B_z & = -c\pd_x E_y' +c\pd_y E_x' - v_{\mathrm{s},y}\pd_y B_z
 \end{cases},
 \label{eq:OABexpl}
\end{equation}
\begin{equation}
 \begin{cases}
  \pd_t E_x' = c\pd_y B_z -c\pd_z B_y + v_{\mathrm{s},y}\pd_y E_x' - v_{\mathrm{s},y}\pd_x E_y' - 4\pi J_x' \\
  \begin{aligned}
  \pd_t E_y' = & -c\pd_x B_z +c\pd_z B_x \\
     & - v_{\mathrm{s},y}\pd_x E_x' - v_{\mathrm{s},y}\pd_y E_y' - v_{\mathrm{s},y}\pd_z E_z' -4\pi J_y'  
  \end{aligned} \\
 \pd_t E_z' = c\pd_x B_y -c\pd_y B_x + v_{\mathrm{s},y}\pd_y E_z' - v_{\mathrm{s},y}\pd_z E_y' - 4\pi J_z'
 \end{cases},
 \label{eq:OAEexpl}
\end{equation}
where we have neglected all terms of order $\om^2$. This form of Maxwell's equations is easy to handle numerically, since it can be written as
\begin{equation}
 \pd_t\vecB = -v_{\mathrm{s},y}\pd_y\vecB + \text{(other terms)},
\end{equation}
\begin{equation}
 \pd_t\vecE' = \pm v_{\mathrm{s},y}\pd_y\vecE' + \text{(other terms)},
\end{equation}
where the $\pm$ in the equation for $\vecE'$ is a $-$ for the $y$-component and a $+$ for the other components. These equations present a manifestly advective term in the $y$-direction, which can then be treated with any of the available schemes for numerical advection (see Appendix \ref{app:methods} for further details). 

In our approach, we choose to keep the particle position in the lab frame, but to solve for the particle momenta $\vecu'$ in the comoving frame. In the nonrelativistic limit, this simply reduces to separating out the shearing velocity $\vecv_\mathrm{s}$ from velocity fluctuations superimposed to this background. This choice serves three purposes: first, we avoid the representation of a constant background by a finite amount of computational particles, greatly reducing computational noise (as also noted by \citealt{kunz2014code}); second, by evolving the particle velocity in the comoving frame, we can directly collect the current $\vecJ'$, justifying the use of the comoving current as a source term in Maxwell's equations; third, we avoid the need to modify the $y$-velocity of particles crossing $x$-boundaries, thus simplifying the application of boundary conditions.

To derive our particle equations of motion, we start from equation \eqref{eq:rotframeu}; we then boost $\vecu$ to the frame moving with $\vecv_\mathrm{s}$,
\begin{equation}
 \begin{cases}
  u_x' = u_x \\
  u_y' = \gamma_\mathrm{s} u_y - \gamma_\mathrm{s} v_{\mathrm{s},y} \gamma \\
  u_z' = u_z
 \end{cases},
\end{equation}
where $\gamma_\mathrm{s}=1/\sqrt{1-v_\mathrm{s}^2/c^2}$. If we again assume  $v_\mathrm{s}^2\ll c^2$, this essentially becomes a Galilean transformation, and we have $\gamma_\mathrm{s}\simeq 1$ and $u_y'=u_y - \gamma' v_{\mathrm{s},y}$, with $\gamma' \simeq \gamma$. We then substitute into the equation of motion \eqref{eq:rotframeu} to obtain
\begin{equation}
\begin{aligned}
\frac{\rmd\vecu}{\rmd t} \simeq & \frac{\rmd\vecu'}{\rmd t} + \gamma'\frac{\rmd\vecv_\mathrm{s}}{\rmd t} + \vecv_\mathrm{s}\frac{\rmd\gamma'}{\rmd t} \\
= & \frac{q}{m}\left(\vecE+\frac{\vecu'}{c\gamma'}\btimes\vecB + \frac{\vecv_\mathrm{s}}{c}\btimes\vecB\right) \\
 & + 2\left(\vecu'\btimes\vecom + \gamma'\vecv_\mathrm{s}\btimes\vecom +\gamma' s\om^2 x \hatvece_x\right),
 \label{eq:OAu}
\end{aligned}
\end{equation}
and for the particle position we have
\begin{equation}
 \frac{\rmd\vecx}{\rmd t} = \frac{\vecu'}{\gamma'} + \vecv_\mathrm{s}.
 \label{eq:OAx}
\end{equation}

The terms containing $\gamma'$ on the left-hand side of equation \eqref{eq:OAu} can be expanded: the first term is
\begin{equation}
 \gamma'\frac{\rmd\vecv_\mathrm{s}}{\rmd t} = -\gamma' s\om\frac{\rmd x}{\rmd t}\hatvece_y = -s\om u_x'\hatvece_y,
\end{equation}
and the second term can be treated in the following way: if $\gamma'\simeq\gamma$,
\begin{equation}
  \frac{\rmd\gamma'}{\rmd t} \simeq \frac{\rmd\gamma}{\rmd t} = \frac{\vecu}{\gamma}\bcdot\frac{\rmd\vecu}{\rmd t} \simeq
  (\vecv' + \vecv_\mathrm{s})\bcdot \frac{\rmd\vecu}{\rmd t}
\end{equation}
where $\vecv'=\vecu'/\gamma'$. Now using equation \eqref{eq:rotframeu} together with $\vecE'=\vecE+(\vecv_\mathrm{s}/c)\btimes\vecB$ we get
\begin{equation}
 \begin{aligned}
  \frac{\rmd\gamma'}{\rmd t} \simeq & (\vecv' + \vecv_\mathrm{s})\bcdot \frac{q}{m}\left(\vecE' + \frac{\vecv'}{c}\btimes\vecB\right) \\ 
  & + (\vecv' + \vecv_\mathrm{s})\bcdot\left( 2\vecu\btimes\vecom +2\gamma s\om^2 x \hatvece_x\right).
 \end{aligned}
\end{equation}
Multiplying by $\vecv_\mathrm{s}$ and neglecting all terms of order $\om^2$ (for consistency with the assumption $\gamma_\mathrm{s}\simeq 1$) we thus have
\begin{equation}
 \vecv_\mathrm{s}\frac{\rmd\gamma'}{\rmd t} \simeq \frac{q}{m}\left(\vecv'\bcdot\vecE'\right)\vecv_\mathrm{s}.
\end{equation}
Finally, on the right-hand side of equation \eqref{eq:OAu} we can expand the term
\begin{equation}
 \gamma'\vecv_\mathrm{s}\btimes\vecom = -\gamma' s\om^2 x \hatvece_x,
\end{equation}
which cancels exactly with the rotational-gravitational term in equation \eqref{eq:OAu}. The final form of our momentum equation is therefore
\begin{equation}
\begin{aligned}
 \frac{\rmd\vecu'}{\rmd t} = & \frac{q}{m}\left(\vecE'+\frac{\vecu'}{c\gamma'}\btimes\vecB\right) \\
 & + 2\vecu'\btimes\vecom + s\om u_x'\hatvece_y - \frac{q}{m}\left(\frac{\vecu'}{\gamma'}\bcdot\vecE'\right)\vecv_\mathrm{s},
 \end{aligned}
 \label{eq:OAu_1}
\end{equation}
which is exactly the same as equation \eqref{eq:riquelmeu} from the \KSBSC~system plus the additional term\footnote{This additional term is not a peculiarity of our formulation; it is simply neglected in \cite{riquelme2012} and \cite{inchingolo2018}. This term represents additional inertial forces proportional to the energy gain of a particle due to work exerted by electric fields (\citealt{arendtjr1998}).} $(q/m)(\vecu'\bcdot\vecE')\vecv_\mathrm{s}/\gamma'$. Note that the correspondence with the \KSBSC~momentum equation is not a coincidence: in both cases, the evolution of the particle velocity is calculated by separating out $\vecv_\mathrm{s}$.

To summarize our choice of equations, we have (dropping all primes for simplicity)
\begin{equation}
 \begin{cases}
  \pd_t B_x = -c\pd_y E_z +c\pd_z E_y - v_{\mathrm{s},y}\pd_y B_x \\
  \pd_t B_y = c\pd_x E_z -c\pd_z E_x - v_{\mathrm{s},y}\pd_y B_y - s\om B_x \\
  \pd_t B_z = -c\pd_x E_y +c\pd_y E_x - v_{\mathrm{s},y}\pd_y B_z
 \end{cases},
 \label{eq:OABfinal}
\end{equation}
\begin{equation}
 \begin{cases}
  \pd_t E_x = c\pd_y B_z -c\pd_z B_y + v_{\mathrm{s},y}\pd_y E_x - v_{\mathrm{s},y}\pd_x E_y - 4\pi J_x \\
  \begin{aligned}
  \pd_t E_y = & -c\pd_x B_z +c\pd_z B_x \\
  & -v_{\mathrm{s},y}s\pd_x E_x -v_{\mathrm{s},y}\pd_y E_y - v_{\mathrm{s},y}\pd_z E_z -4\pi J_y \\
  \end{aligned} \\
  \pd_t E_z = c\pd_x B_y -c\pd_y B_x + v_{\mathrm{s},y}\pd_y E_z - v_{\mathrm{s},y}\pd_z E_y - 4\pi J_z
 \end{cases},
 \label{eq:OAEfinal}
\end{equation}
\begin{equation}
 \grad\bcdot\vecB=0,
 \label{eq:OAdivBfinal}
\end{equation}
\begin{equation}
 \grad\bcdot\left(\vecE-\frac{\vecv_\mathrm{s}}{c}\btimes\vecB\right)=4\pi\rho,
 \label{eq:OAdivEfinal}
\end{equation}
for the electromagnetic fields, and
\begin{equation}
\begin{aligned}
 \frac{\rmd\vecu}{\rmd t} = & \frac{q}{m}\left(\vecE+\frac{\vecu}{c\gamma}\btimes\vecB\right) \\
 & + 2\vecu\btimes\vecom + s\om u_x\hatvece_y - \frac{q}{m}\left(\frac{\vecu}{\gamma}\bcdot\vecE\right)\vecv_\mathrm{s},
 \end{aligned}
 \label{eq:OAufinal}
\end{equation}
\begin{equation}
 \frac{\rmd\vecx}{\rmd t} = \frac{\vecu}{\gamma} + \vecv_\mathrm{s},
 \label{eq:OAxfinal}
\end{equation}
for the particle motion. Equations \eqref{eq:OABfinal}--\eqref{eq:OAxfinal} constitute the paradigm for the \emph{kinetic shearing box with orbital advection} to be employed in our 3D simulations, and will be referred to as the \KSBOA~system from now on. This set of equations requires simpler boundary conditions than the \KSB~case and can be directly applied to 3D studies, contrary to the \KSBSC~framework. The numerical approach to solve these equations is described in detail in Appendix \ref{app:methods}.

Finally, we note that equation \eqref{eq:OAxfinal} can result in superluminal particle motion, especially when $u/\gamma\sim c$, since the velocity addition with $\vecv_\mathrm{s}$ is nonrelativistic. To avoid this issue, we can apply the proper relativistic boost back to the observer's frame to ensure that the velocity on the right-hand side is always subluminal,
\begin{equation}
\frac{\rmd x}{\rmd t} = \frac{u_x}{[\gamma+u_y v_{\mathrm{s},y}]\gamma_\mathrm{s}},
\label{eq:OAxproperx}
\end{equation}
\begin{equation}
\frac{\rmd y}{\rmd t} = \frac{u_y+\gamma v_{\mathrm{s},y}}{\gamma+u_y v_{\mathrm{s},y}},
\label{eq:OAxpropery}
\end{equation}
\begin{equation}
\frac{\rmd z}{\rmd t} = \frac{u_z}{[\gamma+u_y v_{\mathrm{s},y}]\gamma_\mathrm{s}}.
\label{eq:OAxproperz}
\end{equation}
The numerical solution of these equations is slightly more complicated, and requires a simple nonlinear iteration (see Appendix \ref{app:methods3D} for details). In large-scale simulations, however, this difference in computational cost may be significant. We thus resort to employing equations \eqref{eq:OAxproperx}--\eqref{eq:OAxproperz} only when superluminal motion is detected during a run. Note however that these equations of motion are not entirely consistent with the derivation of the momentum equation \eqref{eq:OAufinal} given above, where $\rmd\vecx/\rmd t=\vecu/\gamma+\vecv_\mathrm{s}$ was assumed. We have verified that this discrepancy does not impact our results significantly.

\section{Choice of physical parameters and computational considerations}
\label{sec:parameters}

In section \ref{sec:intro}, we have discussed the scale hierarchy that characterizes multi-scale systems such as collisionless accretion disks. In realistic scenarios, the various length scales represented in Figure \ref{fig:scales} typically differ by orders of magnitude; in simulations, we are bound to employ drastically a reduced scale separation, due to computational constraints. Here, we illustrate how the choice of physical parameters influences the computational cost of simulations, and the limitations that affect such numerical experiments.

The simulations presented in the next sections always start from a weak vertical magnetic field $B_{z,0}$ in a box filled with uniform, thermal pair plasma (with particle mass $m=m_i=m_e$, number density $n_0=n_{i,0}=n_{e,0}$) with dimensionless temperature $\theta_0=kT_0/(mc^2)$ (with $T_0=T_{i,0}=T_{e,0}$) and zero bulk speed (measured in the frame moving with~$\vecv_\mathrm{s}$). In addition to $\theta_0$, the other basic dimensionless parameters we must choose in our simulations are the initial Alfv\'en speed $v_{\mathrm{A},0}/c=B_{z,0}/\sqrt{4\pi n_0 m_e(1+ m_i/m_e)c^2}$ (normalized over the speed of light, and calculated with the nonrelativistic expression) and the cyclotron-to-rotational frequency ratio $\omrat = qB_{z,0}/(mc\om)$. Setting these quantities also automatically produces the value of the total (counting both species) thermal-to-magnetic pressure ratio $\beta_0=2n_0kT/[B_{z,0}^2/(8\pi)]=2\theta_0/(v_{\mathrm{A},0}^2/c^2)$. To choose the free simulation parameters, we need to consider i) the underlying assumptions of our model, i.e.\ a nonrelativistic shearing box, and ii) the astrophysical environments we target, i.e.\ accretion disks around SMBHs. To respect the nonrelativistic shearing-box assumptions, we need $v_{\mathrm{A},0}/c\ll1$; for conditions appropriate to astrophysical accretion disks, we need $\theta_0<1$ and $\omrat\gg1$. Setting these free parameters produces an ordering of microscopic and macroscopic scales; to appropriately model accretion around SMBHs, we should in principle respect the ordering depicted in Figure~\ref{fig:scales}, but we will show that this is not always straightforward.

In terms of the free parameters $\theta_0$, $v_{\mathrm{A},0}/c$, and $\omrat$, the length scales shown in Figure~\ref{fig:scales}, normalized to the total (including both species) plasma skin depth~$c/\omp$, are given by
\begin{equation}
    \frac{\lambda_\mathrm{D}}{c/\omp} = \sqrt{\theta_0},
\end{equation}
\begin{equation}
    \frac{\rho_\mathrm{C}}{c/\omp} = \frac{\sqrt{\theta_0}}{v_{\mathrm{A},0}/c},
\end{equation}
\begin{equation}
    \frac{\lmri}{c/\omp} = 2\pi\frac{\omega_\mathrm{C}}{\Omega_0},
\end{equation}
\begin{equation}
    \frac{H}{c/\omp} = \frac{\omega_\mathrm{C}}{\Omega_0}\frac{\sqrt{\theta_0}}{v_{\mathrm{A},0}/c},
\end{equation}
and $r_0/(c/\omp)\sim H/(c/\omp)$ if we assume a thick accretion disk, i.e.\ setting the pressure scale-height $H$ automatically sets the distance $r_0$ from the central object\footnote{This last assumption indicates an underlying discrepancy in the application of the unstratified shearing-box model to thick accretion disks, where vertical stratification would in principle be required.}. Imposing the aforementioned typical values $\theta_0<1$, $v_{\textrm{A},0}/c\ll1$ thus produces the nonrelativistic ordering of microscopic scales $\lambda_\textrm{D}<c/\omp<\rho_\textrm{C}$. Next, we need to choose the separation between the kinetic scales and the macroscopic scales $\lmri$ and~$H$. These are related by 
\begin{equation}
    \frac{\lmri}{\rho_\mathrm{C}} = 2\pi\frac{\omega_\mathrm{C}}{\Omega_0}\frac{v_{\mathrm{A},0}/c}{\sqrt{\theta_0}},
\end{equation}
\begin{equation}
    \frac{H}{\rho_\mathrm{C}} = \frac{\omega_\mathrm{C}}{\Omega_0},
\end{equation}
with 
\begin{equation}
    \frac{H}{\lmri} = \frac{\sqrt{\theta_0}}{2\pi v_{\mathrm{A},0}/c} = \frac{1}{2\pi\beta_0}.
\end{equation}
We thus see that it is easy to impose $\lmri/\rho_\mathrm{C}\gg1$, $H/\rho_\mathrm{C}\gg1$ by carefully choosing $\omrat\gg1$, but the separation $H/\lmri$ depends instead on the previous choice of $\theta_0$ and ${v_{\mathrm{A},0}/c}$. Hence, care is needed in the choice of plasma temperature and Alfv\'en speed to ensure $\rho_\mathrm{C}<\lmri<H$ while also keeping the correct microscopic ordering. We note that the free dimensionless parameters also determine the macroscopic-to-microscopic temporal-scale separation,
\begin{equation}
    \frac{P_0}{\omp^{-1}} = 2\pi \frac{\omrat}{v_{\mathrm{A},0}/c},
\end{equation}
\begin{equation}
    \frac{P_0}{\omega_\mathrm{C}^{-1}} = 2\pi \frac{\omega_\mathrm{C}}{\Omega_0},
\end{equation}
where $P_0=2\pi/\om$ is the orbital period, and the ordering $\omp^{-1}<\omega_\mathrm{C}^{-1}<P_0$ must be respected. 

When choosing parameters for numerical simulations, a realistic scale separation and correct scale ordering are generally hard to achieve. The computational cost of a typical shearing-box simulation directly derives from the relations expressed above, with the addition that the vertical box size $L_z/\lmri$ and the final simulation time $t_\mathrm{end}/P_0$ enter as additional macroscopic scales that should respect the ordering $\lmri<L_z<H$ and $P_0<t_\mathrm{end}$. As we will show in section \ref{sec:2D}, physically valid simulations indeed require box sizes of at least a few $\lmri$ in each spatial direction, as well as $t_\mathrm{end}\gtrsim10P_0$. In this work, we will always employ a spatial resolution $\Delta x\simeq c/\omp$; while this choice does not typically resolve the Debye length $\lambda_\mathrm{D}$ at the beginning of the simulation, in all cases the plasma is heated quickly enough to increase the effective Debye length to scales larger than the grid spacing well before the nonlinear MRI stage. We have also checked that our results do not change significantly with higher resolution. The choice of temporal resolution immediately follows as $\Delta t\simeq r_\mathrm{CFL}\omp^{-1}$ (where $r_\mathrm{CFL}\simeq 2^{-1/2}$ and $r_\mathrm{CFL}\simeq2^{-1/3}$ are the 2D and 3D Courant-Friedrichs-Lewy factors, respectively).

For 2D simulations, we will employ the \KSBSC~paradigm and typically allow for $L_z>H$ to avoid excessive computing costs (see below) when conducting the parameter-space scan presented in section~\ref{sec:paramspace2D}. We have verified that the qualitative behavior of the results is the same when employing parameters that instead respect the constraint $L_z \le H$. The computational cost for our 2D runs can thus be estimated using
\begin{equation}
    N_g \equiv \frac{L}{\Delta x} \simeq \frac{L}{c/\omp}  = \frac{L}{\lmri}\frac{\lmri}{\rho_\mathrm{C}}\frac{\rho_\mathrm{C}}{c/\omp}\propto \frac{L}{\lmri}\frac{\omega_\mathrm{C}}{\Omega_0},
\end{equation}
for the number of required grid points $N_g$ along a generic spatial direction, and
\begin{equation}
    N_t \simeq \frac{t_\mathrm{end}}{r_\mathrm{CFL}\omp^{-1}}  = \frac{t_\mathrm{end}}{P_0}\frac{P_0}{r_\mathrm{CFL}\omp^{-1}}\propto \frac{t_\mathrm{end}}{P_0}\frac{\omrat}{v_{\mathrm{A},0}/c},
\end{equation}
for the number of time steps $N_t$ needed to reach a final time~$t_\mathrm{end}$. We can observe that, in terms of the free parameters, the required grid size increases linearly (in each direction, i.e.\ as $N_g^2$ in 2D runs) with~$\omrat$. In 2D runs, the typical choice for the free parameters is $\omrat=10\mbox{--}30$ (but we have run simulations with much larger values), $v_{\mathrm{A},0}/c=0.01$, $\theta_0=1/32$ (i.e.\ $\beta_0=624$ for both species), which translate into $\lmri/\Delta x\simeq 60\mbox{--}180$ and $P_0/\Delta t\simeq 8{,}000\mbox{--}25{,}000$. With these values, the microscopic scales are rather well separated, $\rho_\textrm{C}/(c/\omega_\mathrm{p})\simeq 17.8$;
however, the macroscopic scales are much less distinguishable, $H/\lmri\simeq2.8$; in addition, the separation between macroscopic and microscopic scales is large but still far from realistic, $\lmri/\rho_\textrm{C}\simeq10.6$. For a domain size $L=8\lmri$ in each dimension and a final simulation time $t_\mathrm{end}=15P_0$, these parameters imply evolving a grid of size up to $\sim1500\times1500$ cells for $\sim400{,}000$ time steps. Achieving better scale separation is in principle possible, e.g.\ by increasing~$\omrat$; however, the required grid size and simulation duration can become prohibitive well before reaching realistic values (e.g.\ an estimated $\omrat\sim10^7$ around~M87*).

With the insight gained from our 2D campaign, we will run 3D simulations employing the \KSBOA~formulation and demand $L_z=H$, for complete consistency with the spatial scale ordering. We will then vary $L_x/L_z$ and $L_y/L_z$ independently, taking care to ensure that the velocity offset $v_\mathrm{s}(L_x)=s\om L_x<c$ at $x$-boundaries\footnote{This is not necessary in 2D: as mentioned by \cite{riquelme2012}, in the shearing-coordinate frame the global profile of the shearing velocity is $\vecv_\mathrm{s}/c=-\tanh(s\om x/c)\hatvece_y$, such that $v_\mathrm{s}<c$ regardless of the simulation size. This does not apply to our 3D simulations, which do not employ shearing coordinates.}.
The required grid size in each direction is thus
\begin{equation}
    N_g \simeq \frac{L}{c/\omp}  = \frac{L}{H}\frac{H}{\lmri}\frac{\lmri}{\rho_\mathrm{C}}\frac{\rho_\mathrm{C}}{c/\omp}\propto \frac{L}{H}\frac{\omega_\mathrm{C}}{\Omega_0}\frac{\sqrt{\theta_0}}{v_{\mathrm{A},0}/c},
\end{equation}
while the number of time steps is computed as in the 2D case. Since in general $\sqrt{\theta_0}/(v_{\mathrm{A},0}/c)>1$, it is clear that 3D runs will be more expensive than 2D simulations not simply due to the dimensionality but also in terms of the effect of the free parameters. For example, a large-box 3D simulation with $\omrat=15$, $v_{\mathrm{A},0}/c\simeq0.007$, $\theta_0=1/128$ in a box of size $L_z=2\lmri$, $L_y=2L_x=8\lmri$, run until $t_\mathrm{end}=15P_0$, implies evolving a grid of $\sim256\times1024\times128$ cells for $\sim 250{,}000$ time steps.

Finally, we note that the reasoning outlined above only relates to the \emph{initial conditions} of each simulation; in practice, as the system evolves, macroscopic and microscopic scales tend to become better separated as the magnetic field is amplified and $\rho_\textrm{C}$ is reduced. The MRI dynamics also helps to better resolve the kinetic scales (e.g.\ the skin depth), since these will become larger as plasma is heated during the system evolution.

\section{Two-dimensional simulations of the MRI in collisionless pair plasmas}
\label{sec:2D}

\begin{deluxetable}{cccc}[h!]
\tablecaption{List of 2D simulations}
\tablenum{1}
\tablehead{\colhead{Box size $(L_x\times L_z)$} & \colhead{$\omega_\mathrm{C}/\Omega_0$} & \colhead{$\beta_0$} & \colhead{$\lambda_\mathrm{MRI}/\rho_\mathrm{C}$}} 
\startdata
$2\times2\lmri^2$ & 6 & 624 & 2.13  \\
$2\times2\lmri^2$ & 30 & 624 & 10.67  \\
$2\times2\lmri^2$ & 120 & 624 & 42.65  \\
$2\times2\lmri^2$ & 240 & 624 & 85.30  \\
$4\times4\lmri^2$ & 30 & 624 & 10.67  \\
$8\times8\lmri^2$ & 6 & 624 & 2.13  \\
$8\times8\lmri^2$ & 15 & 624 & 5.33  \\
$8\times8\lmri^2$ & 30 & 0.1 & 842.30  \\
$8\times8\lmri^2$ & 30 & 1 & 266.36  \\
$8\times8\lmri^2$ & 30 & 5 & 119.12  \\
$8\times8\lmri^2$ & 30 & 78 & 30.16  \\
$8\times8\lmri^2$ & 30 & 624 & 10.67  \\
$8\times8\lmri^2$ & 30 & 2{,}496 & 5.33  \\
$8\times8\lmri^2$ & 60 & 624 & 21.33  \\
$16\times4\lmri^2$ & 30 & 624 & 10.67 \\
$16\times16\lmri^2$ & 30 & 624 & 10.67 \\
$32\times8\lmri^2$ & 30 & 624 & 10.67  \\
$32\times32\lmri^2$ & 30 & 624 & 10.67  \\
\enddata
\tablecomments{In all cases, $v_{\mathrm{A},0}/c=0.01$; the numerical resolution is such that $\Delta x\simeq c/\omega_\mathrm{p}$, and we employ 25 particles per cell per species.}
\label{tab:2D}
\end{deluxetable}

In this section we present the results of several PIC simulations of the collisionless MRI in 2D. The parameters for each simulation are specified in Table \ref{tab:2D}. In all runs we solve the \KSBSC~system with the relativistic PIC code \textsc{Zeltron} (\citealt{cerutti2013}). Our initial setup consists of a 2D (in the $xz$ poloidal plane) box filled with thermal plasma and a weak vertical magnetic field determined by our choice of initial parameters $\omrat$, $v_{\mathrm{A},0}$, and $\beta_0$ (or equivalently $\theta_0$). We always employ 25 particles per cell per species, and adopt a resolution such that $\Delta x\simeq c/\omega_\mathrm{p}$. Our largest 2D simulations employ computational grids with $4{,}480^2$ cells, for a total of over 1 billion particles. Finally, note that in our runs we do not perturb the initial conditions, such that the MRI grows from random, base-level fluctuations (i.e.\ numerical noise).

In this section we first briefly describe the typical stages of the evolution of the collisionless MRI in two dimensions (section~\ref{sec:evol_phases}). We then provide a physical explanation for all the features observed in this evolution (section~\ref{sec:explanation2D}). We subsequently present an exploration of the physical-parameter space characterizing kinetic shearing-box simulations of the MRI (section~\ref{sec:paramspace2D}). Then, we discuss the characteristics of MRI-driven turbulence (section \ref{sec:spectra2D}), particle energization (section \ref{sec:ene2D}), angular-momentum transport (section \ref{sec:stresses2D}), and the zero net-flux case (section \ref{sec:zeroflux}). From these runs, we identify a suitable set of parameters to be employed in the much more expensive 3D simulations presented in section~\ref{sec:3D}.

\subsection{Typical evolution of the 2D collisionless MRI}
\label{sec:evol_phases}

\begin{figure*}
\centering
\includegraphics[width=0.96\textwidth, trim={0mm 0mm 0mm 0mm}, clip]{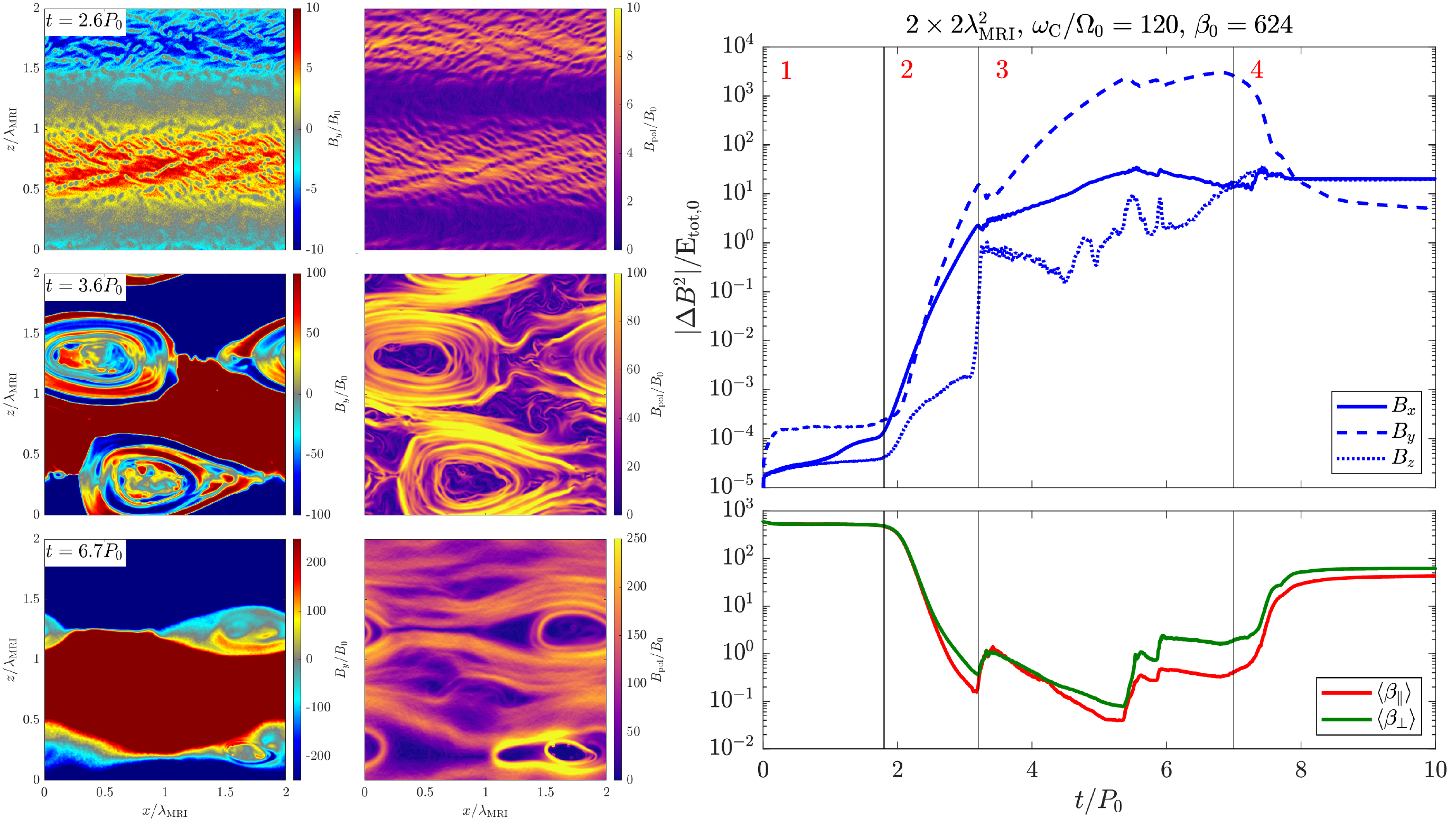}
\caption{Evolution of a typical PIC simulation of the pair-plasma MRI in a small 2D box ($2\times2\lmri^2$). Left column: spatial distribution of the out-of-plane magnetic field $B_y$ and of the in-plane field strength $B_\mathrm{pol}=\sqrt{B_x^2+B_z^2}$ at subsequent times during the simulation. Right column: evolution of the change in magnetic energy in all three components of $\vecB$ (top) and evolution of the volume-averaged $\beta_\|=\ppar/[B^2/(8\pi)]$ and $\beta_\perp=\pperp/[B^2/(8\pi)]$ (bottom). The beginning of different phases characterizing the system evolution (described in the text) is marked with vertical black lines.}
\label{fig:evol_phases_small}
\end{figure*}

Figure \ref{fig:evol_phases_small} describes the evolution of a typical two-dimensional, PIC MRI simulation of limited box size. The top-right panel shows evolution of the change (with respect to the initial value) in magnetic energy for each component of $\vecB$ (normalized to the total energy at $t=0$). This run employs the parameters $\omrat=120$, $\beta_0=624$, $v_{\textrm{A},0}/c=0.01$ in a ``small'' box of size $2\times2\lmri^2$. The evolution proceeds through several stages, with the beginning of each phase marked by vertical black lines in Figure \ref{fig:evol_phases_small}. These phases qualitatively characterize all 2D simulations with finite net flux, independently of the system size and parameters chosen for the run (aside from a less clear distinction between phases in larger systems --- see the explanation below).

\begin{figure*}
\centering
\includegraphics[width=0.96\textwidth, trim={0mm 0mm 0mm 0mm}, clip]{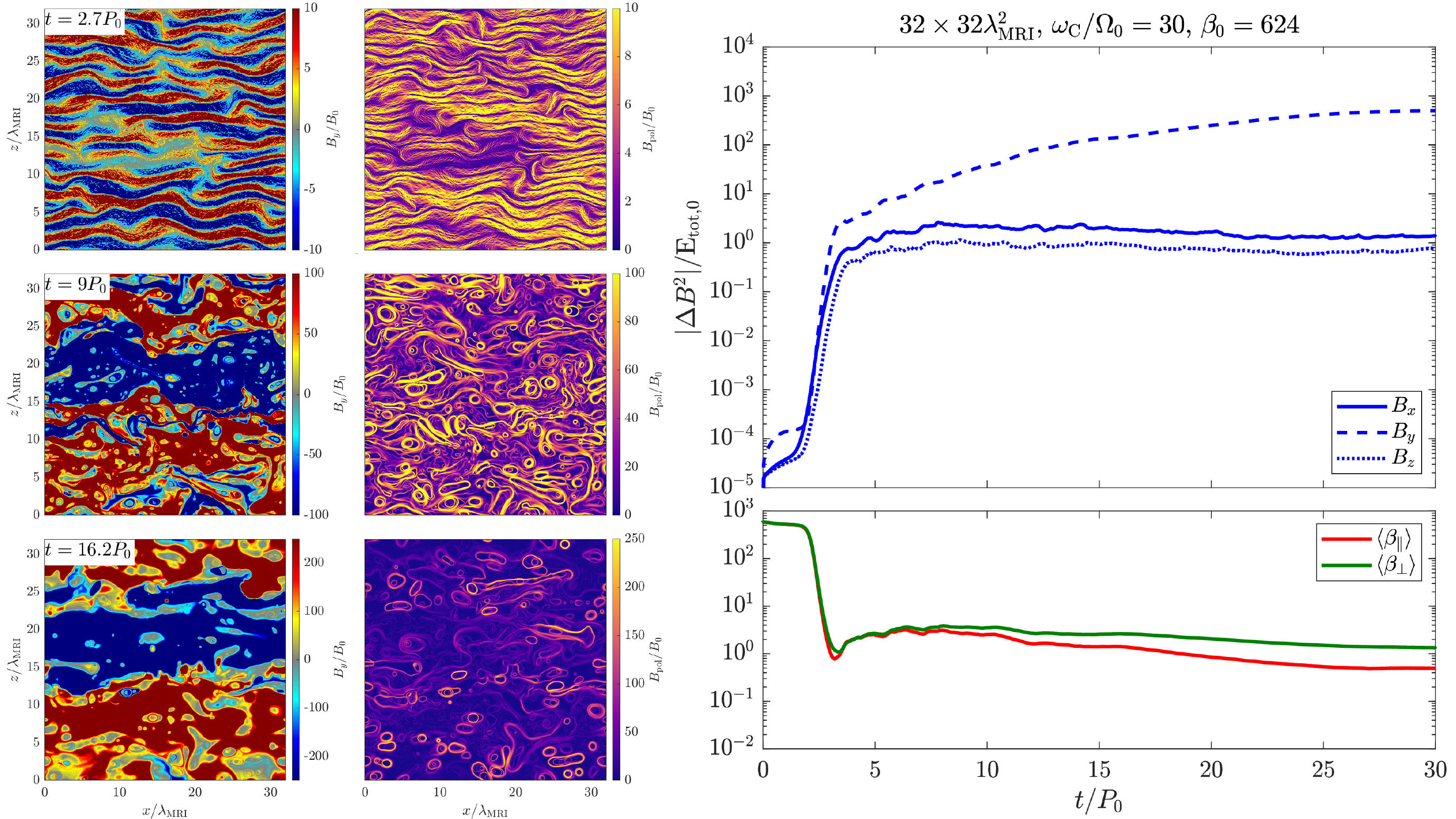}
\caption{As in Figure \ref{fig:evol_phases_small} but for a large 2D box ($32\times32\lmri^2$).}
\label{fig:evol_phases_large}
\end{figure*}

The magnetic energy grows from a pre-instability state (phase 1), which lasts for ${\sim}2P_0$, to the linear MRI growth (phase 2), where the magnetic-field strength increases exponentially by several orders of magnitude. During the linear stage, the initially vertical magnetic field acquires radial and toroidal components, resulting in bent magnetic field lines (in the $xz$-plane) that self-organize into ``channel'' flows. This is visible in Figure \ref{fig:evol_phases_small} (top left) which shows that both the in-plane magnetic field $B_\mathrm{pol}=\sqrt{B_x^2+B_z^2}$ and the out-of-plane $B_y$ present channel structures. These channels are of vertical size $\sim\lmri$, and the radial and toroidal fields are spatially uniform inside each channel; both field components reverse sign at the channel interfaces. 
The channels represent an exact nonlinear, equilibrium solution of the MHD equations (\citealt{goodmanxu1994}); this state is however unstable to linear perturbations, which can lead to further evolution through secondary (parasitic) instabilities developing on top of the primary MRI  (\citealt{goodmanxu1994,latter2009,pessahgoodman2009}). These include tearing modes, which we see developing in our simulation (Figure \ref{fig:evol_phases_small}): throughout the linear stage, the magnetic-field strength in each channel increases, and the thickness of the interfaces where $\vecB$ changes sign decreases; current sheets that form at these interfaces thin out until, at $t\simeq3 P_0$, we observe an abrupt drop in the magnetic-field energy. This moment marks the onset of magnetic reconnection, mediated by the tearing instability in the thinning current sheets, and the beginning of phase 3 of the evolution (Figure \ref{fig:evol_phases_small}, center left).

Phase 3 is the most interesting stage of the evolution. This phase is in principle representative of the typical state of turbulent accretion disks, where particles (electrons and ions) can experience acceleration. For electrons, this acceleration may become strong enough to reach radiative regimes (e.g.\ \citealt{ryan2017,dexter2021}). The nonlinear stage in phase 3 is characterized by the competition between the MRI, which actively strengthens the magnetic field in the channels, and magnetic reconnection, which depletes magnetic energy in the current sheets, thereby energizing the plasma. In addition to the reconnection process, substantial kinking of the current sheets can in principle disrupt the channels and bring the system to a saturated, fully turbulent state, which should remain sustained throughout phase~3. However, here we do not observe the development of such a turbulent state. In fact, this state is generally not achieved in small-box runs, where channels do not experience violent disruption but rather persist throughout several reconnection events (Figure \ref{fig:evol_phases_small}, bottom left), marked by repeated drops in the magnetic-energy evolution during phase~3. The continuous action of the MRI during this phase is indicated by subsequent stages of magnetic-energy growth, which, however, proceed with decreasing growth rates. This magnetic-field amplification is particularly strong along the (out-of-plane) $y$-direction; the in-plane components show a much smaller increase.

Finally, at $t\simeq7 P_0$, we observe that the increase in magnetic energy is halted, and the system rapidly relaxes to a quiescent state that remains unchanged endlessly (phase~4). This results from the MRI growth rate decreasing and the unstable MRI modes migrating to longer and longer wavelengths. Eventually, these modes grow to length scales larger than the box size, and the MRI stops. Previous works have shown that magnetic fields then reorganize into flux tubes that are stable to axisymmetric perturbations and impede further MRI growth (\citealt{riquelme2012,hoshino2013}); in our runs we observe the same behavior during phase~4 (not shown in the left panels of Figure \ref{fig:evol_phases_small}). In realistic environments, the MRI is not expected to be suppressed over such short time scales; as we discuss in the next sections, this effect results from the typical (unrealistic) choice of parameters employed in PIC runs.

As mentioned above, the nonlinear stage of the MRI (phase 3) is expected to achieve a fully turbulent state, which is not observed in any small-box run. A regime of developed turbulence can be attained only by increasing the domain size. Figure \ref{fig:evol_phases_large} shows the typical evolution of a large-box PIC MRI simulation in 2D, with parameters $\omrat=30$, $\beta_0=624$, $v_{\mathrm{A},0}/c=0.01$ and a box size $32\times32\lmri^2$, i.e.\ 16 times larger than the previous run. The left panels show that the (much more numerous) initial channels formed during the linear stage break up into smaller structures: this break-up corresponds to the beginning of the nonlinear state (i.e.\ phase 3 in the evolution). The magnetic-energy evolution (top right) shows that, within 30 orbits (notice the difference in time scales between this and the previous run), the system is slowly reaching the end of phase~3; the energy in $B_y$ is still increasing, indicating that the MRI has not been completely halted yet. However, over sufficiently long time scales, the behavior of the MRI in the large- and small-box cases is similar: in our simulations we have observed that arbitrarily large systems eventually evolve to the same final quiescent state where the MRI has stopped, with large-scale magnetic loops that persist indefinitely. The remarkable difference between the two cases is thus represented by the presence of turbulence when employing large domain sizes. Notice also how the presence of turbulence in the large-box case masks subsequent cycles of MRI growth/reconnection, which can instead be clearly identified in the small-box run.

The two simulations discussed in this section are representative of the state of the art of fully kinetic, two-dimensional MRI simulations in pair plasmas with initial vertical field. In fact, both simulations employ much larger scale separation --- in terms of $\omrat$ and/or box size --- than any previous work. Still, several questions on the physical evolution of such a system remain open: it is so far unclear what exactly causes the subsequent MRI growth-reconnection cycles during phase 3 in the small-box case (and the lack of such a clear distinction between cycles in the large-box case); it is also not apparent how the choice of physical parameters drives the transition to a turbulent state and what causes the decay of turbulence over time; moreover, it remains unexplained why, even in the large-box case, the $B_y$ energy grows much larger than the other components, without reaching a clear saturated state such as that observed in typical MHD simulations (e.g.\ \citealt{baistone2013}). In the next section, we provide a physical explanation for all these observations.

\subsection{Physics of the 2D collisionless MRI}
\label{sec:explanation2D}

In principle, the desirable outcome of any MRI simulation is a saturated, turbulent nonlinear state that lasts for long times. However, no 2D PIC simulation of the collisionless MRI with nonzero net flux presented in literature (including this work) shows this behavior.
While the linear MRI stage is generally well captured (for reasonable simulation parameters), from the onset of magnetic reconnection the overall behavior of 2D PIC runs seems to diverge from 3D MHD expectations. Figures \ref{fig:evol_phases_small} and \ref{fig:evol_phases_large} in particular show that the magnetic energy increases steadily, instead of remaining roughly at the level reached at the end of the linear stage. Furthermore, Figure \ref{fig:evol_phases_large} points out that in large systems a quasi-constant, saturated level can in fact be maintained at least for the energy in $B_x$ and $B_z$ (i.e.\ the in-plane components), but not for $B_y$.
We explain this issue, together with the overall 2D MRI dynamics in PIC, with the following reasoning.

At the end of the linear stage, to maintain a saturated state the magnetic-field amplification by the MRI should be approximately balanced by dissipative mechanisms. Strong magnetic fields induced by the MRI are in principle rapidly destroyed via reconnection, which (generally) leads to turbulence and small-scale magnetic structures, which can then seed the MRI again in a continuous cycle. In 3D ideal- or resistive-MHD simulations with sufficiently large box size, this is realized without issues (see \citealt{latter2009} for a discussion on MHD simulations with varying box size). However, 2D PIC simulations have reduced dimensionality, and also allow for the presence of pressure anisotropy. The latter is typically not included in standard MHD (with exceptions, e.g.\ \citealt{sharma2006}). The 2D geometry and the development of pressure anisotropy both play an important role in axisymmetric PIC MRI simulations.

Figures \ref{fig:evol_phases_small} and \ref{fig:evol_phases_large} (bottom-right panels) show the evolution of the volume-averaged $\beta_\|=\ppar/[B^2/(8\pi)]$ and $\beta_\perp=\pperp/[B^2/(8\pi)]$, where $\ppar$ and $\pperp$ are the parallel and perpendicular (to the magnetic field) thermal plasma pressure. We observe that nonnegligible pressure anisotropy develops during the MRI evolution, starting from the linear phase: a strong increase in $B$ is accompanied by an increase in perpendicular particle energy, due to the invariance of the magnetic moment $\mu=mu_\perp^2/(2B)$.
This anisotropy alone can drive three distinct mechanisms: i) an increase in the growth rate of tearing modes (e.g.\ \citealt{chenpalmadesso1984}); ii) the excitation of mirror modes (e.g.\ \citealt{kunz2014}), which can also couple with tearing modes, further increasing the tearing growth rate (\citealt{altkunz2019,winartokunz2022}); and iii) a decrease in the growth rate of the MRI (\citealt{quataert2002}).
During the linear stage, the first two mechanisms cause the rapid onset of magnetic reconnection: the enhancement of tearing modes occurs concurrently to the natural thinning of current sheets that form at the interface of channel flows. At the same time, the MRI growth rate is reduced both by pressure anisotropy (i.e.\ the third mechanism above) and by the fact that, by this time, $v_\mathrm{A}$ has increased substantially and the most-unstable MRI wavelength $\lmri=2\pi v_\mathrm{A}/\om$ is much larger (possibly larger than the box size) than in the beginning of the simulation. As a result, as reconnection is initiated, the MRI drastically (but temporarily) slows down.

We note that the excitation of the mirror instability mentioned above should in principle limit the growth of pressure anisotropy, since in realistic systems mirror modes can act on time scales comparable to those of the MRI. This represents a crucial point in typical PIC MRI modeling: the usual choice of physical parameters is such that, in simulations, mirror modes grow unrealistically slowly. This can be simply estimated by demanding that the mirror-instability growth rate be comparable to that of the fastest-growing MRI mode, i.e.\ $\omega_\mathrm{C}\Lambda_\mathrm{m}^2 \sim 3\om/4$ where $\Lambda_\mathrm{m}\equiv \pperp/\ppar-1-1/\beta_\perp$ (\citealt{kunz2014}). During the linear stage, we measure a typical peak value $\Lambda_\mathrm{m}\sim10^{-1}$, implying that $\omrat\sim10^2$ would be required for fast mirror modes; the measured typical value of $\Lambda_\mathrm{m}$ is even lower during the nonlinear phase, achieving $\Lambda_\mathrm{m}\leq10^{-2}$, which requires $\omrat\sim10^4$ to satisfy the condition above. As discussed in section \ref{sec:parameters}, this regime simply cannot be achieved realistically in simulations. As a consequence, the MRI dynamics in PIC runs will inevitably result in the development of larger pressure anisotropy than expected with realistic physical parameters\footnote{It is important to emphasize that with our typical choice of parameters, mirror instabilities will not be completely absent in the simulations we perform. Characteristic mirror ``stripes'' are indeed visible inside the channels during the linear stage (e.g.\ Figure \ref{fig:evol_phases_small}, top left). Our (necessarily) unrealistic scale separation simply downsizes the importance of these instabilities on the overall dynamics. This must be taken into account when interpreting the physical results of our simulations.}.

A large pressure anisotropy, combined with the migration of unstable wavelengths to larger scales, could in principle suppress the MRI from proceeding further already at the end of the linear stage. Fortunately, magnetic reconnection can at least partially counter this process, both by decreasing the field strength and by decreasing pressure anisotropy via parallel particle acceleration at reconnection sites. This is exactly what we observe for both the small- and the large-box cases right after the onset of reconnection (Figures \ref{fig:evol_phases_small} and~\ref{fig:evol_phases_large}, bottom right). In all our runs, we consistently find that reconnection events push the anisotropy in the direction opposite to that induced by the MRI, at times achieving $\ppar\simeq\pperp$. We also note that an additional (but probably largely unimportant) mechanism of anisotropy reduction may stem from the conservation of the invariant $\ppar B^2/\rho^3$ (e.g.\ \citealt{sharma2006} and references therein), which further increases $\ppar$ when magnetic energy is depleted during reconnection (\citealt{hoshino2013}).

As reconnection dissipates magnetic energy and reduces pressure anisotropy, the system returns to conditions that allow for the MRI to develop. At the same time, tearing modes are progressively quenched due to the thickening of currents sheets at channel interfaces.
This temporarily halts the reconnection process. The MRI is therefore free to start again, commencing a new exponential-growth stage (as can be observed during phase 3 of the evolution in Figure \ref{fig:evol_phases_small}, top right). However, recall that by this time $v_\mathrm{A}$ is much larger than in the initial state, and the most-unstable MRI modes may no longer fit inside the simulation box. The MRI growth rate is then appreciably smaller than the asymptotic value~$3\om/4$. This new MRI phase thus amplifies magnetic fields more slowly, but the overall dynamics is similar to that of phase~2: $\pperp>\ppar$ anisotropy is again driven due to $\mu$ conservation; current sheets become thinner at the channel interfaces; magnetic reconnection eventually disrupts the sheets, and so on. During each cycle, the MRI growth rate decreases as $v_\mathrm{A}$ increases further; in addition, while reconnection tends to reestablish isotropy, in our 2D runs we observe that the $\pperp>\ppar$ anisotropy accumulates across subsequent cycles, increasing over time. Reconnection and (unrealistically slow) mirror modes thus fail to keep the MRI-driven anisotropy limited. At each subsequent cycle, the accumulated effect of the increase in $\lmri$ and in the pressure anisotropy pushes the MRI to smaller and smaller growth rates. Eventually, all unstable MRI wavelengths grow beyond the box size, and the process cannot remain self-sustaining. The MRI completely stops and reconnection dissipates magnetic energy at the remaining current sheets. The system then reaches the quiescent state described in this and previous works (\citealt{riquelme2012,hoshino2013}).

The reasoning outlined above explains why we observe the MRI slowing down and eventually stopping altogether in 2D simulations of any size: the lack of saturation in magnetic energy (especially in $B_y$) and unregulated pressure anisotropy, over time, cause the migration of MRI wavelengths to scales larger than the box size. This could be mitigated if magnetic-field saturation were achieved after the linear stage, and if pressure anisotropy could be efficiently limited. This is indeed better verified in larger boxes: indeed, even though the end-state of the evolution is the same, qualitative differences between the nonlinear evolution of the MRI arise between large and small simulations. The differences are due to three main reasons: i) In small boxes a large number of tearing modes are suppressed due to the limited separation between current-sheet length (${\sim}L_x$) and thickness (${\sim}c/\omega_\mathrm{p}$), which slows reconnection and allows for magnetic fields to grow stronger via the MRI. ii) Large boxes also allow for the formation of a larger number of channels along $z$, and for the excitation of the DKI in long current sheets, which is instead suppressed in small simulations (as noted by \citealt{inchingolo2018} as well as in section \ref{sec:aspectratio2D} of this work). The DKI in larger boxes significantly contributes to breaking up and mixing the more numerous channel flows, inducing a turbulent state. iii) Relatedly, the lack of developed turbulence in small boxes results in less efficient reduction of pressure anisotropy via turbulent magnetic-field dissipation and pitch-angle particle scattering. Magnetic fields and pressure anisotropy thus grow much faster when the MRI evolves in smaller boxes, resulting in a more rapid migration of unstable modes to wavelengths larger than the box size. This migration is slower in larger boxes, which can also fit larger wavelengths; as a consequence, the nonlinear turbulent stage is sustained for far longer.

The arguments above still do not explain why the MRI, in all 2D simulations, fails to saturate at an approximately constant level during the nonlinear stage, but rather induces a continuous growth in the magnetic energy (specifically in $B_y$, see Figure \ref{fig:evol_phases_large}, top right). This occurs irrespective of the box size, and represents a major discrepancy with 3D MHD simulations. This issue was not addressed in previous works, but it has in fact a clear explanation: in 2D axisymmetric simulations, the out-of-plane, toroidal magnetic field $B_y$ \emph{cannot be subjected to magnetic reconnection}. This is a fundamental problem for maintaining a saturated nonlinear state\footnote{Especially because the MRI is particularly efficient at amplifying magnetic fields in the $y$-direction, as can be inferred by simply observing the \KSBSC~system --- the only first-order (in $\om$) source term appears in the equation for $B_y$.}: without reconnection along $y$, the amplification of $B_y$ via the MRI is practically unimpeded. The only mechanism counteracting this effect is represented by advection, mixing, and reconfiguration of opposite-sign $B_y$ regions; but as observed in Figure \ref{fig:evol_phases_large} (left panels), even for large system sizes the long-term evolution of the system inevitably consists of macroscopic $B_y$ structures emerging from the turbulent state as the MRI is halted.

The reasoning outlined above leads us to conclude that, in 2D, achieving a minimally sustained turbulent state is in fact possible, but only in simulations with sufficiently large boxes (in accordance with \citealt{inchingolo2018}). However, our analysis also shows that the ultimate fate of any 2D simulation is to reach a nonlinear stage that does not truly saturate, but rather features a strong, unimpeded increase in $B_y$ eventually resulting in the suppression of the MRI as unstable modes migrate to longer and longer wavelengths. We note that the absence of saturation for $B_y$ was previously observed by \cite{riquelme2012}, who reported that 2D MHD simulations also display a similar behavior. In section \ref{sec:3D}, we will show that the issues outlined above can be ameliorated in 3D simulations, where reconnection along $y$ is allowed.

\subsection{Parameter-space exploration}
\label{sec:paramspace2D}

Having established the general features of 2D kinetic MRI simulations, we now present a detailed parameter-space scan where we exhaustively explore the effect of four specific parameters. As illustrated in the previous sections, we expect 2D simulations to be fundamentally different from 3D runs; our exploration is nevertheless instructive, since many features of the 2D dynamics are still present in 3D (see section \ref{sec:3D}). In particular, 2D results help us choose a reasonable set of parameters to be employed in 3D runs, which are much more expensive and are necessarily more limited in the parameter space that can be explored. The parameters we explore are:
\begin{itemize}
    \item The separation between macroscopic and microscopic temporal scales, parametrized by the ratio between the initial cyclotron frequency and the rotation frequency, $\omega_\mathrm{C}/\om$; previous works have explored the range $\omega_\mathrm{C}/\om\simeq10\mbox{--}20$ (\citealt{riquelme2012,hoshino2013,hoshino2015,inchingolo2018}). Here, we will employ a much larger range of values.

    \item The separation between physical domain size and (macro-) instability scales in terms of the box size (e.g.\ in the vertical direction, $L=L_z$) compared to the initial most-unstable MRI wavelength $\lambda_\mathrm{MRI}$. Previous 2D works have advocated the need to employ values $L/\lambda_\mathrm{MRI}>8$ to reach convergence in the results of kinetic MRI simulations (\citealt{inchingolo2018}).
    
    \item The initial plasma temperature, or, equivalently, the initial thermal-to-magnetic pressure ratio $\beta_0= 8 \pi p/B_0^2$ (with $p=p_e+p_i$ the total pressure). This was previously investigated in the range $\beta_0\simeq100\mbox{--}6{,}000$ (\citealt{hoshino2013}).
    
    \item The aspect ratio $L_x/L_z$ of the simulation box. In particular, we will explore the case of boxes elongated along $x$. This point was not explored in previous kinetic-MRI works.
\end{itemize}

In this section, we will focus on assessing convergence in the results based on the magnetic-energy amplification. In light of the discussion presented in section~\ref{sec:evol_phases}, we opt for excluding the $B_y$ component from the convergence analysis, since the lack of reconnection along $y$ in 2D simulations fundamentally affects the evolution of magnetic fields in the direction perpendicular to the simulation plane.
The analysis of other important physical processes (turbulence, nonthermal particle acceleration, angular momentum transport) will be presented in section \ref{sec:spectra2D} and onward.

\begin{figure*}
\centering
\includegraphics[width=1\textwidth, trim={0mm 0mm 80mm 0mm}, clip]{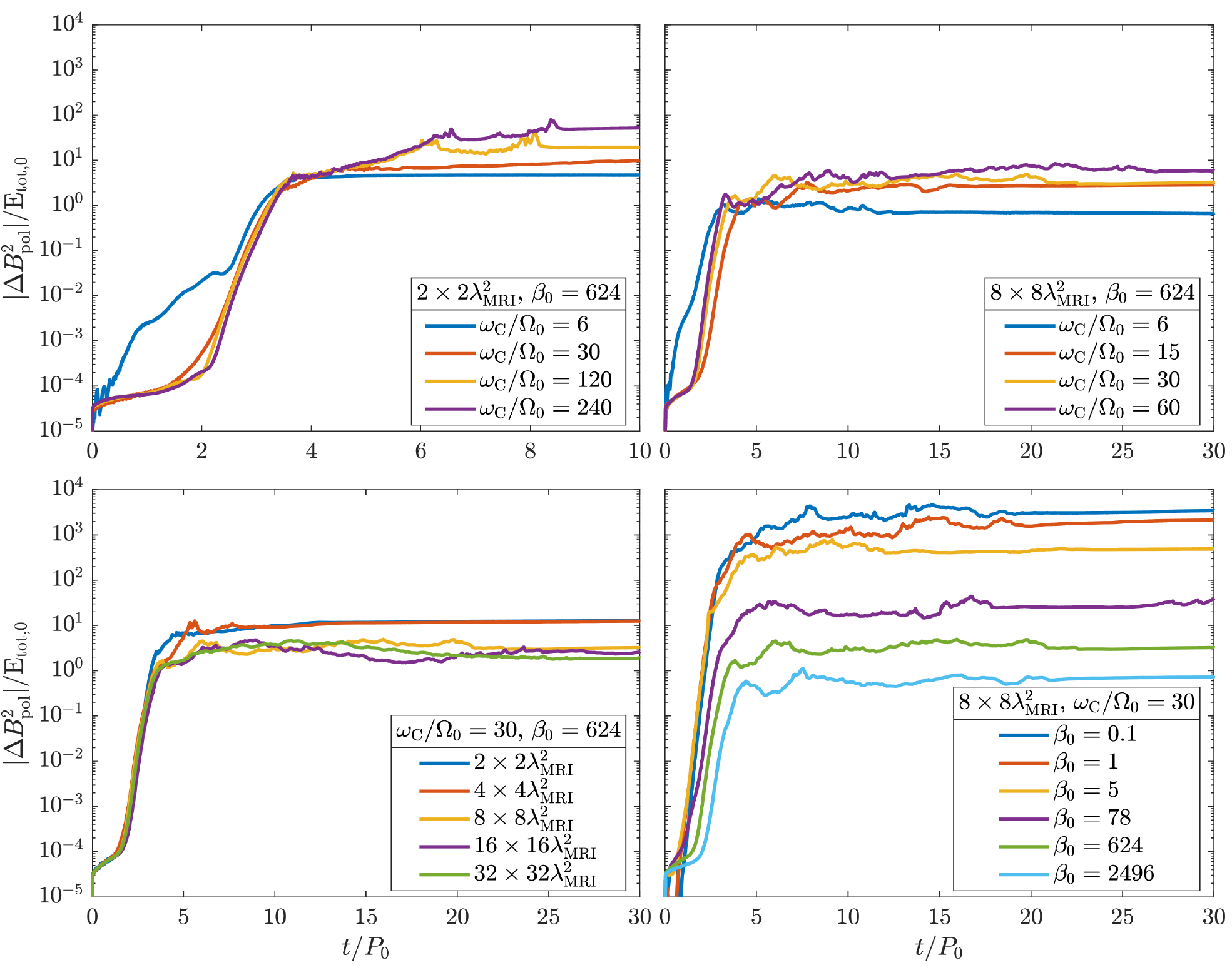}
\caption{Evolution of the change in the in-plane magnetic energy for our parameter-space exploration of the 2D pair-plasma MRI. Top left and top right: varying $\omrat$ at a fixed $\beta_0$ for a small- and a medium-size simulation box, respectively. Bottom left: varying the box size at a fixed $\omrat$ and $\beta_0$. Bottom right: varying $\beta_0$ at a fixed $\omrat$ for a medium-size box.}
\label{fig:2Dcomp}
\end{figure*}

\subsubsection{Effect of the temporal-scale separation ($\omrat$)}
\label{sec:omrat2D}

We begin our investigation by evaluating the impact of the $\omrat$ ratio. At first, we will focus on a limited box size; although small-box runs differ substantially from the large-box case (see section~\ref{sec:explanation2D}), they are still useful to understand the effect of scale separation when only MRI and reconnection are present, and fully developed turbulence is excluded. Moreover, the small-box case allows us to explore a much wider parameter space, since larger $\omrat$ values imply larger $\lmri/(c/\omega_\textrm{p})$ ratios and therefore more resolution needed per $\lmri$ (see section~\ref{sec:parameters}). Limited box sizes will be the necessary condition for affordable 3D runs and it is therefore interesting to analyze this case in 2D before performing 3D simulations. Later, we will consider larger boxes in 2D resulting in a more realistic system evolution.

For this first analysis, our box has size $2\times2 \lmri^2$. We run four 
simulations with $\omrat=6,30,120,240$, keeping the initial $\beta_0=624$ (including both electrons and positrons; the corresponding temperature is $\theta_0=1/32$ for each species) and the initial Alfv\'en speed $v_{\textrm{A},0}/c=0.01$. The large range of frequency ratios we explore allows us to assess the effect of the separation between fluid (of size $\lmri$) and kinetic (of size $\rho_\textrm{C}$) scales when these differ substantially. In particular, our choice of parameters results in $\lmri/\rho_\textrm{C}\simeq 2,10,40,80$ for the four runs.

In Figure \ref{fig:2Dcomp} (top left) we show the evolution of the change in magnetic energy (excluding the energy in $B_y$ as mentioned in the previous section), normalized by the initial total energy of the system, for the four runs until $t=10P_0$. All simulations present the common features outlined in section \ref{sec:evol_phases}, i.e.\ four typical phases of evolution. Each phase develops similarly as $\omrat$ increases, but fundamental differences are present. In particular, the linear phase is well captured only for $\omrat\ge30$. As we increase $\omrat$ from 6 to 240, we observe that the saturation level of the magnetic energy (during the nonlinear stage) increases by roughly one order of magnitude. Larger $\omrat$ values also correspond to an increase in the duration of the nonlinear stage (before the flattening of the curves). 
The cases $\omrat=120$ and $\omrat=240$ exhibit very small differences in the results; the nonlinear stage has approximately the same duration, and the saturation level is approximately the same (at least in the early nonlinear stage, before the MRI slows down). This indicates that, with a fixed box size, the results are well converged\footnote{Recall that in these plots we have excluded the $B_y$ component from the total magnetic energy.} with respect to an increase in $\omrat$, even though the latter is still orders of magnitude far from actually realistic values (e.g.\ $\omrat\sim10^7$ around M87*).

The small-box case is not representative of a realistic MRI evolution, as it lacks important dynamics such as fully developed turbulence. For this reason, we run another series of simulations, employing a larger box of size $8\times8 \lmri^2$ and the values $\omrat=6,15,30,60$ (all other parameters are the same as for the small-box case). The results are shown in Figure \ref{fig:2Dcomp} (top right; note the longer timescales on the $x$-axis). Qualitatively, this medium-box case presents the same trend displayed by the previous set of runs: the linear phase is well captured  for $\omrat\ge15$, and the nonlinear saturation level and nonlinear stage duration increase with~$\omrat$. Additionally, this case displays a full break-up of channels and a turbulent behavior throughout the nonlinear stage. We observe that the magnetic-energy saturation is roughly the same for $\omrat\ge15$; during the nonlinear stage, the results appear well converged between $\omrat=30$ and $\omrat=60$, and only diverge at late times (when the MRI is slowing down). We conclude that, when fixing the box to larger sizes, the results converge more quickly, as $\omrat$ increases, than in the small-box case.

This analysis shows that, when keeping the box size fixed, solely increasing the temporal-scale separation produces a nonlinear MRI evolution that converge for ``large enough'' $\omrat$, both for small box sizes ($\sim2\times2\lmri^2$) and for larger boxes ($\gtrsim8\times8\lmri^2$). However, the converged state is qualitatively different in the small-box and in the large-box case, and the value of $\omrat$ at which convergence is reached also depends on the box size. These considerations are likely to carry over to 3D simulations, which will be explored in section~\ref{sec:3D}.

\subsubsection{Effect of the box size ($L/\lmri$)}
\label{sec:boxsize}

In this section, we explore the effect of an increasing box size~$L/\lmri$. We run five simulations, progressively enlarging the simulation domain from $2\times2\lmri^2$ up to $32\times32\lmri^2$. In all cases we fix $\omrat=30$, $\beta_0=624$, and $v_{\mathrm{A},0}/c=0.01$. Figure \ref{fig:2Dcomp} (bottom-left panel) reports the change in magnetic energy for all runs until $t=30 P_0$. We observe that larger domains tend to increase the duration of the nonlinear stage; for a domain size $\gtrsim16\times16\lmri^2$, the system is not yet completely quiescent at 30 orbits. In addition, the saturation level is roughly the same for system sizes of at least $8\times8\lmri^2$, which suggests that convergence has been reached. This observation appears consistent with the conclusions presented in \cite{inchingolo2018}, but there are fundamental differences between their considerations and the ones presented here. 

First, our runs employ a larger $\omrat=30$ ($\omrat=10$ was chosen for the simulations in \citealt{inchingolo2018}). This difference may appear unimportant, but we have shown in Figure \ref{fig:2Dcomp} (top right) that $\omrat<15$ does not reliably produce converged magnetic-field amplification for a box size of $8\times8\lmri^2$. Second, \cite{inchingolo2018} claim that for a box size of at least $8\times8\lmri^2$, the \emph{total} magnetic energy (including the out-of-plane $B_y$) converges; we do not observe such convergence in our results simply by increasing the box size. Rather, we see that the \emph{in-plane} magnetic-field strength reaches convergence for sufficiently large boxes. Adding the out-of-plane field to this analysis does not show convergence even for very large values of~$\omrat$.
Third, \cite{inchingolo2018} argue that an indicator of convergence is represented by a volume-averaged $v_\mathrm{A}/c<1$ (using the nonrelativistic expression for $v_\mathrm{A}$), which they observe for box sizes of at least $8\times8\lmri^2$. However, their analysis is limited to rather short time scales; in all our 2D numerical experiments (regardless of the box size), we invariably find that, on average, $v_\mathrm{A}/c>1$ over sufficiently long times, due to a lack of saturation in $B_y$ (as we have argued in section~\ref{sec:explanation2D}). This specific issue can be mitigated in 3D simulations, as we will show in section~\ref{sec:2Dvs3D}.

From the analyses carried out here and in the previous section, we conclude that a combination of ``large enough'' frequency ratio and system size is necessary for two reasons, namely i) to reach convergence (at least in terms of the in-plane magnetic-field amplification), i.e.\ to obtain results that become insensitive to further increase in both $\omrat$ and $L/\lmri$; and ii) to include all the relevant physics (such as turbulence) with enough scale separation. Solely varying one of these parameters does not ensure that the results will be converged and also physically valid. From our results, it appears that $\omrat=30$ and a system size of $8\times8\lmri^2$ suffice in respecting this criterion while attaining satisfactory convergence in the results. However, we will show in section \ref{sec:ene2D} that extra care must be taken when choosing simulation parameters if one considers their effect on particle energization. Finally, we note that the aforementioned $\omrat$ and box size may still impose excessive computational costs in 3D runs. However, we will show in section \ref{sec:3D} that the different nature of the 3D MRI dynamics may allow for less stringent conditions on the physical parameters.

\subsubsection{Effect of the initial plasma-$\beta$}

\begin{figure*}
\centering
\includegraphics[width=1\textwidth, trim={0mm 50mm 0mm 0mm}, clip]{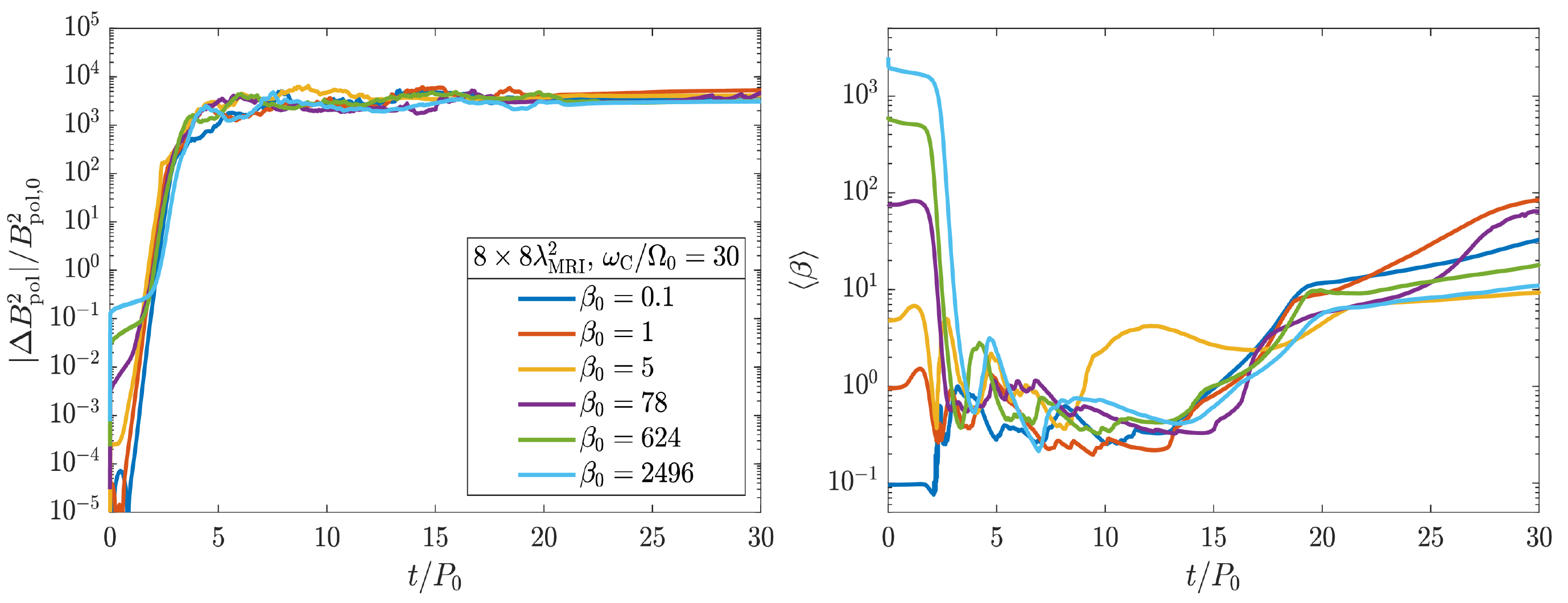}
\caption{Comparison of 2D pair-plasma simulations of the MRI with variable $\beta_0$ in a medium-size box and fixed $\omrat$. Left: in-plane magnetic-energy variation normalized over the initial field strength. Right: corresponding evolution of the volume-averaged $\beta$ in time.}
\label{fig:8x8_beta}
\end{figure*}

In this section, we assess the effect of $\beta_0$ on the system evolution. This was previously investigated by \cite{hoshino2013} who explored the range $\beta_0\simeq 100\mbox{--}6{,}000$; here we run six simulations with $\beta_0=0.1,1,5,78,624,2{,}496$. For consistency with the parameters employed in the previous sections, we choose $\omrat=30$, $v_{\mathrm{A},0}/c=0.01$, and a box size of $8\times8\lmri^2$. Fixing these parameters implies that we are bound to change $\beta_0$ by changing the initial plasma temperature $\theta_0$ correspondingly (see section~\ref{sec:parameters}); to keep the initial temperature sufficiently far from relativistic values, we take a maximum $\beta_0=2{,}496$ (corresponding to $\theta_0=1/8$).

The results of the six runs are shown in Figure \ref{fig:2Dcomp} (bottom right). The magnetic-field strength at saturation appears to decrease steadily as $\beta_0$ increases; however, this is simply an effect of our choice of normalization over the initial total energy. In fact, we observe that the total (i.e.\ between initial state and the saturated nonlinear stage) magnetic-field amplification $B^2/B_0^2$ is roughly the same in all these simulations\footnote{This observation agrees with the results presented in the previous section: the magnetic-field amplification depends on $\omrat$, which does not change during this scan.}; since the energy of the system is dominated by the thermal energy, the ratio between magnetic energy and total initial energy will decrease with increasing $\theta_0$ (and therefore $\beta_0$).

To better evaluate convergence in this case, we thus consider the quantity $B_\mathrm{pol}/B_{\mathrm{pol},0}$. This is shown in Figure \ref{fig:8x8_beta} (left panel), which demonstrates that the amplification of the in-plane magnetic field is well converged across all simulations, and does not depend on the initial $\beta_0$.
In the same Figure (right panel) we also plot the value of the volume-averaged $\beta$ in time. Regardless of $\beta_0$, all simulations reach $\langle\beta\rangle\lesssim1$ at the beginning of the nonlinear stage, followed by a gradual increase until $\langle\beta\rangle\sim1\mbox{--}10$ at $t\simeq20 P_0$, marking the end of the turbulent stage and the emergence of quiescent magnetic loops. The fact that $\langle\beta\rangle$ at saturation does not depend on $\beta_0$ and is roughly consistent (around $0.1\mbox{--}1$) across simulations, together with the fact that the magnetic-field amplification is also comparable, implies that the the particle energy at saturation is approximately the same for all simulations. This signifies that the particle-energy gain (i.e.\ measured as $\theta/\theta_0$ at saturation) is smaller if the initial $\theta_0$ is larger. This also explains the behavior of the $\beta_0=0.1$ case in which $\langle\beta\rangle$ increases between the initial state and the saturated value, instead of decreasing: in this case, $\theta/\theta_0$ is larger than $B^2/B^2_0$ and thus $\beta/\beta_0>1$ at saturation. This dynamics can be understood by the following argument: since the magnetic energy at saturation is the same for all cases, the amount of energy available for dissipation via reconnection (and thus convertible into kinetic energy) is the same. The net increase in particle energy will thus be equal across all cases, but the ratio between final and initial energy will be larger if $\theta_0$ (hence $\beta_0$) is smaller.

These results show that no substantial differences arise for values $\beta_0\simeq0.1\mbox{--}2{,}500$, at least in terms of global quantities such as the magnetic-field amplification and the plasma-$\beta$ at saturation. The system evolution appears very similar particularly during the nonlinear phase, in which the value of $\beta$ may influence mechanisms such as turbulent particle energization. Our results are in good agreement with those presented in \cite{hoshino2013}, with the caveat that, in their case, $v_\mathrm{A,0}$ was also varied alongside $\theta_0$ to change the initial $\beta_0$. We have also explored the case $\beta_0\sim 0.1$, while \cite{hoshino2013} employed a minimum value $\beta_0\sim96$. Our small-$\beta_0$ case is particularly interesting as it produces the same results found for much larger $\beta_0$ cases; in literature, it is instead generally thought that a small $\beta_0$ should completely impede the development of the MRI. However, this only applies when the vertical box size is comparable to or smaller than $H$: indeed, choosing $H/L_z=1$ and varying $\theta_0$ (and hence $\beta_0$) while keeping $\omrat$ and $v_\mathrm{A,0}$ fixed could produce a situation in which the MRI cannot develop because the vertical box size becomes smaller than $\lmri$. In our case, however, we have ignored this constraint and taken $L_z=8\lmri$ (which can result in $L_z\gg H$, e.g.\ in our $\beta_0=0.1$ case). Thus, in all our simulations the vertical box size can still fit channel modes and the MRI can develop regardless of the initial $\beta_0$. Here, we choose to violate $H/L_z\ge1$ because when this condition is imposed (e.g.\ in our 3D simulations, section \ref{sec:3D}), the chosen value of $\beta_0$ results in additional constraints on length and time scales (see section \ref{sec:parameters}). We have however verified that our results do not qualitatively change when respecting the $H/L_z\ge1$ condition.

\subsubsection{Effect of the box aspect ratio}
\label{sec:aspectratio2D}

\begin{figure*}
\centering
\includegraphics[width=1\textwidth, trim={0mm 0mm 20mm 0mm}, clip]{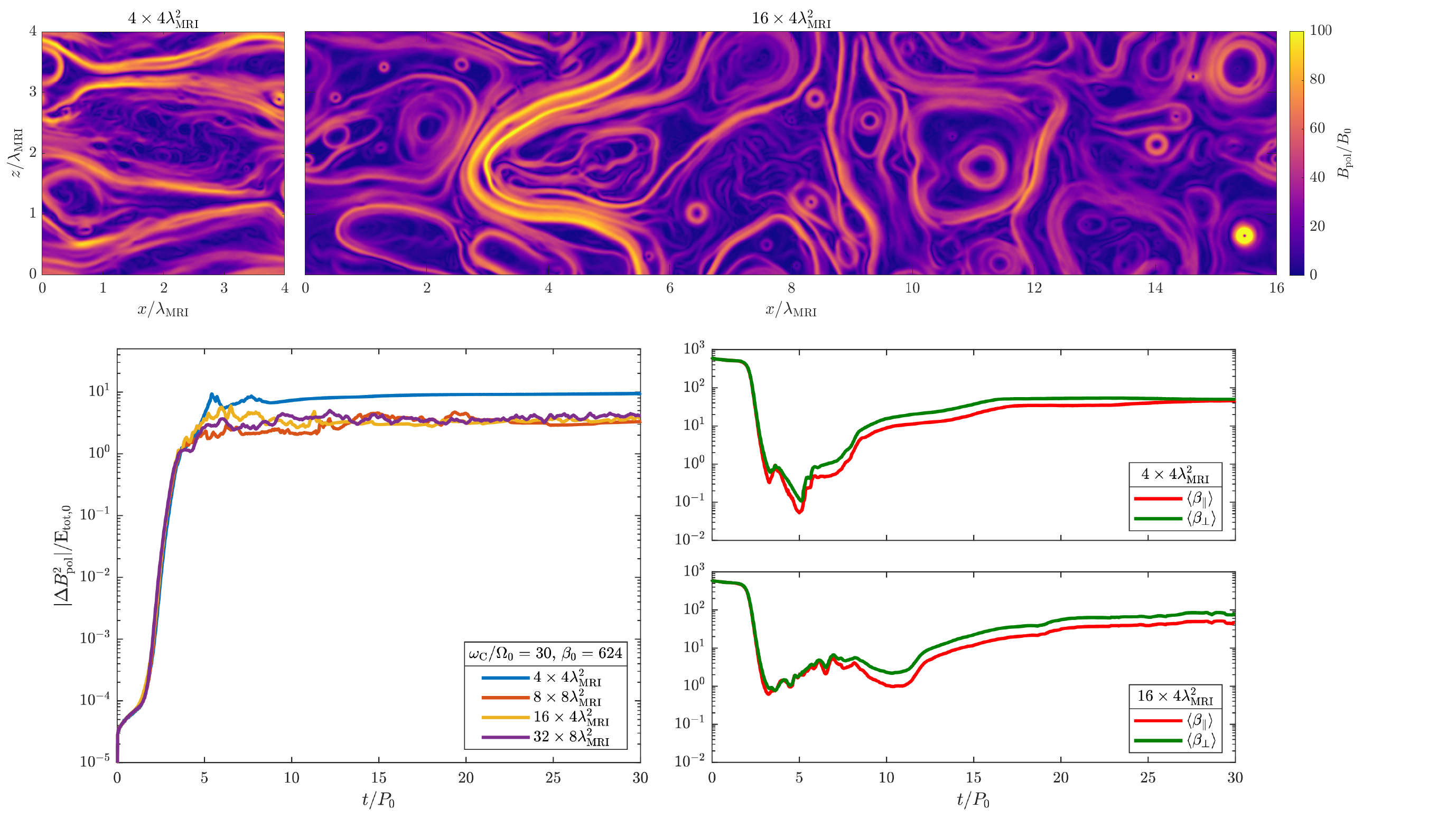}
\caption{Comparison of 2D pair-plasma simulations of the MRI with variable box aspect ratio. Top: spatial distribution of the in-plane field strength at $t=4.2P_0$ for a simulation with box size $4\times4\lmri^2$ (left) and $16\times4\lmri^2$ (right), with $\omrat=30$, $\beta_0=624$. The elongated-box case shows turbulent structures that do not develop in the corresponding square-box case. Bottom left: evolution of the change in magnetic energy for runs with variable box size and aspect ratio at fixed $\omrat$ and $\beta_0$. Bottom right: evolution of the volume-averaged $\beta_\|$ and $\beta_\perp$ for the square-box case $4\times4\lmri^2$ and for the elongated-box case $16\times4\lmri^2$.}
\label{fig:aspectratio}
\end{figure*}

Finally, we analyze the effect of the box aspect ratio ($L_x/L_z$) in 2D MRI simulations, which has not been explored in previous kinetic studies. This analysis is motivated by multiple reasons: first, 3D MHD simulations have shown that domains with aspect ratio $L_x/L_z\ne1$ may produce substantially different results in terms e.g.\ of the efficiency of angular-momentum transport (e.g.\ \citealt{bodo2008,latter2009} and references therein); second, as we (and previous works) have shown, turbulence can more easily develop in large-box simulations that fit a larger number of channel modes (via large $L_z/\lmri$ ratios) and that allow for the DKI to contribute to channel break-up and mixing (via large $L_x/(c/\omega_\textrm{p})$ ratios). However, if enough channel modes fit in $L_z$, it may not be necessary to further enlarge the simulation box in both $x$ and $z$; the development of the DKI may be achieved solely by enlarging the box along $x$, since current sheets develop only along this direction (in~2D). This is attractive, since it allows for simulations with large $L_x$ (resulting in developed turbulence) without the need for a correspondingly large~$L_z$. This is also of great importance for 3D simulations where we will demand that $H/L_z=1$: as discussed in section \ref{sec:parameters}, this condition implies that the chosen value of $L_z/\lmri$ will determine the computational cost (in terms of resolution and time steps). Keeping a reduced $L_z$ while still including all the relevant physics by enlarging $L_x$ (and $L_y$, in 3D) can greatly reduce the computational effort required.

To assess the possible advantages of aspect ratios $L_x/L_z>1$, we compare the square-box cases of size $4\times4\lmri^2$ and $8\times8\lmri^2$ with simulations where we choose $L_x/L_z=4$, resulting in boxes of size $16\times4\lmri^2$ and $32\times8\lmri^2$. We fix the simulation parameters to $\omrat=30$, $\beta_0=624$, and $v_{\mathrm{A},0}/c=0.01$. The results are presented in Figure \ref{fig:aspectratio}. The top panels show a snapshot of the in-plane field strength at $t=4.2 P_0$ (shortly after the beginning of the nonlinear stage) for a square-box simulation of size $4\times4\lmri^2$ and for an elongated-box run of size $16\times4\lmri^2$. Turbulent structures are visible in the distribution of $B_\mathrm{pol}$ when $L_x/L_z>1$; such structures do not develop in the square-box case, indicating that a limited $x$-extent inhibits important dynamics leading to the break-up and mixing of channel flows. This difference clearly appears in the bottom-left panel of the same Figure, which shows the evolution of the in-plane magnetic energy for all four runs. While the linear MRI stage is well captured in all cases, the nonlinear stage proceeds substantially differently for a $4\times4\lmri^2$ box, with the system displaying the small-box behavior discussed in section~\ref{sec:omrat2D}. The evolution of the elongated box of size $16\times4\lmri^2$, instead, is much more similar to the larger-box case $8\times8\lmri^2$, resulting in a saturated, turbulent nonlinear stage. Similarly, the elongated case $32\times8\lmri^2$ shows a much longer nonlinear stage (which lasts until the end of the run) than the corresponding square-box case.

As a further confirmation of the presence of additional channel break-up and mixing dynamics, in Figure \ref{fig:aspectratio} (bottom-right panels) we plot the evolution of the volume-averaged $\beta_\|$ and $\beta_\perp$ for the cases $4\times4\lmri^2$ and $16\times4\lmri^2$. In section \ref{sec:explanation2D}, we argued that the duration and saturation level of the nonlinear stage are heavily influenced by the efficiency of reconnection to limit magnetic-field amplification and pressure anisotropy; here, we observe that the magnetic energy saturates at lower values, and that the pressure anisotropy is indeed better quenched in the elongated-box case, which allows for stronger reconnection and more developed turbulence. The migration of MRI modes toward long wavelengths is therefore slower in elongated boxes. The square-box case, instead, prevents fast DKI modes and very rapidly builds strong magnetic fields and large pressure anisotropy right after the first onset of reconnection. These results prove that simulations in a box of limited vertical extent can still reproduce all the MRI physics of interest (including a saturated turbulent state) simply by increasing the domain's spatial extent in the direction(s) normal to~$z$. For 3D simulations, this will be particularly important to choose the optimal box geometry for physically meaningful simulations that are also computationally affordable.

\subsection{Spectra of 2D MRI-driven turbulence}
\label{sec:spectra2D}

\begin{figure}
\centering
\includegraphics[height=0.80\columnwidth, trim={0mm 0mm 10mm 0mm}, clip]{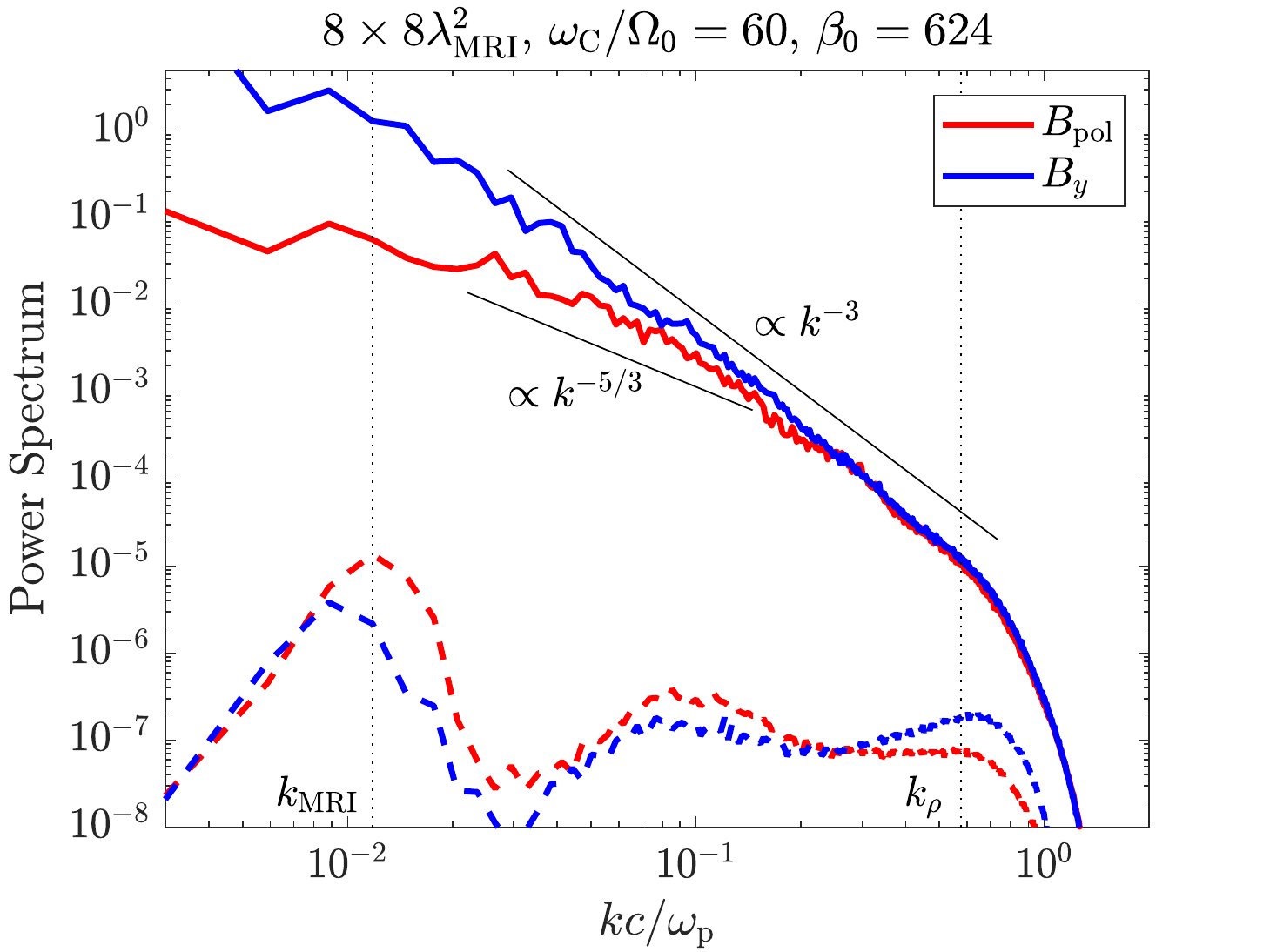}
\caption{Isotropic power spectrum of the in-plane (red lines) and out-of-plane (blue lines) magnetic field for a 2D pair-plasma MRI simulation with parameters $\omrat=60$, $\beta_0=624$, and box size $8\times8\lmri^2$. The spectra during the linear stage (dashed lines) are peaked around the most-unstable MRI wavenumber $k_\mathrm{MRI}$. At late times during the nonlinear stage (solid lines) the spectra feature characteristic slopes indicating an inertial range, and a clear spectral break in the vicinity of the average Larmor-radius wavenumber $k_\rho$.}
\label{fig:spectra2D}
\end{figure}

In this section, we briefly analyze the power spectrum of the magnetic-field component in the simulation plane ($B_\mathrm{pol}$) as well as in the out-of-plane direction ($B_y$).
This analysis provides a clear indication of turbulent activity during the system evolution, and allows us to infer whether MRI-driven turbulence develops at all.

In Figure \ref{fig:spectra2D} we plot the isotropic power spectrum for a simulation run in a domain of size $8\times8\lmri^2$ with $\omrat=60$,  $\beta_0=624$, and $v_{\mathrm{A},0}/c=0.01$. We choose this particular set of parameters, among the various cases explored in the previous sections, as it represents a good compromise between box size and separation between macroscopic and microscopic scales. In the plot, the dashed colored lines represent the spectrum of $B_\mathrm{pol}$ (red) and $B_y$ (blue) measured during the linear MRI phase. Both spectra are strongly peaked around the most-unstable wavenumber $k_\mathrm{MRI}$ (marked by a vertical dotted line), indicating active MRI growth at the fastest rate. A second, less pronounced peak is visible at this time in both spectra around $kc/\omp\simeq0.1$; this is compatible with the most-unstable wavenumber of mirror modes, suggesting that mirror activity is present at this time in the system evolution (as also discussed in section \ref{sec:explanation2D} and as noted by \citealt{inchingolo2018}). We repeat the measurement of magnetic-field spectra at a late time $t\simeq10 P_0$ (solid lines), when the system has entered the nonlinear stage. These spectra feature characteristic slopes that identify an inertial range: at length scales larger than the typical Larmor radius, i.e.\ $k<k_\rho\equiv1/\rho_\mathrm{C}$, the $B_\mathrm{pol}$ spectrum has a shallow slope that could be comparable with $k^{-5/3}$, which hints at a cascade process akin to the typical strong-turbulence MHD picture (e.g.\ \citealt{goldreichsridhar1995}). At $k\simeq k_\rho/2$, this spectrum changes to a slope close to $k^{-3}$ or slightly steeper, indicating the transition from Alfv\'enic turbulence to a kinetic range. The $B_y$ spectrum is instead characterized by this $\propto k^{-3}$ slope (or steeper) at all scales, in agreement with previous 2D studies (\citealt{inchingolo2018}). Power in both spectra falls rapidly at large $k$ as the cascade reaches sub-Larmor scales, where strong magnetic energy dissipation is expected to occur. We have verified that numerical dissipation at the grid scale is minimal; we elaborate on this in Appendix~\ref{app:energycons}.

The measured spectra suggest that the system has indeed reached a state of well-developed turbulence during the nonlinear stage. In general, we measure the same spectral slopes in the majority of the simulations explored in our parameter-space scan. In a few cases, however, the results deviate from this expectation: in particular, we find that (square) system sizes $\leq 4\times4\lmri^2$ and scale separations $\omrat\leq15$ do not produce spectra with such clear power-law behavior, hinting at a turbulent cascade that is not fully developed.
In the previous sections, we have argued that such cases lack the action of certain physical mechanisms such as drift-kink instabilities that promote the transition to a turbulent state. Our analysis supports this statement, as well as our claim from section \ref{sec:aspectratio2D} that an elongated simulation box facilitates the development of turbulence by giving
the additional instabilities sufficient room to grow: indeed, we find that, e.g.,\ the case with box size $16\times4\lmri^2$ produces magnetic-field spectra with characteristic turbulent slopes, while the $4\times4\lmri^2$ case does not.

Even though our results are in good agreement with the only other (to date) spectral analysis of the 2D, fully kinetic MRI presented in literature (\citealt{inchingolo2018}), a few additional observations are worth reporting here. First, the spectra we show in Figure \ref{fig:spectra2D} are not necessarily representative of the whole nonlinear stage. Measuring the spectra at other times reveals additional features, e.g.\ a progressive shift of the spectral break in $B_\mathrm{pol}$ toward smaller wavenumbers. This is expected in numerical simulations of turbulence due to the continuous plasma heating that pushes $\rho_\mathrm{C}$ closer to the macroscopic scales (\citealt{zhdankin2018}; see section \ref{sec:ene2D}), but the scale separation we include in our runs is rather limited from the start, which allows for tracking this phenomenon only over relatively short timescales. Additionally, as we have illustrated in section \ref{sec:explanation2D}, our 2D simulations are characterized by the absence of reconnection along $y$, which probably alters the development of the $B_y$ spectrum with respect to the 3D case. This issue, combined with the fact that the MRI continuously slows down (and then stops; see section \ref{sec:evol_phases}) in our simulations, produces late-time power spectra which are no longer representative of MRI turbulence and differ substantially from those shown here. Finally, we note that over long times we observe a progressive pile-up of energy at large $k$ values (not shown in Figure \ref{fig:spectra2D}), which we attribute to numerical noise accumulating on short wavelengths; this complicates the long-term spectral analysis.

Our results also differ from the typical spectra seen in 3D MHD and hybrid simulations. For example, \cite{kunz2016} and \cite{walker2016} measured a $\propto k^{-2}$ spectrum for $B_y$ in the inertial range, which they attribute to sharp reversals of $B_y$ over short length scales. It is entirely possible that the difference between our 2D results and 3D MHD and hybrid results originates from our fully kinetic, pair-plasma setup. Moreover, the simulations conducted in \cite{kunz2016} also differ in the background magnetic-field configuration, which has zero net flux. Our initial setup includes a nonzero net-flux magnetic field, which changes the MRI evolution drastically (see section \ref{sec:zeroflux}). This may produce additional differences in the turbulent spectra. In section \ref{sec:spectra3D} we will discuss spectra for our 3D simulations.

\subsection{Particle energization in 2D}
\label{sec:ene2D}

\begin{figure*}
\centering
\includegraphics[width=1\textwidth, trim={0mm 0mm 0mm 0mm}, clip]{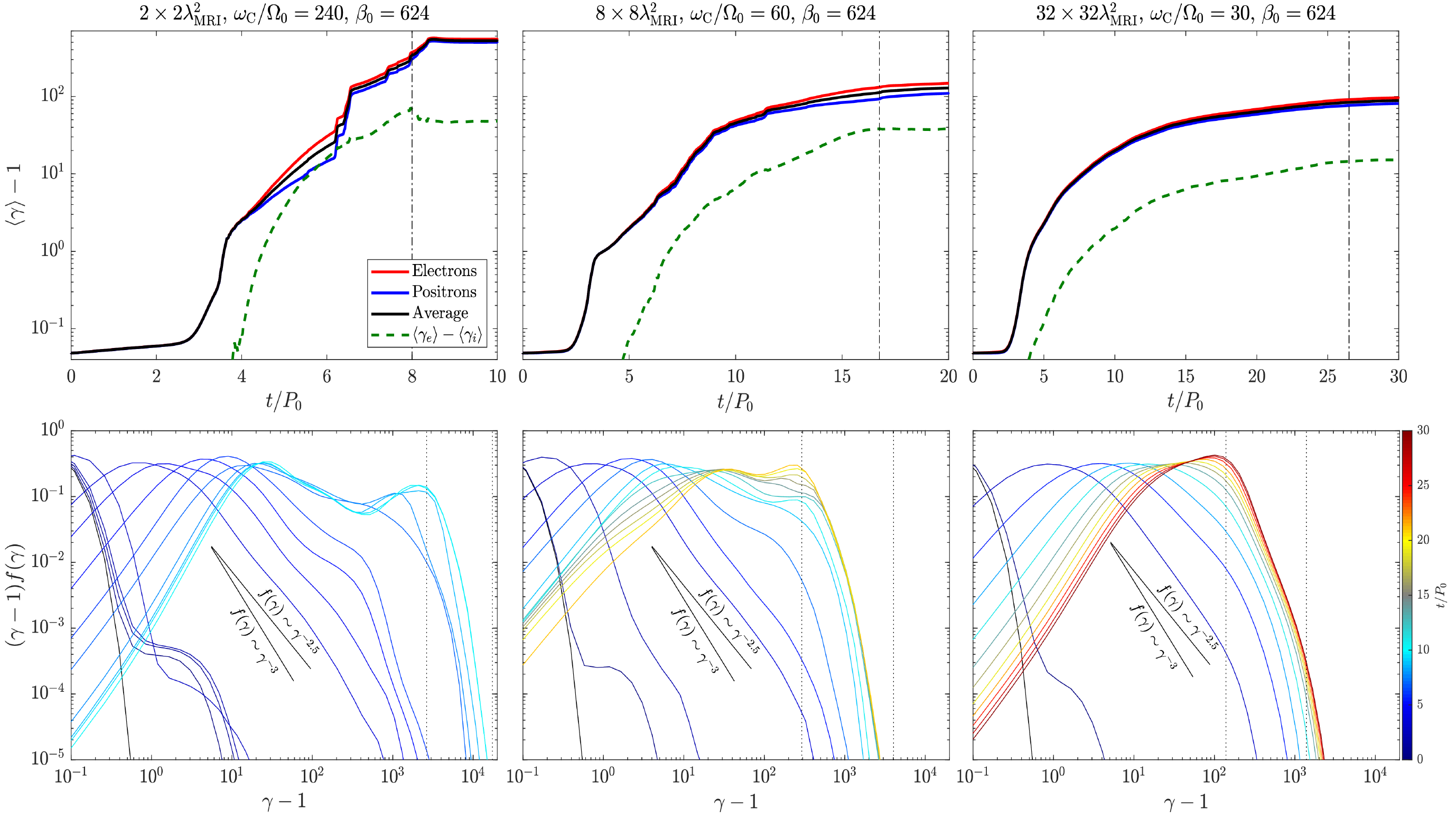}
\caption{Particle energization in 2D pair-plasma MRI simulations. Top row: average Lorentz factor for electrons and positrons for simulations with a small box/large $\omrat$ (left), medium box/medium $\omrat$ (center), large box/small $\omrat$ (right). In each case, the difference between $\langle\gamma\rangle$ of the two species is shown as a dashed green line, and the end of the nonlinear stage is marked with a vertical dash-dotted line (notice the difference in time scales between different runs). Bottom row: evolution of energy distributions in time (including both species) for the three runs. In each case, the distributions are shown until the same final time as in the top row. The dotted lines indicate the average and cutoff Lorentz factors $\gamma\sim qB\lmri/(mc^2)$ (with $B=B_\mathrm{c}$ or $B=B_\mathrm{e}$; see text) characterizing the quiescent end state of each run.}
\label{fig:energy2D}
\end{figure*}

We now turn our attention to the question of particle energization in the 2D collisionless MRI. With our PIC approach, we can measure nonthermal effects, which are not included in MHD models. Our pair-plasma simulations are not representative of electron-ion plasma expected to make up typical RIAFs, but we can still infer qualitative features in the mechanisms driving particle energization during the MRI development that potentially carry over to the case of large~$m_i/m_e$.

Prior to conducting any analysis, we emphasize a few key points raised in earlier sections. First, it is now clear that in all of our 2D simulations (regardless of system size and parameters chosen), the MRI eventually stops and the system reaches a quiet steady state. We should bear in mind that such a ``quiet'' state consisting of magnetic loops is not physically realistic, and is a product of the 2D geometry. 
Second, in the ideal situation where the MRI has evolved into a fully turbulent state, we do not expect the particle energy to saturate in time at a well-defined value. This turbulent state can be regarded as  analogous to driven-turbulence simulations (e.g.\ \citealt{zhdankin2018conv}) where energy is constantly injected into the system. In our case, the injection mechanism is the MRI on large scales, and in addition, heating and dynamo processes self-consistently regulate the energy injection in our simulations.
Third, the rate of particle energy gain will obviously depend on the physical conditions under which the MRI develops. We have already shown that the system evolution is drastically different for small/large boxes and for small/large $\omrat$ (see section \ref{sec:evol_phases}). We will need to consider such differences to attribute the measured energization to the correct mechanism.
Finally, we remark that due to the fundamentally different physical processes at play (e.g.\ lack of reconnection along $y$ in 2D), we expect the results of 3D simulations to potentially differ from those presented here. These differences will be discussed in section~\ref{sec:ene3D}.

Figure \ref{fig:energy2D} shows the evolution of the average Lorentz factor (top row) in three simulations that are representative of the parameter-space scan presented in section~\ref{sec:paramspace2D}: the small-box/large-frequency ratio case $2\times2\lmri^2$, $\omrat=240$ (left column); the medium-box/medium-frequency ratio case $8\times8\lmri^2$, $\omrat=60$ (central column); and the large-box/small-frequency ratio\footnote{Note that the ``small'' $\omrat$ employed here is still approximately twice as large as the largest value employed in previous works on the pair-plasma~MRI.} case $32\times32\lmri^2$, $\omrat=30$ (right column). In all cases, $\beta_0=624$ and $v_{\mathrm{A},0}/c=0.01$. In each case, the energy evolution is shown until shortly after the moment when the MRI wavelength grows beyond the box size. This moment depends on the simulation parameters and differs in the three cases. For each simulation we show the evolution of electron and positron energies separately, as well as their average and difference.

The first feature we observe is a differential energization between electrons and positrons, which becomes substantial at late times. Although surprising (compared to standard PIC simulations), this is a genuine physical phenomenon in the shearing-box paradigm. Our setup involves a simple (and by itself, symmetric) electromagnetic configuration, i.e.\ a uniform vertical field; but differently from standard PIC simulations, the addition of a unidirectional (along $y$) background rotational profile results in a symmetry-breaking effect. Charged particles in this situation ought to behave differently depending on the sign of their charge, which determines how their cyclotron gyromotion combines with the background rotation. Analytically, we can observe that the shearing-box equations include extra forces acting on particles in addition to the usual Lorentz force. These forces cause drifts in the particle motion, which depend on the sign of the charge and of the mean magnetic field. We have found that this additional dynamics plays a fundamental role in the energization process, with electrons consistently gaining more energy than positrons by a factor $\sim3$ in our largest simulations, for our specific choice of an initial $\vecB$ aligned with the object's axis of rotation. This phenomenon is exactly reversed when reversing the orientation of the vertical field: we have indeed verified that simulations in which $B_{z,0}$ is antiparallel to the rotation axis result in positrons gaining more energy than electrons. The differential-heating mechanism reported here has not been discussed in previous works, but it is in fact a crucial point to consider in future electron-ion simulations, where the heating ratio between species is a primary analysis target, since it relates directly to observations. In the following, we will focus on analyzing particle energization on average (including both species); we discuss the causes and the importance of this differential heating in detail in Appendix \ref{app:diffene}.

Comparing the three cases shown in Figure \ref{fig:energy2D}, we observe that the average $\gamma$ at the end of each run consistently reaches values $\gtrsim10^2$; however, the energization proceeds very differently for different simulation parameters, as is evident from the slope of each curve over time. In the small-box case (left column in Figure \ref{fig:energy2D}), the heating rate is evidently much larger during brief, ``bursty'' periods that correspond to large-scale reconnection events, starting at $t\simeq 3.5P_0$, which marks the onset of reconnection after the linear stage. Each event is followed by a ``quiet'' phase where particles gain energy steadily at a much slower rate. During each quiet phase, the energization rate is very similar, and the transition to and from reconnection is visible. Conversely, the large-box case (right column) shows a smooth increase in the kinetic energy, without the energization ``bursts'' observed in the small-box case. The heating rate here progressively decreases as time passes; the energization proceeds much more slowly (notice the difference in time scales between different cases) and around $t\simeq30P_0$, when the MRI is slowing down and halting, the average Lorentz factor reaches a plateau. The medium-box case (central column) shows features that appear as a combination of the other two cases: the energy increases very rapidly at a few moments in time, with smooth phases of slower heating in between.

Regardless of the specific processes contributing to particle energization, analyzing the energy distribution functions in each case reveals several similarities. Figure \ref{fig:energy2D} (bottom row) shows the energy distributions (including both species) at several different times, until the moment the MRI is halted in each of the three cases (again, notice the difference in time scales). In all cases, at the onset of magnetic reconnection (marking the transition to the nonlinear stage), the distribution abruptly jumps to much larger average energies and a clear nonthermal component arises. The nonthermal population appears to follow a power-law distribution with index $\sim2.5\mbox{--}3$ (both slopes are shown in the plot to guide the eye), but this value changes substantially as time passes. As the system evolves, the spectra become shallower and the main part of the distribution resembles a Maxwellian; in addition, in all cases we observe the progressive formation of a second peak in the distribution at high energies, and a stabilization of the cutoff $\gamma$ at a value that grows larger with larger~$\omrat$.

That the energy distributions continue to evolve as long as the MRI is active is unsurprising, since the system includes a free energy source. However, the shift from a power-law distribution to a double-peaked shape is peculiar. Measuring the peak and cutoff $\gamma$ across our large parameter-space reveals a common feature: these values correspond almost exactly to the Lorentz factor of particles with a gyroradius of size $\sim\lmri$, when measuring the magnetic-field strength at the center and at the edge (which we indicate with $B_\mathrm{c}$ and $B_\mathrm{e}$, respectively) of the large-scale, stable magnetic loops that form at the end of the system evolution (the corresponding values of $\gamma$ are indicated in Figure \ref{fig:energy2D} by black dotted lines). These loops originate from plasmoids created via reconnection, well before the end of the nonlinear stage when the MRI starts to slow down. Particles are gathered inside the loops as these continue to merge; when the MRI is completely halted, these loops have grown to size $\sim\lmri$ and contain almost all particles inside. The energy distribution evolves to reflect this state: particles trapped inside the loops, where $B\sim B_\mathrm{c}$, tend to increase their energy such that $mc^2\gamma/(qB_\mathrm{c})\sim\lmri$ (e.g.\ \citealt{hillas1984}), which creates the peaked distribution we observe. The maximum particle energy will be then limited by the strong magnetic field at the loop edge such that $mc^2\gamma/(qB_\mathrm{e})\sim\lmri$, with  $B_\mathrm{e}\sim10B_\mathrm{c}$. 
This creates the energy cutoff that we measure in our runs.

\begin{figure}
\centering
\includegraphics[height=0.8\columnwidth, trim={10mm 0mm 10mm 0mm}, clip]{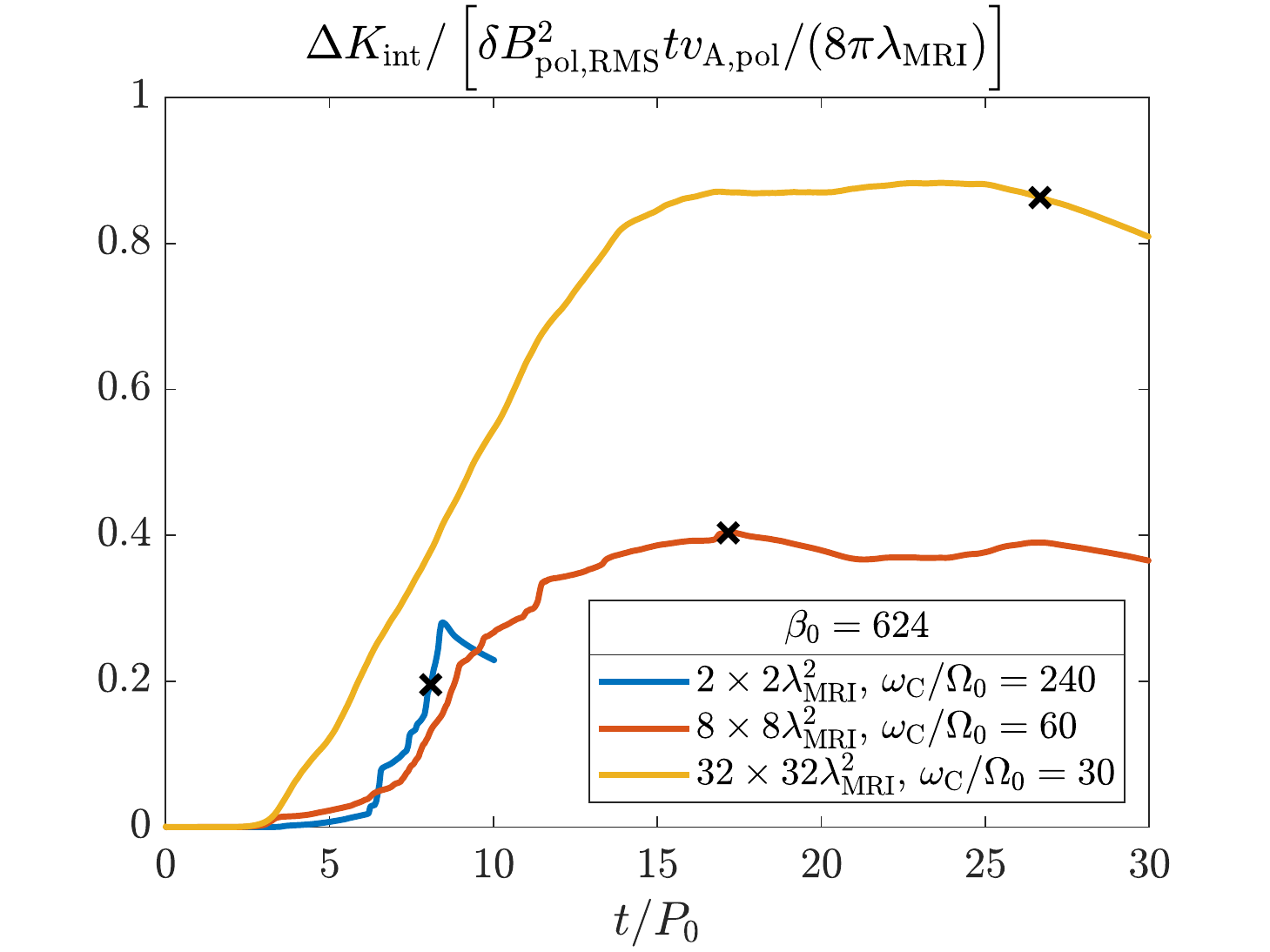}
\caption{Estimate for injection of magnetic energy and conversion into kinetic energy assuming Alfv\'enic turbulence, for the three simulations presented in section \ref{sec:ene2D}. In each case, the end of the nonlinear phase is marked with a black cross.}
\label{fig:energy_injection}
\end{figure}

As discussed in earlier sections, the main difference between the MRI evolution in the three cases is the presence of sustained turbulence in the nonlinear stage. In the small-box case, turbulence cannot develop and the energization is mediated by intermittent reconnection events, separated by ``quiet'' periods where magnetic fields are amplified in coherent channel structures and particles can increase their perpendicular momentum adiabatically via $\mu$ conservation. In large boxes, there is no clear distinction between subsequent (discrete) reconnection events --- turbulence results from the breakup and mixing of channel flows, and more chaotic processes come into play. To verify quantitatively that turbulent heating is the main driver of particle energization in the large-box case, we follow an approach similar to that described by \cite{zhdankin2018}. Assuming Alfv\'enic turbulence, the energy injection from electromagnetic fields to particles can be written as 
\begin{equation}
    \frac{\rmd K_\mathrm{int}}{\rmd t} \sim \frac{\delta B^2_{\mathrm{RMS}}}{8\pi} \frac{v_\mathrm{A}}{\ell_\mathrm{inj}},
\end{equation}
where $K_\mathrm{int}\equiv2nmc^2(\langle\gamma\rangle-1)$ (counting both particle species) is the average internal (kinetic) energy and $\delta B^2_{\mathrm{RMS}}$, $v_\mathrm{A}$, and $\ell_\mathrm{inj}$ are the characteristic root-mean-square magnetic energy, the corresponding Alfv\'en speed, and the injection length scale of the turbulent cascade. In our case, the relevant magnetic field and Alfv\'en speed are those corresponding to the late-time, saturated state achieved in the $xz$-plane, because the out-of-plane $B_y$ only acts as a guide field and does not participate in the cascade process (due to the the lack of reconnection along $y$). For the injection scale, we select $\ell_\mathrm{inj}=\lmri$: during the turbulent nonlinear stage, in all cases we measure the largest magnetic-field structures to be of size comparable to the \emph{initial} MRI wavelength (despite the system evolution pushing the MRI to much larger scales); this is an indication that the system retains some memory of the initial conditions. In Figure \ref{fig:energy_injection}, we show this measurement of energy injection (integrated in time) for the three simulations presented above. We observe that the large-box case reaches, at late times, a saturated state that agrees very well with our simple estimate. This appears to indicate that Alfv\'enic turbulence well describes the turbulent-cascade process at play. For the small- and medium-box cases, the physics involved in the nonlinear stage is substantially different, and our estimate fails to accurately describe the energy-injection process. In 3D, we expect the physics of this injection to include additional dynamics along $y$, which allows $B_y$ to participate in the cascade process. The absence of this mechanism in 2D is a potential source of discrepancy between 2D and 3D results.

\subsection{Stresses and angular-momentum transport in 2D}
\label{sec:stresses2D}

\begin{figure*}
\centering
\includegraphics[width=1\textwidth, trim={0mm 0mm 30mm 0mm}, clip]{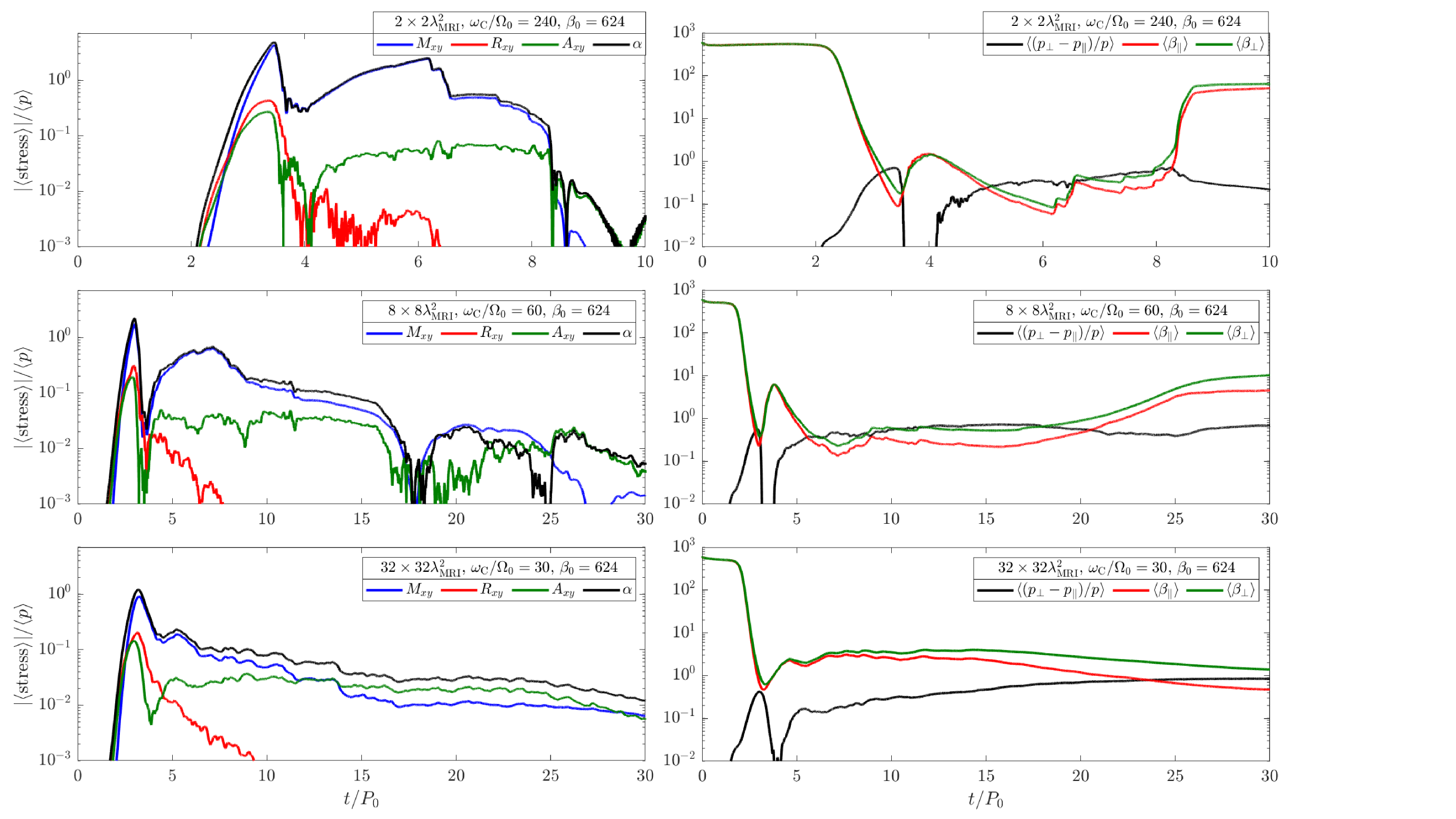}
\caption{Left column: evolution of the $\alpha$-parameter and of the volume-averaged Maxwell, Reynolds, and anisotropic stresses for a small box/large $\omrat$ (top), medium box/medium $\omrat$ (center), and large box/small $\omrat$ (bottom) simulation of the 2D pair-plasma MRI. Right column: evolution of the volume-averaged pressure anisotropy and $\beta_\|$, $\beta_\perp$ for the same simulations.}
\label{fig:stresses2D}
\end{figure*}

In this section, we analyze the effective viscosity that originates from the MRI nonlinear dynamics in 2D pair-plasma simulations. We define an effective-viscosity parameter (\citealt{shakurasunyaev1973}),
\begin{equation}
 \alpha \equiv \frac{\langle R_{xy} + M_{xy} + A_{xy}\rangle}{\langle p\rangle},
\end{equation}
where $\langle ... \rangle$ denotes a volume average. Here, $R_{xy} = mn U_x U_y/\Gamma^2$ is the Reynolds stress, with $\Gamma=\sqrt{1+U^2/c^2}$ and $\bb{U}=(U_x,U_y,U_z)$ the spatial part of the bulk 4-velocity of the plasma in the frame moving with velocity $\vecv_\mathrm{s}$; $M_{xy} = -B_xB_y/(4\pi)$ is the Maxwell stress; and $A_{xy} = -(p_\perp-p_{\|})B_xB_y/B^2$ is the anisotropic stress (e.g.\ \citealt{sharma2006}). Our PIC simulations, where the plasma can be relativistically hot, require special care when measuring $\alpha$ and its components. Given a distribution function $f(\vecx,\vecu,t)$, we first compute the bulk velocity $\bb{U}=(1/n)\int\vecu f(\vecx,\vecu,t) \rmd^3 \vecu$ for each particle species separately. Then, we apply a Lorentz boost to obtain $\widetilde{\vecu}$, the spatial part of the 4-velocity of each particle in the frame moving with~$\bb{U}$. We subsequently define a pressure tensor for each particle species as 
\begin{equation}
\textbf{P} = m\int \frac{\widetilde{\vecu}\widetilde{\vecu}}{\widetilde{\gamma}} f(\vecx,\widetilde{\vecu},t) \rmd^3 \widetilde{\vecu},
\end{equation}
where $\widetilde{\gamma}=\sqrt{1+\widetilde{u}^2/c^2}$. The isotropic, parallel, and perpendicular pressure are then $p=\mathrm{tr}(\textbf{P})/3$, $\ppar=\textbf{P}\bb\bdbldot\hat{\bb{b}}\hat{\bb{b}}$, and $\pperp=(1/2)\textbf{P}\bdbldot(\mathbb{I}-\hat{\bb{b}}\hat{\bb{b}})$ respectively, where $\hat{\bb{b}}=\widetilde{\vecB}/\widetilde{B}$ and $\widetilde{\vecB}$ is the magnetic field in the frame moving with~$\bb{U}$. $R_{xy}$ and $A_{xy}$ are computed for each species separately, and then summed across species. 

In Figure \ref{fig:stresses2D}, we show the time evolution of the average $\alpha$ and its components (left column) as well as the volume-averaged $\langle\beta_{\|}\rangle$, $\langle\beta_\perp\rangle$, and the pressure anisotropy $\langle(\pperp-\ppar)/p\rangle$ (right column) for the small-box/high-frequency ratio case $2\times2\lmri^2$, $\omrat=240$ (top row), the medium-box/medium-frequency ratio $8\times8\lmri^2$, $\omrat=60$ (middle row), and the large-box/small-frequency ratio $32\times32\lmri^2$, $\omrat=30$ (bottom row). In all cases, $\beta_0=624$ and $v_{\mathrm{A},0}/c=0.01$. All simulations are characterized by a rapid increase in all components of $\alpha$ during the linear stage, with a peak at $\alpha\simeq1$ right before the onset of reconnection. At this point, different parameters determine strong differences in the results: during the nonlinear stage, we observe that the volume-averaged $M_{xy}$ can grow and remain close to 1 at early times in the small- and medium-box cases; conversely, in the large-box case $M_{xy}$ rapidly settles to $\sim0.1$ after the onset of reconnection and slowly decreases to $\sim0.01$. In all cases, the anisotropic stress $A_{xy}$ achieves the highest (volume-averaged) value $\sim0.1$ at the onset of reconnection, and then firmly settles around $0.02\mbox{--}0.05$ for the whole duration of the nonlinear phase. The Reynolds stress is consistently the least important component of $\alpha$ and remains much smaller than the other two during the nonlinear phase. Overall, the $\alpha$-parameter closely follows the evolution of $M_{xy}$ in the small- and medium-box cases, attaining values $0.1\mbox{--}1$; in the large-box case, $\alpha\simeq0.01\mbox{--}0.1$ and the contribution of $A_{xy}$ is more significant, providing most of the angular-momentum transport in the late nonlinear stage when the magnetic field is dissipating via reconnection and turbulence. We note, however, that in the large-box case $\omrat$ is substantially smaller than in the small-box case (respectively $\omrat=30$ and $\omrat=240$); since the scale separation affects the development of mirror modes (see section \ref{sec:explanation2D}), pressure anisotropy may be exaggerated in the small frequency-ratio case. Larger values of $\omrat$ would be needed to verify the trend of the anisotropic stress in the large-box case, but this involves computational costs beyond our current possibilities. We will investigate the matter more in detail in future work.

These results align well with the discussion provided in the previous sections. The MRI in the large-box case behaves qualitatively differently from the small-box case: the system develops strong turbulence during the nonlinear stage, promoting the continuous, more efficient dissipation of magnetic fields and faster pressure isotropization. In small boxes, this dissipation is primarily mediated by episodic, large-scale reconnection events, with phases of strong magnetic-field amplification driving pressure anisotropy in between. This suggests that, in a large box, $M_{xy}$ and $A_{xy}$ will be reduced in comparison with a small-box case. The right-hand panels of Figure \ref{fig:stresses2D} confirm this reasoning: pressure anisotropy is much more pronounced at the beginning of the nonlinear stage in smaller boxes, while it remains rather limited in the large-box case.

If $\alpha$ is invoked as a coefficient of turbulent angular-momentum transport, then the large-box case where turbulence can develop is more relevant to provide a first-principles measure of the effective turbulent viscosity in a collisionless plasma. In this case, our 2D numerical experiments indicate that the Maxwell stress provides the primary contribution to $\alpha$, with the anisotropic stress achieving smaller values $A_{xy}/M_{xy}\simeq0.2$. This result differs from the findings of \cite{kunz2016}, who reported $A_{xy}/M_{xy}\simeq1$; however, we stress again that 2D runs have the intrinsic limitation of lacking reconnection processes in the dominant magnetic-field component~$B_y$. In 3D, we may then expect large differences with the results presented here: dissipating $B_y$ via reconnection may imply that the Maxwell stress could achieve much smaller values than those measured in 2D, possibly increasing the relative contribution of pressure anisotropy in determining~$\alpha$. On the other hand, in the previous sections we have argued that reconnection along $y$ may help reduce pressure anisotropy by increasing~$\ppar$, hence decreasing~$A_{xy}$. We will discuss the dynamics of angular-momentum transport in 3D in section~\ref{sec:stresses3D}.

\subsection{Net flux vs.\ zero net flux}
\label{sec:zeroflux}

\begin{figure*}
\centering
\includegraphics[width=1\textwidth, trim={0mm 50mm 0mm 0mm}, clip]{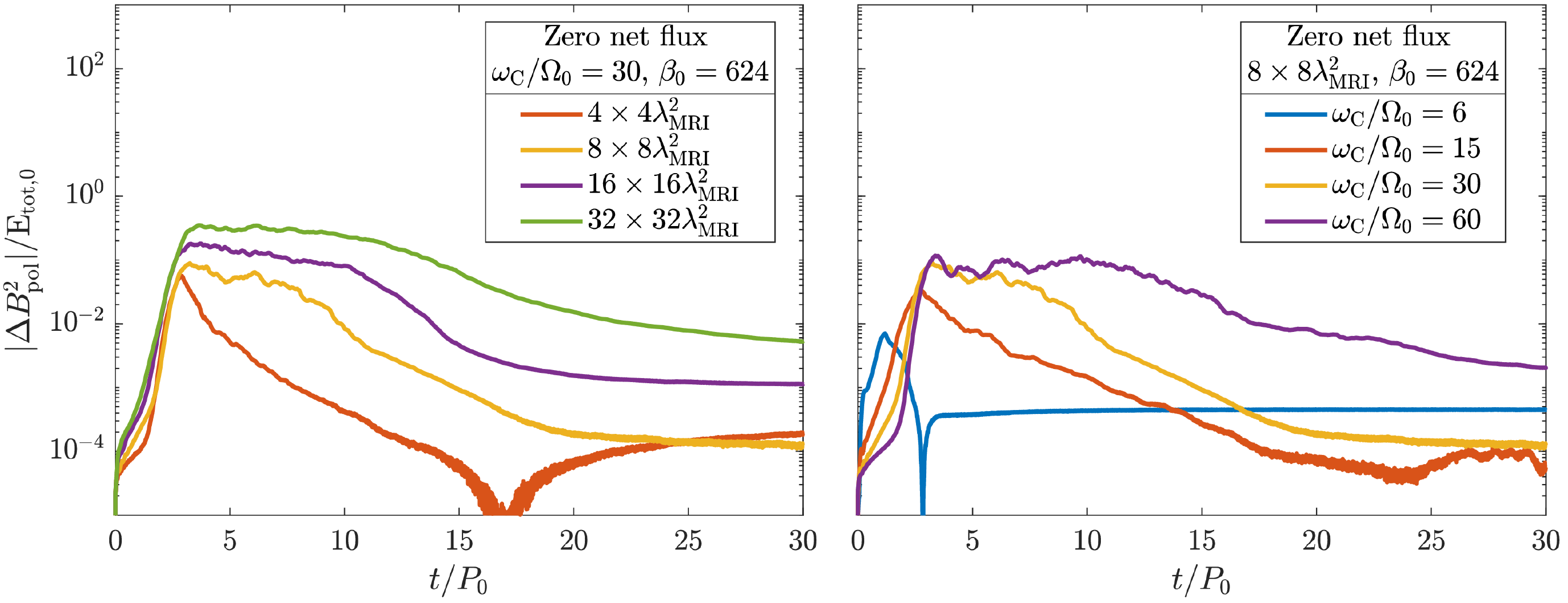}
\caption{Evolution of the in-plane magnetic-energy variation for several 2D pair-plasma simulations of the MRI with zero net magnetic flux. Left: simulations with varying box size and fixed $\omrat$, $\beta_0$. Right: simulations with varying $\omrat$ and fixed box size and $\beta_0$.}
\label{fig:zeroflux}
\end{figure*}

We conclude our investigation of the 2D collisionless-MRI dynamics by briefly considering the case of zero net flux. In contrast with the simulations presented in the previous sections, here the initial magnetic field is set to $\vecB(t=0)=-B_{z,0}\sin\left(2\pi x/L_x\right)\hatvece_z$ such that no net flux is present along $z$. This case was briefly discussed in \cite{riquelme2012} in 2D and was the focus of the only (to date) 3D hybrid study of the MRI (\citealt{kunz2016}).

In Figure \ref{fig:zeroflux} we plot the evolution of the in-plane magnetic energy for a few simulations with exactly the same parameters employed in section \ref{sec:paramspace2D} (aside from the initial magnetic-field configuration). In particular, we compare $B_\mathrm{pol}$ in several cases: first, by keeping a fixed $\omrat=30$, $\beta_0=624$, and $v_{\mathrm{A},0}/c=0.01$ (the latter two measured with the maximum value of~$B$), in larger and larger simulation boxes $L_z/\lmri=L_x/\lmri=4,8,16,32$ (left panel); then, by fixing the box size $L_z/\lmri=L_x/\lmri = 8$, $\beta_0=624$, and $v_{\mathrm{A},0}/c=0.01$, and increasing the frequency ratio $\omrat = 6,15,30,60$ (right panel). From this comparison, we observe that the general statements made for the nonzero net-flux case hold here as well: the duration of the turbulent, saturated nonlinear stage that follows the linear growth of the MRI increases both with box size and with~$\omrat$. Over time, the turbulence rapidly decays as magnetic fields grow and pressure anisotropy develops (not shown), pushing the MRI wavelength to scales larger than the box size. However, the zero net-flux case is also characterized by striking differences, with the most prominent being a much more efficient breakup and mixing of channel flows. This was also observed by \cite{riquelme2012}, who attributed this behavior to the fact that the instability develops nonuniformly in different regions of the domain; in particular, the most-unstable MRI wavelength $\lmri=2\pi v_{\mathrm{A}}/\om$ is shorter where $B_z$ is weaker and longer where $B_z$ is stronger, producing a spatial inhomogeneity in the dominant instability wavevector. Channels forming inside regions of weaker/stronger $B_z$ are therefore spatially nonuniform, and tend to be destroyed much more quickly. Hence, differently from the net-flux case, the break-up of channels formed during the linear stage is not solely mediated by the onset of tearing modes (and concurrently of kink instabilities): the dynamics does not need to ``wait'' for the magnetic field in the channels to grow strong enough (and for current sheets to thin out enough) that reconnection can occur. Instead, the intrinsic inhomogeneity of the background field mediates the mixing and redistribution of magnetic energy. As a consequence, the saturation level reached in all simulations with zero net flux is much smaller than in the corresponding nonzero flux cases.

Figure \ref{fig:zeroflux} also shows important differences in how these simulations tend to converge: while it appears that progressively better convergence is achieved for increasing $\omrat$ at a fixed box size, enlarging the box size does not result in converging results. Larger boxes seemingly produce a steady increase in the magnetic-field amplification, in contrast with the clear convergence observed for the net-flux case. We also note that, since the magnetic field here reverses its direction through the box, we observe no sign of the differential heating of electrons vs.\ positrons that we measured in the net-flux case (see section~\ref{sec:ene2D}). Charges with opposite sign are indeed equally likely to experience differential gyromotion and drift towards current sheets, overall resulting in an almost perfectly balanced energization among electrons and positrons.

In this work, we focus on nonzero net-flux simulations primarily because, in any astrophysical accretion disk, the complete absence of vertical magnetic flux is not expected; realistically, any portion of the disk will be threaded by some magnetic field carrying a net flux, however small. Additionally, the behavior of zero net-flux simulations is fundamentally different from the expected development of a ``pure'' MRI: the former situation is more akin to a dynamo problem, and does not feature the same qualitative evolution of the nonzero net-flux case. Our results show that the case of zero net flux differs enough from the case with nonzero magnetic flux to deserve a thorough, dedicated exploration, which we will pursue in future work.

\section{Three-dimensional simulations of the MRI in collisionless pair plasmas}
\label{sec:3D}

\begin{deluxetable}{cccc}[h!]
\tablecaption{List of 3D simulations}
\tablenum{2}
\tablehead{\colhead{Box size $(L_x\times L_y\times L_z)$} & \colhead{$\omega_\mathrm{C}/\Omega_0$} & \colhead{$\beta_0$} & \colhead{$\lambda_\mathrm{MRI}/\rho_\mathrm{C}$}} 
\startdata
$2\times2\times2\lmri^3$ & 15 & 312 & 7.46  \\
$4\times4\times2\lmri^3$ & 15 & 312 & 7.46 \\
$4\times8\times2\lmri^3$ & 15 & 312 & 7.46 \\
\enddata
\tablecomments{In all cases, $v_{\mathrm{A},0}/c\simeq0.007$; the numerical resolution is such that $\Delta x\simeq c/\omega_\mathrm{p}$, and we employ 27 particles per cell per species. }
\label{tab:3D}
\end{deluxetable}

In this section we present 3D PIC simulations conducted with \textsc{Zeltron}, where we have implemented the newly developed \KSBOA~model (see section \ref{sec:equationsKSBOA}). In general, we expect 3D runs to produce results more physically realistic (at least in terms of the magnetic-field evolution and subsequent development of turbulence) than corresponding 2D simulations, for the reasons discussed in section~\ref{sec:explanation2D}. We will therefore select simulation parameters that strictly enforce three physical constraints: i) a sufficiently subluminal velocity offset at the $x$-boundaries, i.e.\ $v_\mathrm{s}(L_x)=s\om L_x<c$; ii) a vertical extent of the simulation box that does not exceed the disk pressure scale height, i.e.\ $L_z/H\leq 1$; and iii) a sufficiently large separation between macroscopic and microscopic scales (at least in the beginning of each run), in terms of~$\lmri/\rho_\textrm{C}$.

Due to the much higher computational cost of 3D runs, in this first study we will explore a limited parameter range inspired by the results of 2D simulations. We will show that 3D runs of relatively small size already provide fundamental new insight into the physics of the collisionless MRI. The 3D simulations discussed in this work are summarized in Table \ref{tab:3D}; all runs are initialized analogously to the 2D case, with a weak vertical magnetic field determined from the initial simulation parameters. We always employ 27 particles per cell per species, and a numerical grid such that $\Delta x\simeq c/\omp$. In future work, we will conduct larger simulations with better scale separation to achieve more realistic regimes.

In the following sections, we will first describe the physics of small- and large-box 3D PIC simulations of the pair-plasma MRI, and how these relate to corresponding 2D simulations. We will then analyze the 3D turbulence, particle energization, and angular-momentum transport that characterize the 3D case.

\subsection{Physics of the 3D collisionless MRI}
\label{sec:explanation3D}

\begin{figure*}
\centering
\includegraphics[width=1\textwidth, trim={0mm 0mm 0mm 0mm}, clip]{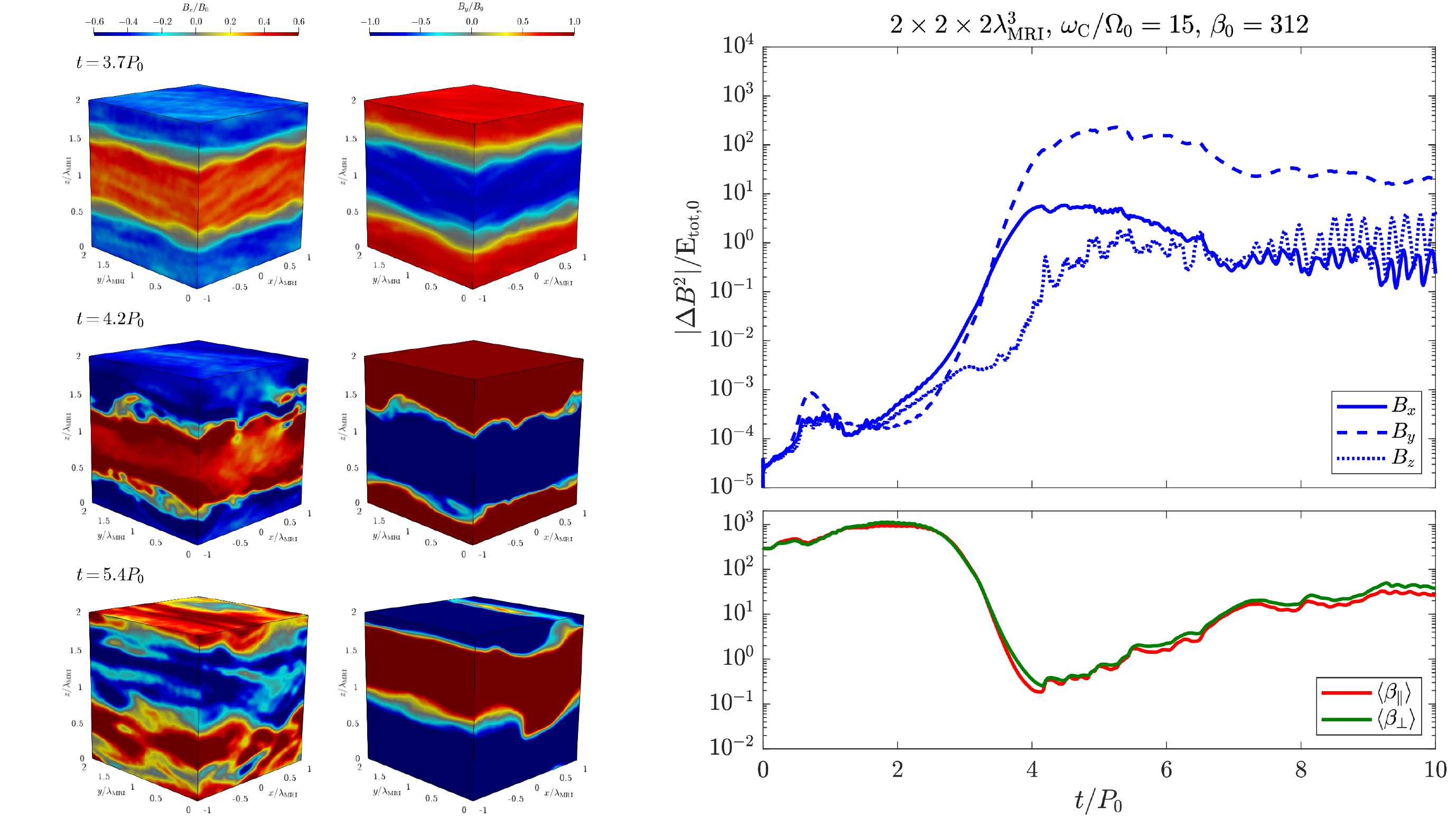}
\caption{Evolution of the pair-plasma MRI in a small 3D box ($2\times2\times2\lmri^3$). Left: spatial distribution of the radial ($B_x$) and toroidal ($B_y$) magnetic field at subsequent times during the simulation. Right: evolution of the change in magnetic energy in all three components of $\vecB$ (top) and evolution of the volume-averaged $\beta_\|$ and $\beta_\perp$ (bottom).}
\label{fig:evol_phases_small_3D}
\end{figure*}

\begin{figure*}
\centering
\includegraphics[width=1\textwidth, trim={0mm 0mm 0mm 0mm}, clip]{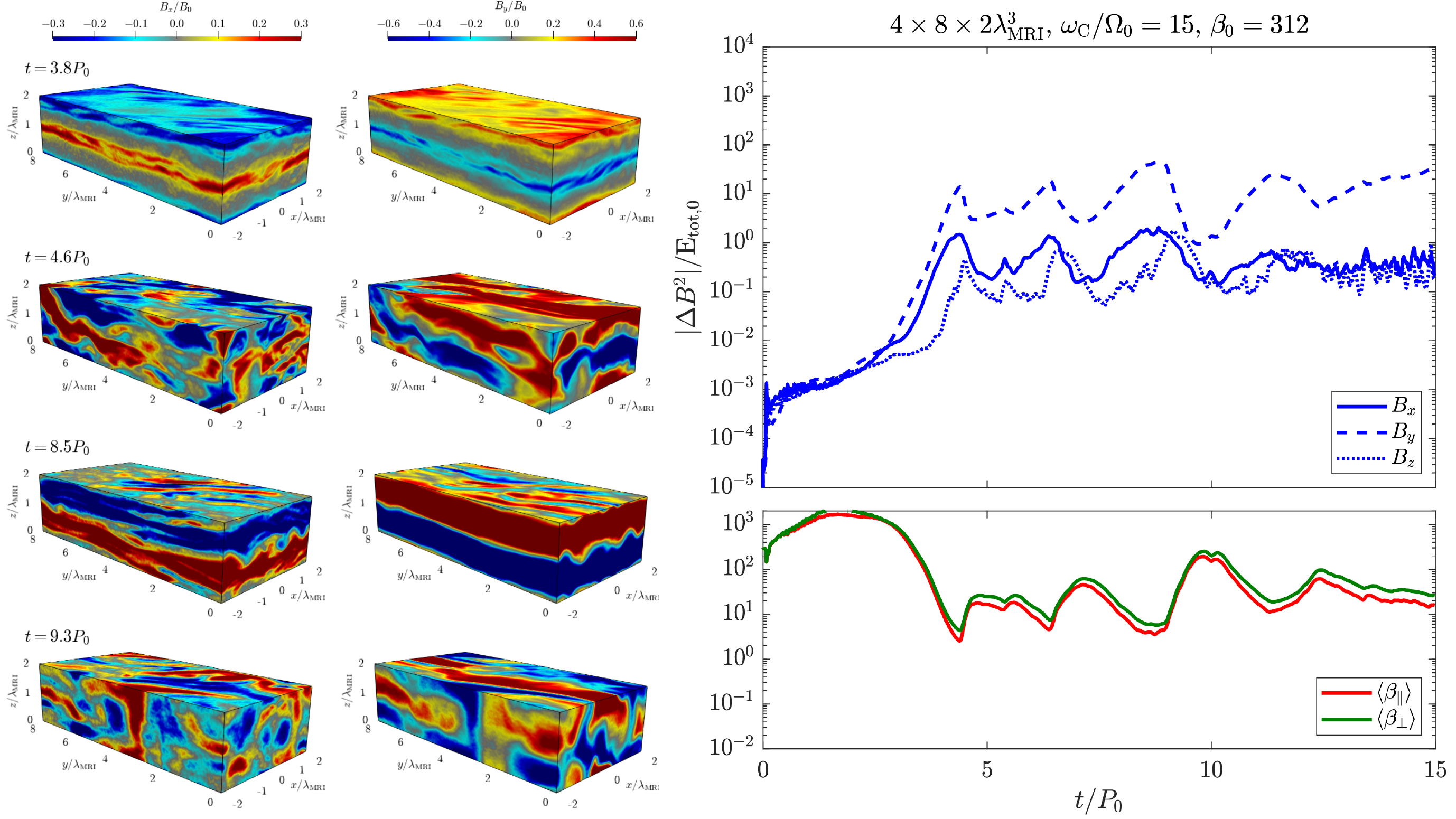}
\caption{As in Figure \ref{fig:evol_phases_small_3D} but for a box size of $4\times8\times2\lmri^3$.}
\label{fig:evol_phases_large_3D}
\end{figure*}

Figures \ref{fig:evol_phases_small_3D} and \ref{fig:evol_phases_large_3D} show the results for two representative 3D simulations of the collisionless MRI in pair plasmas, employing a small ($2\times2\times2\lmri^3$) and a large ($4\times8\times2\lmri^3$) box size, respectively. In these figures, we show subsequent snapshots of the spatial distribution of $B_x$ and $B_y$ (left panels), the evolution of the change in magnetic energy (top-right panel), and of the volume-averaged parallel and perpendicular plasma-$\beta$ (bottom-right panel). These simulations are initialized with parameters such that $L_z=H$, $\lmri/\rho_\textrm{C}\simeq7.5$, and $v_\mathrm{s}(L_x)/c\simeq0.125$ (in the small-box case) and $v_\mathrm{s}(L_x)/c\simeq0.25$ (in the large-box case). In both cases, imposing these conditions results in a frequency ratio $\omrat=15$ and a plasma temperature $\theta_0=1/128$ such that $\beta_0=312$ (including both species), with an initial Alfv\'en speed $v_{\mathrm{A},0}/c\simeq0.007$. 

For the small-box case ($2\times2\times2\lmri^3$) shown in Figure \ref{fig:evol_phases_small_3D}, we observe that the evolution of the magnetic energy (top-right panel) is initially qualitatively similar to that of 2D simulations of comparable $L_x,L_z$ (cf.\ the 2D case of Figure~\ref{fig:evol_phases_small}): from the initial state, the field strength grows exponentially until saturation is reached. The nonlinear stage, however, shows qualitative differences: we observe that the magnetic energy in all components (dominated by $B_y$, which does not saturate in~2D) remains approximately around the saturation level for $\sim2$ orbital periods after the end of the linear stage, before starting to slowly decrease. This behavior is due to the 3D geometry and is illustrated by subsequent snapshots of the $B_x$ and $B_y$ distribution in the domain, shown in Figure \ref{fig:evol_phases_small_3D} (left panels). Around $t\simeq 3.7P_0$ (top-left panels) we observe that channel flows form, during the linear stage, in the $xz$-plane over length scales of~$\lmri$, similarly to the 2D dynamics. However, in 3D these channels also span the entire $y$-extent of the simulation domain. At the channel interfaces, reconnection starts dissipating magnetic energy when current sheets become sufficiently thin and tearing modes are excited (around $t\simeq4.2 P_0$, center-left panels). In 3D, reconnection can take place also along $y$, thus dissipating energy stored in $B_y$ (and not just in $B_x$ and $B_z$ as in~2D). The channel interfaces are indeed heavily perturbed and magnetic reconnection can act along all spatial directions (e.g.\ around $t\simeq5.4 P_0$, bottom-left panels). The saturated energy level of the nonlinear stage we measure between $t\simeq4P_0$ and $t\simeq6P_0$ is maintained due to the balance achieved between injection (via the MRI) and dissipation (via reconnection) of energy into the system. 

Starting from $t\simeq 6P_0$, the system slowly relaxes toward an end state where the MRI slows down and then stops. This final state (not shown in the left panels of Figure \ref{fig:evol_phases_small_3D}) is a two-channel configuration with lower $B_y$ energy. From $t\simeq8P_0$ onward, we observe that the energy in $B_x$ and $B_z$ oscillates around approximately the same values; the spatial distribution of the magnetic field tends to assume a ``shear-wave'' configuration with coherent, large-scale structures aligned with the shear flow appearing cyclically. No reconnection occurs at this point; the system appears to maintain this state endlessly. The dynamics that characterizes this state is different from that of the early nonlinear stage, and it is doubtful whether it can even be associated with the MRI at all. Qualitatively, it appears that MRI modes have migrated to wavelengths longer than the box size, and that the 3D geometry is allowing a quasi-steady state that consists of periodic oscillations. This dynamics is of no relevance for this work, but it is interesting to note the difference between the 3D and 2D end states, with the latter consisting of stable loops in the $xz$-plane.

The evolution of the volume-averaged $\beta_\|$ and~$\beta_\perp$, shown in Figure \ref{fig:evol_phases_small_3D} (bottom right), closely follows that of the magnetic field: in the early nonlinear stage (between $t\simeq4P_0$ and $t\simeq6P_0$), we measure an increase $\langle\beta_\|\rangle\simeq\langle\beta_\perp\rangle\sim0.2\mbox{--}10$, with very brief periods where $\langle\beta_\perp\rangle>\langle\beta_\|\rangle$ and subsequent ``bursty'' reconnection events reestablishing isotropy periodically. At late times, pressure anisotropy accumulates and pushes the MRI growth rate to smaller and smaller values.

The small 3D run discussed above achieves magnetic-energy saturation in the nonlinear stage, thus already improving over 2D runs of much larger size (in $x$ and $z$). However, in such as small simulation we do not obtain a developed turbulent state, consistently with the results of the sole 3D fully kinetic study previously presented (\citealt{hoshino2015}). For this reason, we now focus on the larger simulation shown in Figure \ref{fig:evol_phases_large_3D} with box size $4\times8\times2\lmri^3$. This case is chosen as representative of a situation where the numerical domain size is rather limited in $z$ (in our case, by the requirement $L_z \leq H$) --- hence a small number of channel modes are expected to grow during the linear stage --- but the larger extent in $x$ and $y$ now allows for additional instability mechanisms which cannot be included in the previous $2\times2\times2\lmri^3$ case. For 2D simulations, in section \ref{sec:aspectratio2D} we have argued that boxes elongated in $x$ promote the development of turbulence by allowing for drift-kink modes that can disrupt the channels; here, we observe exactly this effect in~3D. The top-right panel of Figure \ref{fig:evol_phases_large_3D} shows that the system evolution now proceeds through a linear stage where the magnetic field is amplified roughly 10 times less than in the previous cubic-box case. The onset of magnetic reconnection results in a nonlinear stage that is qualitatively different from that of the small-box case: repeated cycles of period $\sim2P_0$ are observed where the magnetic energy in all components grows and is rapidly dissipated before growing again. Each growth phase corresponds to the formation of macroscopic channels, and each dissipation phase results in the complete breakup of said channels, which lose coherence and decay into ``turbulent'' structures (whether this can be truly called a turbulent state will be discussed in section~\ref{sec:spectra3D}). This behavior is visible in the spatial distribution of $B_x$ and $B_y$ at subsequent times in Figure \ref{fig:evol_phases_large_3D} (left panels). 

Reconnection in this elongated-box simulation is evidently much more active than in the cubic-box case. We observe that the macroscopic current sheets violently kink in both $x$ and $y$, eventually resulting in the emergence of the turbulent state. The bottom-right panel of Figure \ref{fig:evol_phases_large_3D} shows that the stronger reconnection dynamics also heavily impacts the evolution of $\beta$: because of the smaller magnetic-field amplification (with respect to the cubic-box case), the nonlinear stage is characterized by a volume-averaged $\langle\beta\rangle\sim10$, which can grow up to $\sim100$ during the dissipation phase of the channel-breakup cycles. Pressure anisotropy is well regulated, during the early nonlinear stage, by subsequent reconnection events, which periodically reestablish isotropy; in the late nonlinear stage, a finite (tens of percent), unquenched pressure anisotropy is retained, marking the transition to the same end state described above for the small-box simulation.

The numerical experiment reported in Figure \ref{fig:evol_phases_large_3D}, however limited in box size and scale separation, provides a first, striking evidence that large 3D PIC simulations of the collisionless MRI possess qualitative features analogous to corresponding MHD simulations. In contrast to the 2D case, where important physical processes are impeded by the reduced dimensionality, here we observe an MRI behavior akin to that exhibited by 3D MHD runs, with the cyclic creation (via the MRI) and destruction (via reconnection) of channel flows. In the next subsections, we discuss more quantitative comparisons with 2D simulations as well as the effect of choosing different aspect ratios in 3D runs.

\subsubsection{2D vs.~3D MRI dynamics}
\label{sec:2Dvs3D}

\begin{figure*}
\centering
\includegraphics[width=1\textwidth, trim={0mm 50mm 0mm 0mm}, clip]{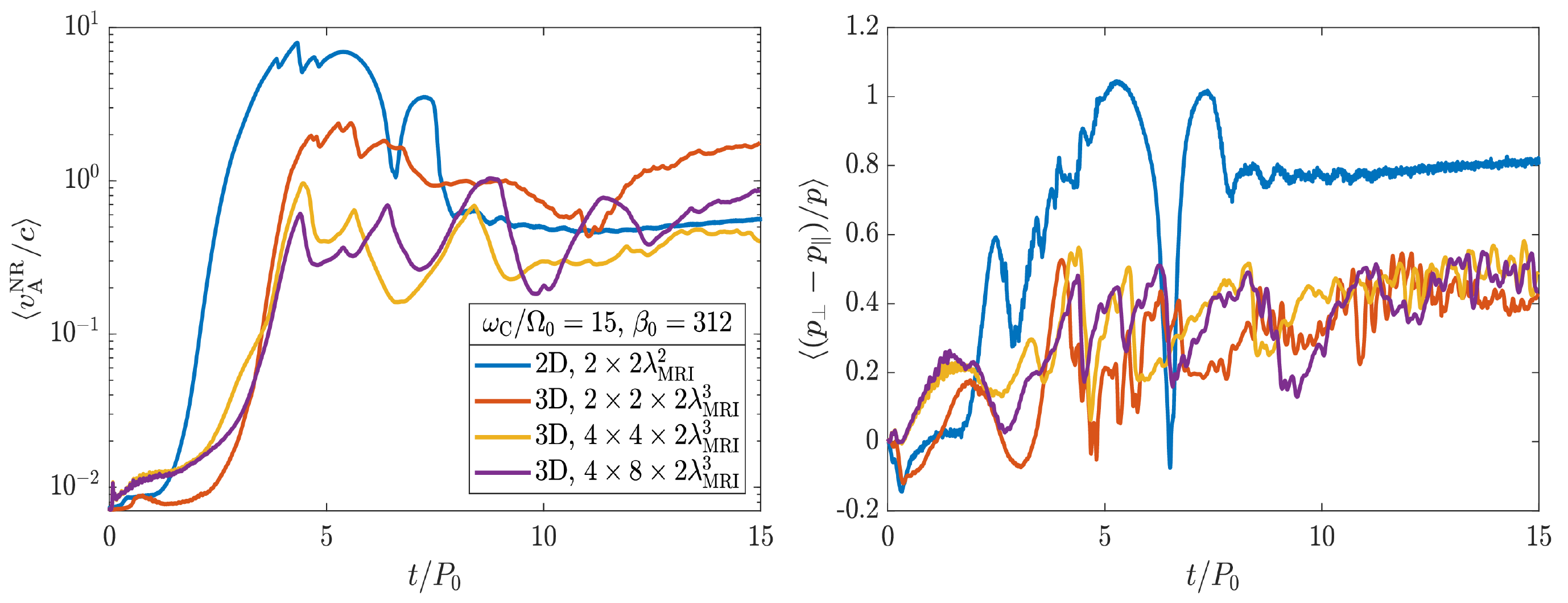}
\caption{Left: evolution of the volume-averaged Alfv\'en speed (calculated with the nonrelativistic expression) for several 3D PIC simulations of the pair-plasma MRI, compared with a 2D simulation using the same initial physical parameters. In the 3D runs, the constraint $v_{\rm A}^\mathrm{NR}/c<1$ is much better respected. Right: evolution of the volume-averaged pressure anisotropy, which influences the MRI growth rate, in the same runs. The 3D runs develop smaller anisotropy over longer times.}
\label{fig:2Dvs3D}
\end{figure*}

Figure \ref{fig:2Dvs3D} shows the evolution of the volume-averaged Alfv\'en speed (left panel), calculated with the nonrelativistic expression, and of the volume-averaged pressure anisotropy $(\pperp-\ppar)/p$ (right panel) for the two runs presented in the previous section (with box size $2\times2\times2\lmri^3$ and $4\times8\times2\lmri^3$, respectively), as well as for an analogous run with box size $4\times4\times2\lmri^3$. In the same plots, these 3D runs are compared with a small-box 2D simulation (with box size $2\times2\lmri^2$) employing the same physical parameters. We observe that the 3D simulations display a qualitatively different behavior overall.

As was also found in earlier work (\citealt{riquelme2012,inchingolo2018}), in the 2D run the average $v_\textrm{A}^\mathrm{NR}$ vastly exceeds the speed of light during the nonlinear stage. For 2D, \cite{inchingolo2018} attributed this behavior to the limited size of the simulation box, showing that for larger domains $\gtrsim8\times8\lmri^2$ the constraint $v_\mathrm{A}^\mathrm{NR}/c<1$ was fulfilled over short times. However, our 2D campaign (section \ref{sec:2D}) shows that in all cases (irrespective of the box size) the system inevitably evolves such that $v_\mathrm{A}^\mathrm{NR}/c>1$ over sufficiently long times. This occurs because the out-of-plane $B_y$ fails to saturate in 2D, due to a lack of dissipation processes (i.e.\ reconnection) along $y$ (see section~\ref{sec:explanation2D}). The violation of the $v_\mathrm{A}^\mathrm{NR}/c<1$ condition breaks the underlying nonrelativistic shearing-box assumption, and therefore puts into question the physical validity of the 2D results.  In addition, the lack of saturation in 2D heavily contributes to the migration of MRI modes to scales larger than the box size.
Our 3D simulations shown in Figure \ref{fig:2Dvs3D} (left panel) display a different behavior, with the magnetic-field amplification (and therefore $v_\mathrm{A}^\mathrm{NR}/c$) reaching much smaller values. Even a limited-size, cubic-box run already results in $v_\mathrm{A}^\mathrm{NR}/c\sim 1$ (i.e.\ $\sim10$ times smaller than in the corresponding 2D case), and larger and larger 3D boxes progressively improve this result. Our largest two cases, with box size $4\times4\times2\lmri^3$ and $4\times8\times2\lmri^3$, firmly maintain $v_\mathrm{A}^\mathrm{NR}/c\lesssim 1$ for the whole duration of the nonlinear stage.

The fact that in 3D MRI saturates at lower magnetic-field amplitudes implies that unstable MRI wavelengths do not migrate to larger length scales as quickly as in 2D. Thanks to the more efficient reconnection during the nonlinear stage, in 3D the MRI is not aggressively suppressed by the unbound growth of magnetic fields. This supports our statement on the significance of 3D simulations: capturing the MRI and reconnection dynamics along the toroidal direction is essential to model a physically meaningful evolution of the instability, which can then maintain a saturated state for longer times.

2D simulations also display a qualitatively different behavior in the development of pressure anisotropy: Figure \ref{fig:2Dvs3D} (right panel) shows the evolution of the volume-averaged $(\pperp-\ppar)/p$ in the same 2D and 3D simulations. In 2D, the anisotropy rapidly increases during the nonlinear stage, and progressively accumulates as the MRI continues to amplify $B_y$ without the dissipative effect of reconnection along $y$, up to a maximum $\langle(\pperp-\ppar)/p\rangle\simeq1$. In all 3D runs, this anisotropy is limited to smaller values (up to a maximum $\langle(\pperp-\ppar)/p\rangle\simeq0.5$) for the whole duration of the nonlinear phase, mildly increasing toward the end of the run. This is verified both in the cubic-box case and in larger boxes, indicating that the additional reconnection along $y$ plays a major role in constraining the development of pressure anisotropy. This anisotropy, eventually, contributes to pushing MRI wavelength to larger scales and to halt the MRI development, but over longer timescales than in corresponding 2D cases.

\subsubsection{Effect of the aspect ratio}
\label{sec:aspectratio3D}

The 2:4:1 aspect ratio employed for the simulation presented in Figure \ref{fig:evol_phases_large_3D} has been chosen by carefully analyzing a number of runs with varying $L_x,L_y,L_z$. As illustrated in section \ref{sec:parameters}, the physical domain size in $z$ (i.e.\ $L_z/\lmri$) heavily impacts the simulation cost, when demanding that the macroscopic scale ordering $H/L_z\ge1$ be respected. In practice, choosing $L_z/\lmri$ and $\lmri/\rho_\mathrm{C}$ determines the required number of grid cells per $\lmri$ and the number of time steps per~$P_0$. Then, the aspect ratios $L_x/L_z$ and $L_y/L_z$ are free parameters, each impacting the overall simulation cost linearly (i.e., quadratically when combined). Ideally, we are free to choose a small vertical extent, e.g.\ $L_z/\lmri=2$, and then elongate the box in $x$ and $y$ to better include the reconnection physics (as shown in sections \ref{sec:aspectratio2D} and~\ref{sec:explanation3D}). However, the $x$-extent of the box cannot be increased indefinitely, since we demand that the condition $v_\mathrm{s}(L_x)/c<1$ be satisfied at the radial boundaries. Our tests indicated that numerical artifacts appear at the boundary when $L_x/L_z\ge4$ (corresponding to $v_\mathrm{s}(L_x)/c\ge0.5$). Hence, we choose $L_x/L_z=2$ in our 3D simulations. Elongating the box in $y$ is less problematic, since no constraints related to the nonrelativistic shearing-box model exist in the toroidal direction\footnote{Note however that curvature effects have been neglected when constructing the shearing-box equations; $L_y/L_z$ can then be larger than $L_x/L_z$, but must remain limited to a certain extent.}. We have chosen $L_y/L_z=4$ in our largest run.

In earlier sections, we have argued that boxes with larger $L_x/L_z$ produce better developed turbulence; Figure \ref{fig:2Dvs3D} indeed demonstrates that as we enlarge a 1:1:1 box to 2:2:1 and then to 2:4:1, we can observe the emergence of the characteristic cycles of creation and destruction of channel flows. Reconnection acts more efficiently when $L_x/L_z>1$, resulting in dynamics more similar to that expected from 3D MHD numerical experiments. It remains to be assessed whether the effect of the aspect ratio changes when macroscopic and microscopic length scales are better separated, and whether even larger $L_y/L_z$ produce significant changes in the results. Due to the exceptional computing costs involved for such simulations, we defer this analysis to future work.

\subsection{Spectra of 3D MRI-driven turbulence}
\label{sec:spectra3D}

\begin{figure}
\centering
\includegraphics[height=0.80\columnwidth, trim={0mm 0mm 10mm 0mm}, clip]{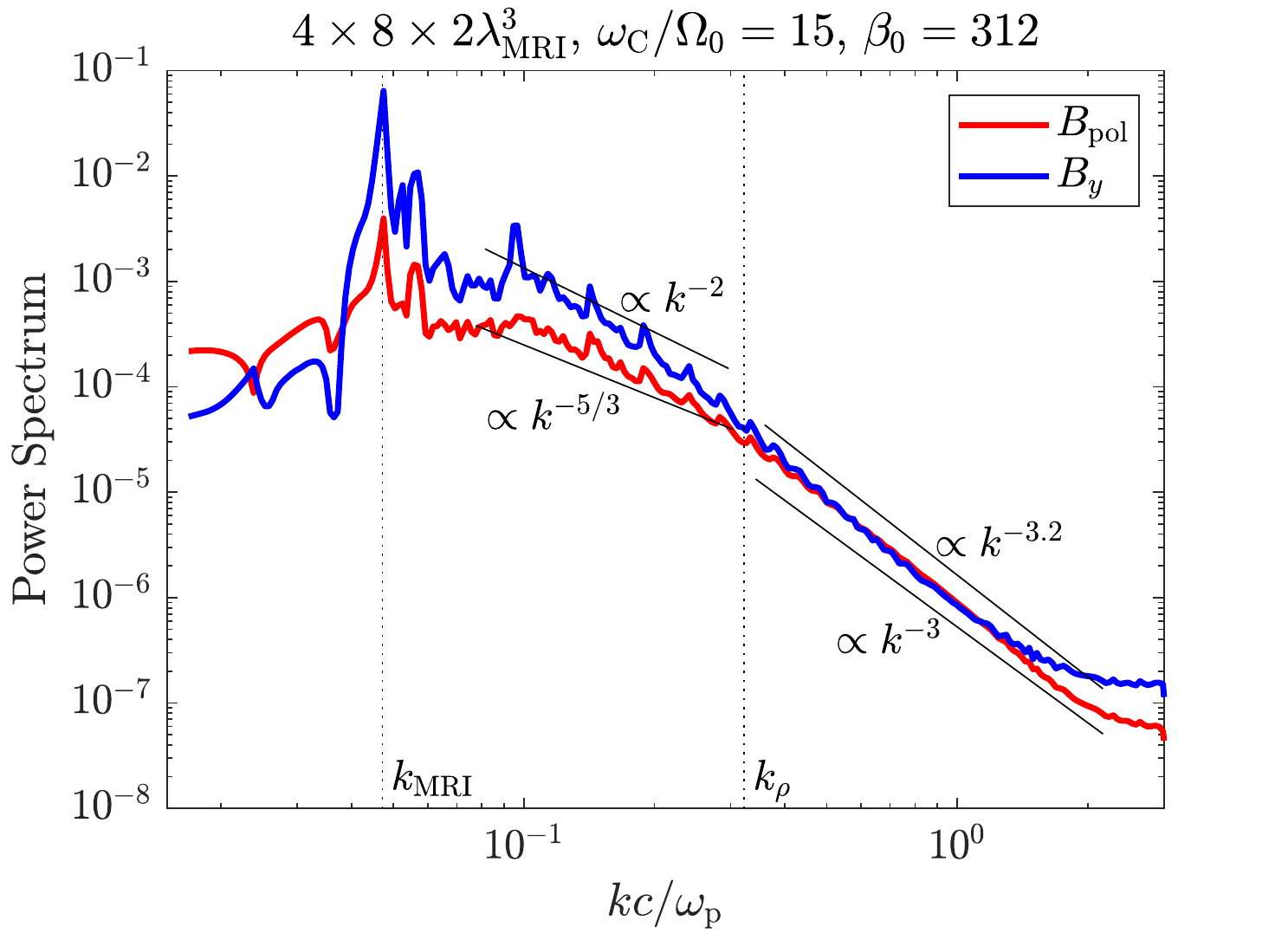}
\caption{Isotropic power spectrum of the poloidal (red line) and toroidal (blue line) magnetic field for the 3D pair-plasma MRI simulation of Figure \ref{fig:evol_phases_large_3D} ($\omrat=15$, $\beta_0=312$, and box size $4\times8\times2\lmri^3$). The spectra are averaged over one channel disruption/creation cycle during the nonlinear stage (between $t\simeq9P_0$ and $t\simeq11 P_0$), showing a persisting peak around the most-unstable MRI wavenumber $k_\mathrm{MRI}$. The spectra feature characteristic slopes indicating an inertial range, and a spectral break in the vicinity of the average Larmor-radius wavenumber $k_\rho$ measured at $t\simeq10 P_0$.}
\label{fig:spectra3D}
\end{figure}

In Figure \ref{fig:spectra3D} we show the isotropic power spectrum of the poloidal ($B_\mathrm{pol}=\sqrt{B_x^2+B_z^2}$) and toroidal ($B_y$) magnetic-field components for the large-box simulation discussed in Section \ref{sec:explanation3D} (with parameters $\omrat=15$, $\beta_0=312$, and box size $4\times8\times2\lmri^3$; see Figure~\ref{fig:evol_phases_large_3D}). This run is characterized by an ``episodic'' behavior during the nonlinear stage, where cycles of channel creation/destruction can be clearly identified. The magnetic-field strength and configuration change drastically throughout each cycle; for this reason, in Figure \ref{fig:spectra3D} we focus on the cycle occurring toward the end of the nonlinear stage (between $t\simeq9 P_0$ and $t\simeq11 P_0$, see Figure \ref{fig:evol_phases_large_3D}), and time-average the power spectra throughout the cycle.

The results show a persistent peak in the spectra around $k_\mathrm{MRI}$, the wavenumber of the most-unstable MRI mode calculated with the initial simulation parameters. This peak is indicative of the fact that the MRI is active through the cycle, although magnetic-field dissipation via reconnection is clearly dominant during the dissipation phase, when channels are being disrupted and magnetic energy is converted into heat. Such dissipation results in a turbulent cascade, which can be identified in the spectra via the presence of an inertial range with characteristic slopes. At moderate wavenumbers, the poloidal-field spectrum features a shallow slope which could be comparable to~$k^{-5/3}$, although the spectra are quite noisy (especially in the inertial range) and a precise measurement of the spectral slopes is hard to obtain. Regardless of the exact power-law index, we clearly observe that a spectral break occurs around $k\simeq k_\rho$, with $k_\rho$ the wavenumber corresponding to the average Larmor radius measured at $t\simeq10 P_0$ (i.e.\ at the midpoint in time through the cycle). In the sub-Larmor range ($k>k_\rho$), the $B_\mathrm{pol}$ spectrum steepens to a slope $\propto k^{-3.2}$; in Figure \ref{fig:spectra3D}, a slope $\propto k^{-3}$ is also indicated to guide the eye. The toroidal-field spectrum similarly features a shallow slope consistent with $k^{-2}$ at $k<k_\rho$ (although this part of the spectrum is also rather noisy), and a spectral break and steepening to $\propto k^{-3.2}$ in the kinetic range.

These 3D spectra present similarities as well as substantial differences with those shown in Figure \ref{fig:spectra2D} for the 2D case. Like in 2D, for the poloidal field we observe an inertial range with a $\propto k^{-5/3}$ slope, indicative of a turbulent MHD-type cascade driven by energy injection at large scales; similarly to 2D, for $B_\mathrm{pol}$ we measure a spectral break and a steepening in the sub-Larmor range, although in 3D this part of the spectra features a slightly steeper $\propto k^{-3.2}$ slope. Especially at kinetic scales, 2D and 3D spectra show substantial differences: the $B_y$ spectrum in 3D presents a spectral break not observed in 2D, and the kinetic range is here much more extended. In addition, we pointed out that in 2D the spectra continuously evolve and, over time, the measured characteristic slopes do not persist. We believe that this is an artifact of the reduced 2D dimensionality; in 3D, we have verified that the spectra shown in Figure \ref{fig:spectra3D} consistently appear over each channel creation/disruption cycle in the nonlinear stage as long as the MRI is active. 

Our 3D results are in good agreement, at least in the inertial range ($k<k_\rho$), with the 3D hybrid simulations of \cite{kunz2016} as well as with 3D MHD results (e.g.\ \citealt{walker2016}). Slight differences with these works arise in the kinetic range ($k>k_\rho$), where we measure a $\propto k^{-3.2}$ slope instead of the $\propto k^{-3}$ reported by \cite{kunz2016}, whose result was however related to the ion sub-Larmor scales in a hybrid simulation, while here we are considering fully kinetic pair plasmas. Future 3D PIC runs with electron-ion plasma will clarify whether this discrepancy is related to our pair-plasma assumption.

\subsection{Particle energization in 3D}
\label{sec:ene3D}

\begin{figure*}
\centering
\includegraphics[width=1\textwidth, trim={0mm 45mm 0mm 0mm}, clip]{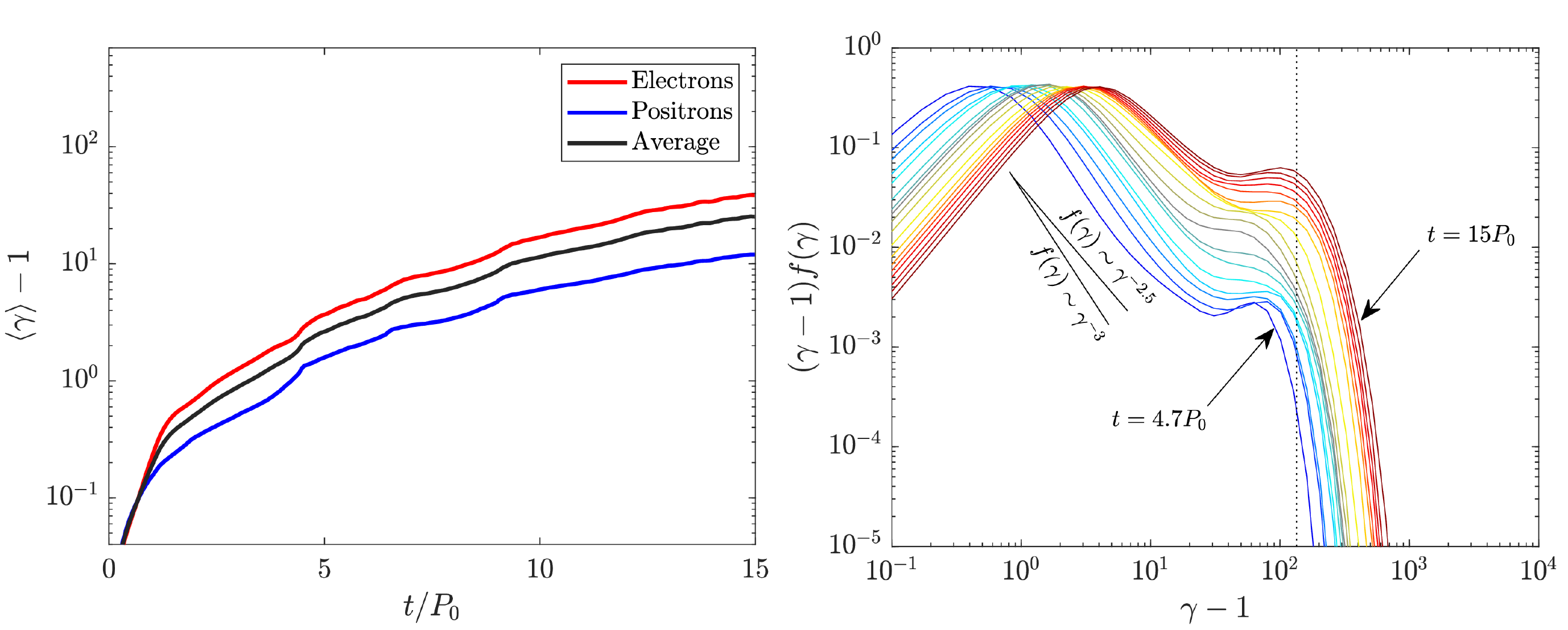}
\caption{Left: evolution of the average Lorentz factor for electrons, positrons, and for the two species combined in the 3D pair-plasma MRI simulation of Figure \ref{fig:evol_phases_large_3D} ($\omrat=15$, $\beta_0=312$, and box size $4\times8\times2\lmri^3$). As in 2D, electrons heat significantly more than positrons. Right: evolution of the particle energy distribution during the nonlinear stage ($t\simeq4.7\mbox{--}15 P_0$) of the same simulation. Right after the onset of reconnection at $t\simeq4.7P_0$, the distribution has developed a nonthermal tail following a power law (the slopes of power laws with index 2.5 and 3 are shown to guide the eye). The nonthermal population persists until the end of the run, while a separated high-energy peak develops around $\gamma\sim100$ (dotted line), which corresponds to the maximum attainable energy due to the system size in the $z$-direction.}
\label{fig:energy3D}
\end{figure*}

In this section, we briefly analyze particle energization in the 3D pair-plasma MRI. We focus on our largest simulation presented in Figure \ref{fig:evol_phases_large_3D} ($\omrat=15$, $\beta_0=312$, and box size $4\times8\times2\lmri^3$) and measure the evolution of the mean Lorentz factor for electrons and positrons separately, as well as the average value for both species. In addition, we compute the energy distribution function at subsequent times during the simulation. The results are shown in Figure~\ref{fig:energy3D}.

Much like in the 2D case (section \ref{sec:ene2D}), we observe that the particle energy constantly increases throughout the run, with the average Lorentz factor reaching values $\sim10\mbox{--}40$ at the end of the simulation (left panel). Also similarly to our 2D simulations, we measure significant differential heating between the two particle species, with electrons reaching energies $\sim4$ times larger than positrons on average. The energy distribution functions measured for both species together (right panel) show that right after the onset of reconnection at $t\simeq4.7P_0$, there is a significant nonthermal component in the energy spectra; this nonthermal tail follows a power law that persists until the end of the run as the plasma heats, while an additional high-energy peak progressively develops around $\gamma\simeq100$ (indicated by a dashed line in Figure~\ref{fig:energy3D}). At the end of the simulation ($t\simeq15P_0$), particles have accumulated around the peak energy, eroding the power-law tail of the spectrum.

These 3D results show both remarkable similarities and qualitative differences with our 2D runs. First, the differential heating already found in 2D simulations arises in 3D essentially in the same way, with electrons gaining more energy than positrons; the mechanism driving this effect is the same as in 2D simulations (on which we elaborate in appendix~\ref{app:diffene}). The plasma heating proceeds, on average, very similarly to our large-box 2D runs (see Figure~\ref{fig:energy2D}), where developed Alfv\'enic turbulence drives a rather smooth increase in particle kinetic energy. In 3D, the ``episodic'' nature of the channel creation/disruption cycles manifests itself in the evolution of~$\langle\gamma\rangle$, where we observe periodic (small) increases corresponding to the repeated onset of reconnection events. The particle energy distribution also evolves similarly to the 2D case, at least in the beginning of the nonlinear stage; like in 2D, the slope of the power-law tail is $\sim 2.5\mbox{--}3$, persisting until the end of the run. A qualitative difference with 2D simulations lies in the end state reached in 3D runs: the high-energy peak shown in Figure \ref{fig:energy3D} here precisely corresponds to the Lorentz factor of particles with $\rho_\mathrm{C}\simeq L_z$ (calculated with the mean magnetic-field strength at saturation). In 2D, we observed that a high-energy peak develops due to the ``quiet-loop'' end state that emerges due to the reduced dimensionality; here, it appears that the system is simply reaching a state where particles accumulate around the highest attainable energy according to the simulation-box size. This evolution is very similar to that of forced-turbulence kinetic simulations, where particles accumulate, at late times, around the maximum energy determined by the box size (\citealt{zhdankin2018conv}).
In our runs, the accumulation of particles around the maximum energy and the shift of the distribution to larger average energies progressively reduce the extent of the power law. The development of the high-energy peak also relates to (but is not a direct consequence of) the differential heating discussed above, with electrons accumulating faster around the highest attainable energy.

Both the differential heating and the emergence of the high-energy peak can be ameliorated with larger boxes and/or better scale separation, which imply higher computational costs; for example, we have conducted preliminary runs with larger domain sizes, confirming that in such cases the high-energy peak shifts to higher~$\gamma$, proportionally to the extent of the smallest side (usually~$L_z$) of the simulation domain. Systematically studying the effect of box size and scale separation on particle energization in 3D involves high computational costs, and we thus defer it to future work.

\subsection{Stresses and angular-momentum transport}
\label{sec:stresses3D}


\begin{figure*}
\centering
\includegraphics[width=1\textwidth, trim={2.5mm 40mm 0mm 0mm}, clip]{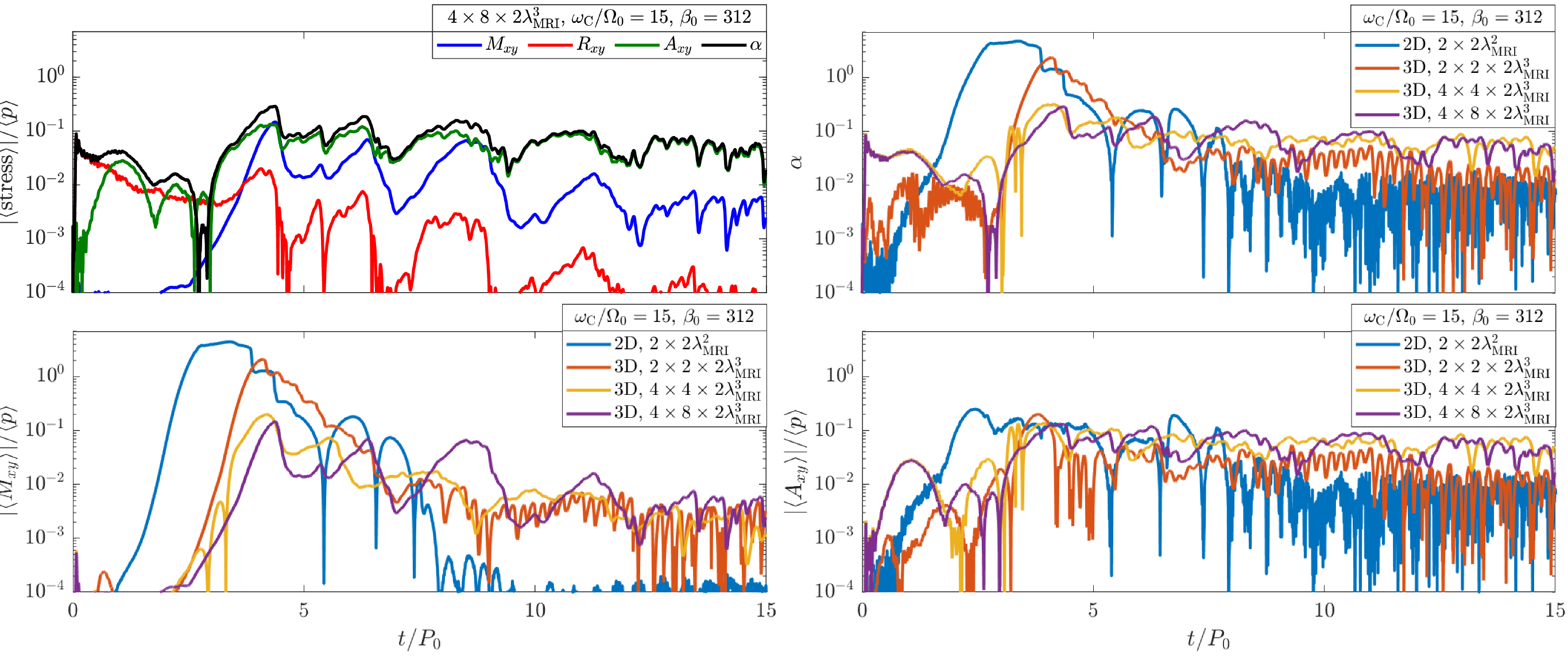}
\caption{Top left: evolution of $\alpha$ and of the volume-averaged Maxwell, Reynolds, and anisotropic stress in the 3D pair-plasma MRI simulation of Figure \ref{fig:evol_phases_large_3D} ($\omrat=15$, $\beta_0=312$, and box size $4\times8\times2\lmri^3$). Top right: comparison of $\alpha$ for 2D and 3D simulations employing the same $\omrat$ and $\beta_0$ and different box sizes. Bottom left and right: comparison of the volume-averaged Maxwell and anisotropic stress for the same 2D and 3D simulations.}
\label{fig:stresses3D}
\end{figure*}

To conclude our first study of 3D MRI simulations, we consider the evolution of the $\alpha$-parameter and of the viscous stresses. Figure \ref{fig:stresses3D} (top left) shows the measured values of $\alpha$ and of the volume-averaged $M_{xy}$, $R_{xy}$, and $A_{xy}$ for the simulation presented in Figure~\ref{fig:evol_phases_large_3D} ($\omrat=15$, $\beta_0=312$, and box size $4\times8\times2\lmri^3$). We clearly observe that, for this run, the anisotropic stress dominates the overall angular-momentum transport; this differs substantially from 2D cases, where $A_{xy}>M_{xy}$ was verified only at late times (when the MRI is halting and pressure anisotropy accumulates). The top-right, bottom-left, and bottom-right panels of Figure \ref{fig:stresses3D} show, respectively, the evolution of $\alpha$ and of the volume-averaged $M_{xy}$ and $A_{xy}$ for the 2D and 3D cases already compared in section~\ref{sec:2Dvs3D}. In all these cases $\omrat=15$ and $\beta_0=312$, while the box size and dimensionality are varied to compare the 2D and 3D dynamics as well as the 3D small- and large-box evolution. We observe that, during the active nonlinear phase, $\alpha$ is smaller for 3D runs than in the 2D case; a further decrease in $\alpha$ occurs from small to large box sizes in~3D runs. Comparing the most important stress components, we notice that $M_{xy}$ follows the same decreasing trend observed for $\alpha$, while $A_{xy}$ remains essentially unchanged across all runs. It is therefore evident that the smaller value of $\alpha\simeq0.1$ achieved in the large-box 3D case (whereas $\alpha\ge1$ in the active nonlinear phase of the 2D run) is solely related to a decrease in the Maxwell stress.

The explanation for this behavior is straightforward: as we have argued in the previous sections, large 3D boxes drive much more efficient reconnection dynamics, which in turn results in a magnetic-field saturation at lower values than both small 3D simulations and arbitrarily large 2D runs. The Maxwell stress, on average, becomes correspondingly smaller and so does~$\alpha$. Since $A_{xy}$ remains somewhat constant across all these cases, our results could in principle suggest that the main driving mechanism for angular-momentum transport in large 3D simulations is represented by pressure anisotropy (this result was also reported by \citealt{hoshino2015}). However, we believe that the large value of $A_{xy}$ observed here may simply be an artifact of the reduced scale separation employed in our 3D numerical experiments. In particular, we measure $\langle A_{xy}\rangle/\langle p\rangle\simeq 0.1$ in these runs, which is nearly twice as large as the value found in 2D simulations with large boxes or large scale separation (compare Figure~\ref{fig:stresses3D}). In such cases, pressure anisotropy is more efficiently limited by developed turbulence and/or faster mirror modes; with the limited scale separation and box size employed in our 3D runs, it is not unreasonable to think that the measured anisotropy is thus exaggerated. 3D Braginskii-MHD and hybrid simulations (\citealt{sharma2006,kunz2016}) have indeed shown that $A_{xy} \sim M_{xy}$ is expected in fluid modeling, but that anisotropic momentum transport does not typically dominate the overall value of $\alpha$ by the large margin we observe here. 
To assess the precise role and trend of anisotropic stresses in 3D fully kinetic simulations, we will conduct large-box studies with better scale separation in future work.

\section{Discussion and conclusions}
\label{sec:conclusions}

In this work, we have presented a comprehensive exploration of the magnetorotational instability (MRI) in collisionless pair plasmas via fully kinetic Particle-in-Cell (PIC) simulations. With a shearing-box setup implemented in our relativistic PIC code \textsc{Zeltron} (\citealt{cerutti2013}), we have carried out a vast array of 2D runs, exploring an unprecedentedly large parameter space. In particular, we have conducted very large-scale 2D simulations with macroscopic-to-microscopic temporal-scale separation up to $\sim 10$ times larger than previous works and with system size up to twice larger than the largest simulations presented in literature. In addition, we have carried out large-scale 3D PIC simulations of the MRI in pair plasmas, achieving for the first time a global mesoscale dynamics akin to that observed in MHD works. To study the axisymmetric MRI with PIC, we have resorted to the established 2D shearing-coordinate approach employed in previous publications (\citealt{riquelme2012,inchingolo2018}); for 3D nonaxisymmetric simulations, we have developed and applied a novel ``orbital-advection'' formulation of the shearing box that simplifies pre-existing methods (\citealt{hoshino2013,hoshino2015}) and is less complicated to implement numerically.

An important result of our study is the demonstration that 2D (axisymmetric) fully kinetic simulations of the MRI are qualitatively different from corresponding 3D kinetic runs. We have pointed out for the first time such differences, and provided a physical explanation for the behavior of the kinetic MRI in 2D. We found that axisymmetric PIC simulations are intrinsically limited in the physical processes they can model, and such a limitation has important consequences for the overall MRI evolution. In particular, the lack of reconnection along the out-of-plane (toroidal) direction results in a lack of saturation of the strongest magnetic-field component, $B_y$, with detrimental consequences for the development and sustainment of turbulence. However, we also found that many important lessons can still be learned from 2D runs which carry over to the 3D case:
for example, with our extensive parameter scans in 2D, we have determined the minimal values of key physical parameters (in terms of the dimensionless scale separation, system size, etc.) that need to be reached for meaningful studies. Thus, 2D simulations give us guidance (starting point) for designing 3D runs. Differences and similarities between the 2D and 3D cases are illustrated in detail below.

By conducting large-scale 3D simulations of the MRI, for the first time we have demonstrated that fully kinetic simulations can reproduce the mesoscale dynamics typically observed in MHD: multi-dimensional, macroscopic channel flows cyclically form and get disrupted by parasitic instabilities, particularly via drift-kink modes. The channel disruption results in developed turbulence, and several channel creation/growth/disruption cycles can be observed during the nonlinear stage in large-scale 3D runs. This result also aligns with previous 3D hybrid simulations (\citealt{kunz2016}). Conversely, channels developing in small-scale runs are not efficiently disrupted and no turbulence develops, in agreement with the sole 3D fully kinetic study presented in literature so far (\citealt{hoshino2015}), which employed a simulation box of limited size.

In addition to studying the general MRI dynamics, we have explored in detail several physical processes that develop concurrently with the linear and nonlinear stage of the main instability. Our results can be summarized as follows:
\begin{itemize}
    \item We have extensively described the typical nonlinear evolution of the MRI in fully kinetic shearing-box simulations, explaining each phase of the evolution and the differences between the small- and large-box cases, as well as between the 2D and 3D cases. We have explained how small and large simulations differ in producing developed turbulent states, and how the choice of physical parameters impacts the overall evolution and duration of the nonlinear MRI. We have also identified the role of pressure anisotropy in these simulations, observing that a large $\Delta p=\pperp-\ppar$ can develop and accumulate throughout the nonlinear stage of the instability; this anisotropy is likely exaggerated by our choice of modest scale separation (but still much larger than that employed in previous studies) that necessarily characterizes PIC simulations. A limited scale separation can result in inefficient mirror modes, which grow much slower than the MRI, allowing for persistently large pressure anisotropy. This anisotropy ultimately participates in pushing MRI modes to wavelengths larger than the box size, halting the MRI dynamics over long times.
    
    \item By exploring a large parameter space, we have assessed the numerical convergence of key quantities (e.g.\ the magnetic-field amplification) in 2D simulations when the physical parameters are varied. We have studied the effect of the separation between the macroscopic ($\Omega_0^{-1}$) and microscopic ($\omega_C^{-1}$) temporal scales, $\omrat$; of the separation between box size and MRI scales~$L/\lmri$; of the initial plasma-$\beta$; and of the box aspect ratio. 
    Increasing $\omrat$ with a fixed box size generally results in a longer duration of the nonlinear stage and in an increase of the magnetic energy at saturation. We have shown that, at some large~$\omrat$, the results (in particular, the saturated magnetic energy) eventually converge; however, this converged nonlinear state is qualitatively different for small and large boxes, and the precise value of $\omrat$ resulting in convergence depends on the box size.
    Similarly, fixing $\omrat$ and only increasing the box size produces results that converge for sufficiently large box sizes; in large boxes, a sustained-turbulence state can develop while ti does not in small simulations. This partly agrees with previous 2D work by \cite{inchingolo2018}, who focused their analysis on increasing box sizes while keeping $\omrat$ fixed. They concluded that convergence in the magnetic-field amplification is attained, in 2D, for box sizes of at least $8\times8\lmri^2$, and that large-box simulations can maintain a volume-averaged $v_\mathrm{A}/c<1$ (calculated with the nonrelativistic expression) throughout the system's evolution, respecting the underlying assumptions of the nonrelativistic shearing box. We have instead demonstrated that 2D simulations, over sufficiently long times, invariably develop an average $v_{\mathrm{A}}/c>1$, owing to the absence of reconnection in the $y$-direction. This does not occur, instead, in our 3D runs of sufficiently large size; we conclude that 3D simulations are of key importance to obtain physically valid results that respect all the underlying assumptions of the shearing-box model.
    Considering the effect of the initial plasma-$\beta$, we found that the results are practically unchanged in the range $\beta_0\simeq0.1\mbox{--}2{,}500$: independent of the initial $\beta_0$, all simulations end up with plasma beta more or less within the same universal final range, of order $0.1\mbox{--}1$ for 2D runs. This result holds as long as the unstable MRI modes can fit in the simulation box, i.e.\ under the constraint $L_z>\lmri$.
    Finally, we have observed that the geometry of the simulation box can impact the development of the nonlinear MRI stage dramatically: boxes with aspect ratio $L_x/L_z>1$ result in developed turbulence much more easily than corresponding boxes with $L_x/L_z=1$, both in 2D and~3D. We have observed that this is caused by additional drift-kink modes that are impeded in small simulation domains, and that can promote channel-disruption events in elongated boxes instead.
    
    \item We have verified that sustained turbulence develops, in our simulations, during the nonlinear MRI evolution. The isotropic power spectra of the poloidal ($B_\mathrm{pol}=\sqrt{B_x^2+B_z^2}$) and toroidal ($B_y$) magnetic field during this phase show the presence of an inertial range with characteristic power laws, indicative of turbulent activity. Both in 2D and 3D, the $B_\mathrm{pol}$ spectrum features a shallow slope roughly consistent with $\propto k^{-5/3}$ at large and intermediate spatial scales, and a spectral break in the vicinity of the average Larmor-radius wavenumber~$k_\rho=\rho_\mathrm{C}^{-1}$. In the 3D case, at length scales below $\rho_\mathrm{C}$ the $B_\mathrm{pol}$ spectrum steepens to a $\propto k^{-3.2}$ slope and a clear kinetic range is present. The $B_y$ spectrum appears to follow a $\propto k^{-3}$ slope at all scales in~2D; in 3D, $B_y$ instead follows a shallower slope compatible with $\propto k^{-2}$ in the inertial range, and a $k^{-3.2}$ slope in the kinetic range. Our 2D results are consistent with previous 2D studies (\citealt{inchingolo2018}); our 3D results for $B_y$, in the inertial range, are in agreement with MHD and hybrid-kinetic simulations (\citealt{kunz2016,walker2016}). However, the latter comparison is complicated by the difference in the underlying models (our fully kinetic PIC vs.\ MHD or hybrid methods). Considering works focusing on pair plasmas, a similarity can be drawn between our model and that employed in \citealt{zhdankin2017,zhdankin2018}, since both cases consider a forced-turbulence system evolution. However, in those works a $\propto k^{-4}$ slope (or steeper) was found for $k>k_\rho$, which differs from our shallower $k^{-3.2}$ result in the kinetic range. This measurement is also hard to compare against analytic expectations, as no previous studies (to the best of our knowledge) focused on the specific case of pair-plasma MRI-driven turbulence we consider. For example, \citealt{loureiroboldyrev2018} considered a low-$\beta$ pair plasma in a tearing-mediated cascade, finding a $\propto k^{-3}$ slope in the kinetic range; our conditions, however, are those of a high-beta plasma and we have no basis to claim that our cascade is tearing-mediated in nature. It is also interesting to note that solar-wind observations commonly report a spectral slope $\propto k^{-7/3}\mbox{--}k^{-8/3}$ (not extremely far from our $\propto k^{-3.2}$ result) at sub-ion-Larmor scales, consistent with gyrokinetic calculations (e.g.\ \citealt{schekochihin2009}). Whether a similarity can be drawn between this and our case will be verified with future electron-ion simulations.
    
    \item Concerning particle energization in 2D and 3D, we found that energy injection in large runs is akin to that brought about by Alfv\'enic turbulence. In these simulations turbulence is well developed, and particle heating proceeds smoothly, with the system achieving a steady-state balance between turbulent magnetic-energy dissipation and particle energization. In contrast, when turbulence is not well developed (e.g.\ in small boxes), particle energization occurs mostly during short ``bursts'' corresponding to large-scale reconnection events.
    Moreover, for the first time we have reported that substantial differential heating can occur between electrons and positrons in MRI simulations. The physical mechanism behind this phenomenon is represented by additional (related to the background differential rotation) drift forces that affect opposite charges differently, causing a symmetry break in the gyromotion of electrons and positrons in a uniform magnetic field. This effect is unrealistically large in our simulations due to our choice of (necessarily) limited scale separation~$\omrat$. We have elaborated on how tuning the values of physical parameters can ameliorate this issue and its implications for the interpretation of the results.
    
    \item The energy distribution functions in our 2D and 3D simulations show the presence of substantial nonthermal particle acceleration during the nonlinear MRI stage. A power-law tail with index $\sim2.5\mbox{--}3$ consistently develops as the MRI transitions to sustained turbulence, both in 2D and in~3D. Our results are in good agreement with previous works; however, here we have demonstrated that in 2D this power-law state is transitory, and that the overall evolution can produce very different energy distributions over time. \cite{inchingolo2018} carried out a similar analysis but focused on the early nonlinear stage, where a power-law index $\sim2$ can indeed be realized; but such a state is clearly still evolving, and changes drastically later on. At late times, the power-law part of the distributions progressively disappears, and a high-energy peak develops. This is completely determined by the artificial 2D end state (with characteristic ``magnetic loops''), driven by the reduced dimensionality of these runs. \cite{riquelme2012} described a similar late-time 2D dynamics, but attributed this evolution to the MRI; we emphasize that such a state is in fact not representative of the response of particles to the MRI, since the latter has slowed down (or completely stopped) by the time when the quiescent end state has developed.
    Conversely, in our 3D simulations the developed nonthermal features are maintained with the same power-law index until the end of the run. However, the nonthermal tail is progressively eroded by the accumulation of particles around the highest energy attainable (which increases with box size). \cite{hoshino2015} observed a similar effect in small-box 3D simulations; this result is also consistent with forced-turbulence PIC simulations (\citealt{zhdankin2018conv}) and points to the need for larger (and more expensive) 3D numerical experiments.
    
    \item In our 2D and 3D simulations, we observed that viscous stresses (in particular, the dominant $xy$-component of the stress tensor) can develop during the nonlinear MRI stage. This can lead to efficient angular-momentum transport: an effective (dimensionless) collisionless viscosity $\alpha=(\langle M_{xy}\rangle+\langle A_{xy}\rangle+\langle R_{xy}\rangle)/\langle p\rangle$ (\citealt{shakurasunyaev1973}) arises both in 2D and 3D runs, mainly due to Maxwell and anisotropic stresses. In 2D, the Maxwell stress $M_{xy}=-B_xB_y/(4\pi)$ is likely exaggerated due to a lack of saturation in $B_y$ caused by the reduced dimensionality; in 3D large-scale runs, where magnetic fields saturate at lower amplitudes, this stress firmly settles on $\langle M_{xy}\rangle/\langle p\rangle\simeq0.01\mbox{--}0.1$, consistently with previous hybrid simulations (\citealt{kunz2016}). The anisotropic stress $A_{xy} = -(p_\perp-p_\|)B_xB_y/B^2$, instead, shows a more complex trend: when keeping the same~$\omrat$, the average pressure anisotropy $(p_\perp-p_\|)$ generally decreases from 2D to 3D and from small to large boxes; but since the average $B^2$ decreases as well, $\langle A_{xy}\rangle/\langle p\rangle$ maintains roughly the same average value $\sim0.1$ in 2D and 3D simulations. As a result, we measure $A_{xy}>M_{xy}$ in our largest 3D runs, which is against the general expectation $A_{xy}\lesssim M_{xy}$. We believe that this enhanced anisotropic angular-momentum transport is related to our (necessarily) limited scale separation and system size; with realistic parameters, efficient mirror modes would rapidly quench pressure anisotropy, likely decreasing~$A_{xy}$. \cite{hoshino2015} measured a similarly enhanced angular-momentum transport due to pressure anisotropy in small-box 3D simulations, which was likely due to the same mechanisms we describe here.
    
\end{itemize}

Although we have addressed several key aspects of the kinetic physics of the MRI, our study still presents several limitations. The most prominent open question is whether 3D PIC simulations could produce results that completely align with hybrid and MHD studies, and how the nonthermal particle acceleration and angular-momentum transport depend on the box size and choice of physical parameters. Here, we have conducted the largest (to date) 3D PIC runs that successfully reproduce the global mesoscale MRI dynamics, but our scale separation and box size remain rather limited. Larger simulations with better separation would be needed, i) to completely characterize the properties of particle energization, e.g.\ whether large-scale runs, where the box size does not rapidly interfere with the nonthermal part of the spectrum, present the same nonthermal features as those observed here; and ii) to understand the general trend of viscous stresses when the scale separation is substantially larger, possibly achieving the expected ordering $A_{xy}\lesssim M_{xy}$ in simulations where pressure anisotropy is well regulated. Moreover, the issue of substantial differential heating in 3D runs remains to be addressed, as it can complicate the analysis of particle energization even in pair-plasma simulations.

Despite these limitations, our work paves the way for several future studies focused on the exploration of other properties of the~MRI. For example, we have reported here the stark contrast between simulations with and without net magnetic flux; this is a well-known aspect of the MRI dynamics that has been extensively studied with MHD shearing-box models (e.g.\ \citealt{gardinerstone2005}). Several works have even pointed out that 3D simulations with and without net magnetic flux behave substantially differently for boxes with different aspect ratios (e.g.\ \citealt{bodo2008,walkerboldyrev2017}); moreover, in ideal MHD it is generally observed that the $\alpha$-parameter representing angular-momentum transport tends to 0 in the case of zero net magnetic flux, unless explicit dissipation coefficients are included (e.g.\ \citealt{fromangpapaloizou2007,fromang2007}). In collisionless models (including ours), such dissipative effects are naturally included (as shown e.g.\ by \citealt{kunz2016}), and the phenomenology of angular-momentum transport can be studied from first principles. All in all, we believe that a systematic PIC study of the zero-net-flux case may be particularly important to understand the exact mechanisms behind the phenomenology of MRI dynamo; we will pursue such a study in the future. Another important analysis to be carried out in future work concerns the initial magnetic-field configuration: here, we have limited ourselves to the purely vertical-field case, but it is well known that different field geometries (especially when an initial $B_y$ is present) can result in qualitatively different global MRI evolution and turbulence properties (e.g.\ \citealt{gardinerstone2005,goedbloedkeppens2022}). This aspect has not been considered in kinetic or hybrid works, but it would be important to study generic field geometries, which are expected in realistic astrophysical scenarios. Finally, we note that studying the kinetic MRI with vertical stratification (as opposed to the case without stratification that we considered) could in principle provide results that are more physically relevant for modeling thick accretion disks (e.g.\ \citealt{hirabayashihoshino2017}) and resolve convergence issues that notably affect MHD studies (e.g.\ \citealt{regevumurhan2008,bodo2014}).

We conclude by remarking that the next step in our line of work concerns electron-ion ($m_i\gg m_e$) simulations. Studying the case of large mass ratio, and especially the energy partition between particle species, is fundamental for the correct interpretation of data from current and future observations targeting RIAFs (\citealt{EHT2019e}). In these collisionless astrophysical environments, electrons and ions are decoupled and a two-temperature state can originate from turbulence and/or radiation-reaction dynamics (\citealt{arzamasskiy2019,zhdankin2019,kawazura2020,kawazura2021,zhdankin2021}). With our simulations, we will be able to study the process of temperature decoupling (and possibly the dynamics of radiation) during the MRI from first principles. Moreover, information from fully kinetic simulations can be used as input for global fluid simulations of accretion around compact objects employing subgrid models to account for the missing microphysics (\citealt{ressler2015,chael2018,scepi2022}). Given the ever-growing interest in understanding the dynamics of collisionless accretion disks around SMBHs, we believe that this work has the potential to substantially impact our insight into the physics of such space environments.

\section*{Acknowledgements}
The authors would like to thank Mario Riquelme, Masahiro Hoshino, Giannandrea Inchingolo, Andrea Mignone, Gianluigi Bodo, Jim Stone, Eliot Quataert, Anatoly Spitkovsky, Matthew Kunz, Elias Most, Sasha Philippov, Nuno Loureiro, and Nicolas Scepi for useful discussions and suggestions throughout the development of this work.

The authors acknowledge support from a NASA ATP grant 80NSSC20K0545.
Additional support was provided by National Science Foundation grants AST 1806084 and AST 1903335 to the University of Colorado.
F.B.\ was also partially supported by a Junior PostDoctoral Fellowship (grant number 12ZW220N) from Research Foundation -- Flanders (FWO).
Support for L.A.\ was provided by the Institute for Advanced Study.
Support for V.Z.\ at the Flatiron Institute is provided by the Simons Foundation.

This work made use of substantial computational resources provided by several entities. We acknowledge support by the VSC (Flemish Supercomputer Center), funded by the Research Foundation -- Flanders (FWO) and the Flemish Government -- department EWI.
An award of computer time was provided by the Innovative and Novel Computational Impact on Theory and Experiment (INCITE) program, using resources (namely Theta) of the Argonne Leadership Computing Facility, which is a DOE Office of Science User Facility supported under contract DE-AC02-06CH11357.
Additional computer time was provided by the Extreme Science and Engineering Discovery Environment (XSEDE), which is supported by National Science Foundation grant number ACI-1548562 \citep{towns2014}, and by the Frontera computing project \citep{stanzione2020} at the Texas Advanced Computing Center (TACC, \href{http://www.tacc.utexas.edu}{www.tacc.utexas.edu}). We acknowledge TACC for providing HPC resources on both Stampede2 and Frontera.
Additional resources (namely Pleiades) supporting this work were provided by the NASA High-End Computing (HEC) Program through the NASA Advanced Supercomputing (NAS) Division at Ames Research Center.

\appendix

\section{Numerical methods for the kinetic shearing box}
\label{app:methods}

\subsection{Shearing-coordinate simulations in 2D}
\label{sec:methods2D}
For two-dimensional simulations, in this work we employ the \KSBSC~formulation, which is preferable since no complicated boundary conditions are required (the simulation domain is periodic in all directions). Numerically, the implementation of the \KSBSC~system of equations in a PIC code requires the solution of modified Maxwell's equations and a peculiar particle-push step. These modifications are however straightforward: for the electromagnetic fields, the additional terms on the right-hand side of equations \eqref{eq:riquelmeB} and \eqref{eq:riquelmeE} are linear and point-wise, and can therefore be treated explicitly with a standard leapfrog algorithm by interpolating in time and space to maintain second-order accuracy (see \citealt{riquelme2012}). The particle-push step is similarly formulated as a modification of a standard Boris scheme that includes here extra (linear) terms. We refer to section \ref{sec:modBoris} for more details on this modified Boris algorithm.

\subsection{Orbital-advection simulations in 3D}
\label{app:methods3D}

For three-dimensional simulations, in this work we make use of the newly developed \KSBOA~framework (section \ref{sec:equationsKSBOA}). The numerical implementation of our method requires solving modified Maxwell's equations which generally include implicit terms, as well as a modified Boris push (similar to that required for 2D runs, see section \ref{sec:methods2D} above). Our numerical approach also requires shearing-periodic boundary conditions on electromagnetic fields, source terms, and particles.

\subsubsection{Solution of Maxwell's equations}
Our version of Maxwell's equations \eqref{eq:OAB}--\eqref{eq:OAE} involves nonlocal, coupled terms on the right-hand side; this complicates the numerical solution, since a simple leapfrog scheme cannot be applied to explicitly push the electromagnetic fields in time. To avoid costly matrix-inversion operations (see e.g.\ \citealt{bacchini2019b}), in our first numerical implementation we split the solution in two explicit steps. First, all terms on the right-hand side of Maxwell's equations that are not related to advection along $y$ are discretized with a central-difference scheme in time and space. The resulting system of discrete equations (excluding the advective terms) is iterated upon with a simple predictor-corrector step, which is applied 3 times\footnote{This approach is essentially an iterative Crank-Nicolson scheme. It is well known that this scheme is not asymptotically stable for an infinite number of iterations (see e.g.\ \citealt{leilerrezzolla2006}); the scheme is, however, stable for 3 iterations, which is our choice for all simulations. This represents a good compromise between stability and accuracy in the solution.}.
At the end of each predictor-corrector iteration, we can include the advection of $\vecE$ and $\vecB$ along $y$ in several ways. In our numerical experiments, we have observed that a simple central-difference scheme applied to these terms inevitably results in instabilities developing at the shearing boundaries; we could therefore opt for a more stable advection scheme among the many available in literature. However, as a first approach we choose an even simpler strategy: to model the advection of field quantities along $y$, we simply shift each component of $\vecE$ and $\vecB$ in space by a distance $v_{\mathrm{s},y}(x)\Delta t$ (with appropriate signs). The shifted arrays are interpolated at the resulting spatial locations, which approximates the advection step. This operation introduces some numerical diffusion, but we observe that this is in fact beneficial as it suppresses numerical instabilities at the shearing boundaries. Our approach represents a cheap, less sophisticated alternative to more classical schemes; while this leaves ample grounds for improvements, it only requires simple spatial interpolation to model advection, and we observe that this choice produces acceptable results for this first study. In future works, we will implement more robust algorithms to accurately capture the field advection and minimize the numerical diffusion introduced by our approach. Finally, Gauss's law (equation \eqref{eq:OAdivE}) is enforced, in our approach, with a Langdon-Marder correction (\citealt{marder1987,langdon1991}); the resulting (small) errors on charge conservation are thus bounded in time.

\subsubsection{Modified Boris push and particle-position update}
\label{sec:modBoris}

To solve the momentum equation \eqref{eq:OAu_1}, we proceed as follows: first we define $ \veceps = q\Delta t\vecE/(2m)$, $\vectau = q\Delta t\vecB/(2mc)$; we then discretize the equation of motion between two consecutive time steps as
\begin{equation}
 \begin{aligned}
  \vecu^{n+1/2}-\vecu^{n-1/2} & = 2\veceps + \frac{\vecu^{n+1/2}+\vecu^{n-1/2}}{\bgam}\btimes\vectau \\
  & + \Delta t\left(\vecu^{n+1/2}+\vecu^{n-1/2}\right)\btimes\vecom + 
  \frac{s\om\Delta t}{2}(u^{n+1/2}_x+u^{n-1/2}_x)\hatvece_y \\
  & + s\om x^n \left(\frac{\vecu^{n+1/2}+\vecu^{n-1/2}}{\bgam}\bcdot\veceps\right)\hatvece_y,
  \end{aligned}
\end{equation}
where $\bgam$ is the Lorentz factor defined at an intermediate time between $n+1/2$ and $n-1/2$. Now we recast this equation as
\begin{equation}
  \vecu^+ - \vecu^+\btimes\tilvectau - \alpha_1 u^+_x\hatvece_y - \alpha_2 (\vecu^+\bcdot\tilveceps)\hatvece_y = 
  \vecu^- + \vecu^+\btimes\tilvectau + \alpha_1 u^-_x\hatvece_y + \alpha_2 (\vecu^-\bcdot\tilveceps)\hatvece_y,
  \label{eq:OAu_disc}
\end{equation}
where we have defined $\vecu^\pm=\vecu^{n\pm1/2}\mp\veceps$,
$\tilvectau = \vectau/\bgam+\Delta t\vecom$,
$\tilveceps = \veceps/\bgam$,
$\alpha_1 = s\om\Delta t/2$,
$\alpha_2 = s\om x^n$,
and $\vecu^+$ is the unknown to be solved for. Assuming that $\bgam$ is known (see next paragraph), equation \eqref{eq:OAu_disc} can be solved explicitly by recasting it as
\begin{equation}
 \textbf{A}\vecu^+ = \bb{C},
\end{equation}
where the right-hand side is the known term $
\bb{C} =  \vecu^- + \vecu^-\btimes\tilvectau + \alpha_1 u^-_x\hatvece_y + \alpha_2 (\vecu^-\bcdot\tilveceps)\hatvece_y$, and 
\begin{equation}
 \textbf{A} = 
 \begin{bmatrix}
 1 & -\tiltau_z & \tiltau_y \\
 \tiltau_z-\alpha_1-\alpha_2\tileps_x & 1-\alpha_2\tileps_y & -\tiltau_x-\alpha_2\tileps_z \\
 -\tiltau_y & \tiltau_x & 1
 \end{bmatrix}.
\end{equation}
The solution is then straightforward,
\begin{equation}
 \vecu^+ = \textbf{A}^{-1}\bb{C},
\end{equation}
with the inverse matrix
\begin{equation}
 \textbf{A}^{-1} = \frac{1}{D}
 \begin{bmatrix}
 1+\tiltau_x^2+\alpha_2(\tiltau_x\tileps_z-\tileps_y) &
 \tiltau_z + \tiltau_x\tiltau_y &
 -\tiltau_y+\tiltau_x\tiltau_z+\alpha_2(\tiltau_y\tileps_y+\tiltau_z\tileps_z) \\
 -\tiltau_z+\tiltau_x\tiltau_y+\alpha_1+\alpha_2(\tiltau_y\tileps_z+\tileps_x) &
 1+\tiltau_y^2 &
 \tiltau_x+\tiltau_y\tiltau_z-\alpha_1\tiltau_y+\alpha_2(-\tiltau_y\tileps_x+\tileps_z) \\
 \tiltau_y+\tiltau_x\tiltau_z-\alpha_1\tiltau_x-\alpha_2(\tiltau_x\tileps_x+\tiltau_y\tileps_y) &
 -\tiltau_x+\tiltau_y\tiltau_z &
 1 + \tiltau_z^2-\alpha_1\tiltau_z-\alpha_2(\tiltau_z\tileps_x+\tileps_y)
 \end{bmatrix},
\end{equation}
where
\begin{equation}
D = 1+\tiltau^2 - \alpha_1(\tiltau_z+\tiltau_x\tiltau_y) - \alpha_2[\tiltau_y(\tilvectau\bcdot\tilveceps) + \tileps_y+\tileps_x\tiltau_z-\tileps_z\tiltau_x].
\end{equation}
After $\vecu^+$ has been calculated, the particle velocity at the next time step is simply
\begin{equation}
\vecu^{n+1/2} = \vecu^+ + \veceps.
\end{equation}

The definition of $\bgam$ is not unique and it is in fact the peculiarity that generally distinguishes relativistic particle pushers (see e.g.\ \citealt{ripperda2018a}). For simplicity, we can define the mid-point Lorentz factor as
\begin{equation}
\bgam = \sqrt{1 + \left[(u^-)^2 + 2\alpha_1 u^-_x u^-_y + 2\alpha_2(\vecu^-\bcdot\veceps) u^-_y/\bgam\right]/c^2},
\end{equation}
which is a choice similar to that used in the standard Boris pusher. This definition is such that in the absence of electric fields, energy is conserved exactly. We can solve the resulting depressed cubic equation for $\bgam$ with any available algorithm, e.g.\ the one described by \cite{tiruneh2020}: define $X = a_1/3$, $Y = -a_2/2$, where
\begin{equation}
 a_1 = -1 - \left[(u^-)^2 + 2\alpha_1 u^-_x u^-_y\right]/c^2,
\end{equation}
\begin{equation}
 a_2 = -2\alpha_2(\vecu^-\bcdot\veceps) u^-_y/c^2,
\end{equation}
such that $\bgam^3 + a_1\bgam +a_2 =0$. The discriminant $Z=X^3+Y^2$ indicates that there is only one real root if $Z>0$,
\begin{equation}
 \bgam = \sqrt[3]{Y+\sqrt{Z}} + \sqrt[3]{Y-\sqrt{Z}},
\end{equation}
and three real roots (of which only one positive and $>1$) if $Z<0$,
\begin{equation}
 \bgam_j = 2\sqrt{-X}\cos\frac{\phi+2(j-1)\pi}{3},
\end{equation}
where $j=1,2,3$ and $\phi=\cos^{-1}\left(Y/\sqrt{-X^3}\right)$.

When the new particle velocity is known, the position push can be carried out in two steps. The position is first updated with the motion with respect to the background, i.e.\
\begin{equation}
\vecx' = \vecx^n + \Delta t \frac{\vecu^{n+1/2}}{\gamma^{n+1/2}},
\end{equation}
after which the current can be collected with the usual Villasenor-Buneman charge deposition (\citealt{villasenorbuneman1992}) using the displacement $\vecx'-\vecx^n$. Then, the position is updated with the shearing-velocity part,
\begin{equation}
\vecx^{n+1} = \vecx' - s\om \frac{x'+x^n}{2} \Delta t\hatvece_y,
\end{equation}
which only modifies the $y$-coordinate, using the $x$-position at the midpoint in time (note that $x^{n+1}\equiv x'$). Depositing the current using $\vecx'-\vecx^n$ instead of $\vecx^{n+1}-\vecx^n$ is essential to recover $\vecJ'$ (in this paradigm, the current due to velocity fluctuations with respect to the background) instead of $\vecJ$.

As noted in section \ref{sec:equationsKSBOA}, if particle velocities are relativistic the particle position update above may result in superluminal motion, since it is not guaranteed that $(\vecu/\gamma+\vecv_\mathrm{s})^2<c^2$. As an alternative strategy, we can employ the proper relativistic velocity boost to update the particle position with a speed that is ensured to be subluminal. Applying a central-difference scheme to equations \eqref{eq:OAxproperx}--\eqref{eq:OAxproperz}, the position update is then formulated as
\begin{equation}
x^{n+1} = x^n + \Delta t \frac{u_x^{n+1/2}}{[\gamma^{n+1/2}+u_y^{n+1/2}v_{\mathrm{s},y}^{n+1/2}]\gamma_\mathrm{s}^{n+1/2}},
\label{eq:OAxproperxdisc}
\end{equation}
\begin{equation}
y^{n+1} = y^n + \Delta t \frac{u_y^{n+1/2}}{\gamma^{n+1/2}+u_y^{n+1/2}v_{\mathrm{s},y}^{n+1/2}} + \Delta t \frac{\gamma^{n+1/2}v_{\mathrm{s},y}^{n+1/2}}{\gamma^{n+1/2}+u_y^{n+1/2}v_{\mathrm{s},y}^{n+1/2}},
\label{eq:OAxproperydisc}
\end{equation}
\begin{equation}
z^{n+1} = z^n + \Delta t \frac{u_z^{n+1/2}}{[\gamma^{n+1/2}+u_y^{n+1/2}v_{\mathrm{s},y}^{n+1/2}]\gamma_\mathrm{s}^{n+1/2}},
\label{eq:OAxproperzdisc}
\end{equation}
where $v_{\mathrm{s},y}^{n+1/2} = -s\om x^{n+1/2}$, $\gamma_\mathrm{s}^{n+1/2}=1/\sqrt{1-(v_{\mathrm{s},y}^{n+1/2}/c)^2}$, and $x^{n+1/2}=(x^{n+1}+x^n)/2$. These equations are less trivial to employ, especially since equation \eqref{eq:OAxproperxdisc} is now nonlinearly implicit in the unknown $x^{n+1}$. Fortunately, the nonlinearity is weak and there are no numerical pathologies on the right-hand side (e.g., the denominator is always $>1$). The position update can then be carried out in the following steps:
\begin{enumerate}
    \item Solve the nonlinear equation \eqref{eq:OAxproperxdisc} for $x'\equiv x^{n+1}$, obtaining $v_{\mathrm{s},y}^{n+1/2},\gamma_\mathrm{s}^{n+1/2}$ in the process. We have found that 3-4 fixed-point iterations with $x^n$ as initial guess typically suffice in converging below an absolute tolerance of $10^{-10}$.
    \item Partially update the $y$-coordinate as
    \begin{equation}
     y' = y^n + \Delta t \frac{u_y^{n+1/2}}{\gamma^{n+1/2}+u_y^{n+1/2}v_{\mathrm{s},y}^{n+1/2}}.
    \end{equation}
    \item Calculate $z'\equiv z^{n+1}$ with equation \eqref{eq:OAxproperzdisc}.
    \item Using $\vecx'$, deposit the comoving-frame current $\vecJ'$.
    \item Complete the update of the $y$-coordinate with the shearing-velocity part,
    \begin{equation}
     y^{n+1} = y' + \Delta t \frac{\gamma^{n+1/2}v_{\mathrm{s},y}^{n+1/2}}{\gamma^{n+1/2}+u_y^{n+1/2}v_{\mathrm{s},y}^{n+1/2}}.
    \end{equation}
\end{enumerate}

\section{Differential heating in kinetic shearing-box simulations of pair plasmas}
\label{app:diffene}

Kinetic shearing-box simulations are characterized by additional drift forces in the particle equations of motion, related to the orbital motion of the accretion disk. These forces are not charge-agnostic --- that is, opposite charges of equal mass experience difference forces acting upon their motion. This causes a symmetry break in the dynamics of electrons and positrons; as a result, in all our simulations (both 2D and 3D) we observe that the two species in a pair plasma do not experience the same energization when solving the shearing-box equations. This appears surprising if compared to standard (non-shearing-box) PIC simulations of pair plasmas, where the two species behave essentially in the same way, and the opposite sign in the charge only serves to maintain global quasi-neutrality. Here, we prove that the differential energization we observe is not a numerical artifact, but a physical effect that characterizes shearing-box PIC simulations.

\begin{figure*}
\centering
\includegraphics[width=1\textwidth, trim={0mm 0mm 0mm 0mm}, clip]{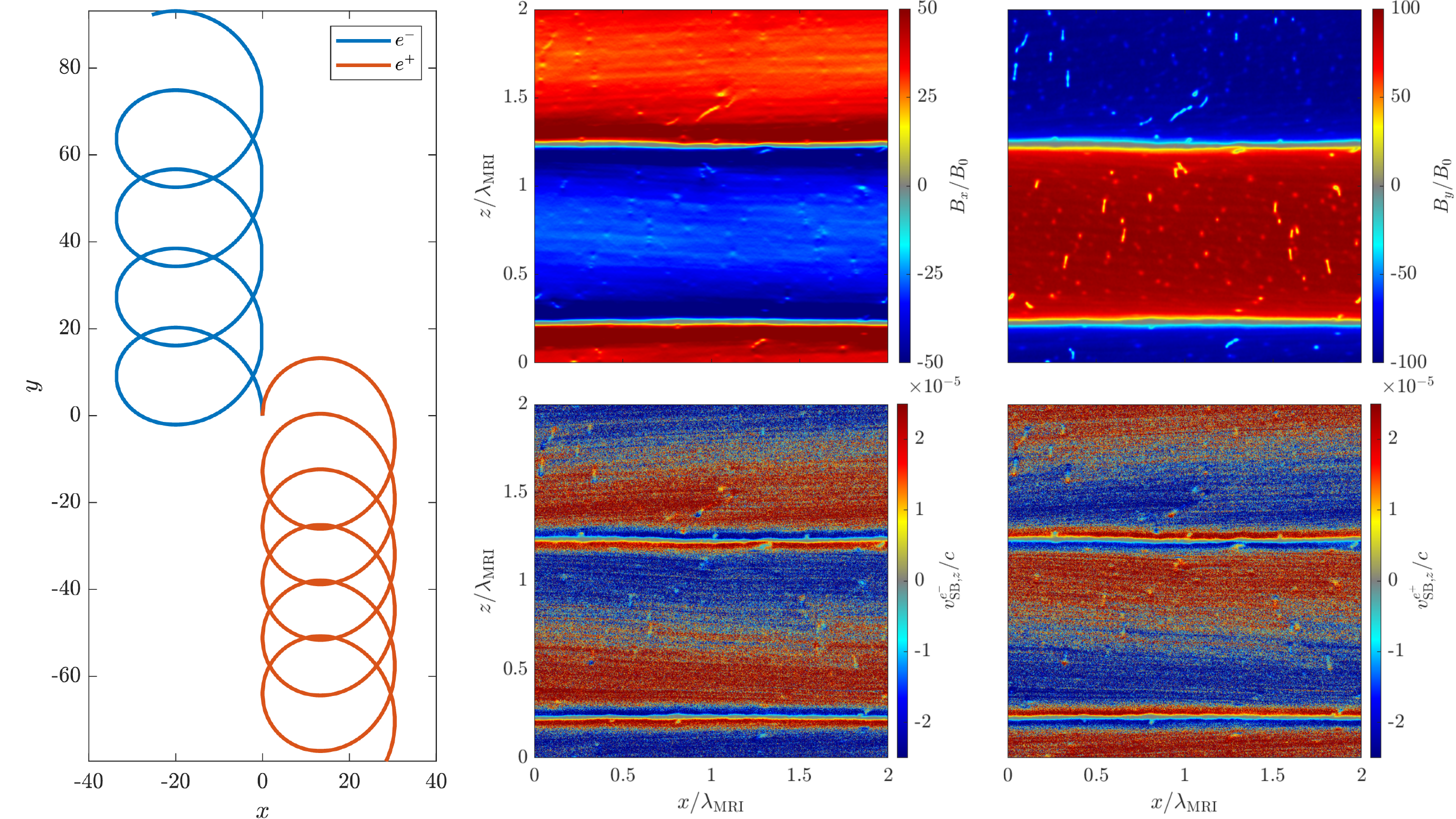}
\caption{Left: trajectories of test particles (an electron and a positron) moving in the $xy$-plane when a uniform and constant magnetic field $\vecB=(0,0,B_0)$ is present. The particle motion is computed by solving the equations employed in the shearing-box paradigm. Right: spatial distribution of the radial and toroidal magnetic field (top row) after the formation of channel flows in a small-box, 2D MRI simulation of size $2\times2\lmri^2$ (with $\omrat=120$, $\beta_0=624$). The corresponding spatial distribution of particle drift velocity along $z$ (bottom row) shows that electrons are pulled toward current sheets, and positrons are pushed away.}
\label{fig:diffdrift}
\end{figure*}

This effect stems from the kinetic shearing-box paradigm, which involves additional forces in the particle equations of motion. In the \KSBSC~system, the momentum equation \eqref{eq:riquelmeu} contains the force term
\begin{equation}
    \vecF_\mathrm{SB} = 2m\vecu\btimes\vecom + ms\om u_x \hatvece_y = 
    \begin{bmatrix}
    2m \om u_y \\ m(s-2)\om u_x \\ 0
    \end{bmatrix},
\end{equation}
which is related to Coriolis and centrifugal forces in the comoving frame where all quantities are measured (the same applies to the \KSBOA~system). This additional force causes a drift of the particle guiding center which is absent in typical non-shearing-box kinetic simulations. In the nonrelativistic limit, this drift has the usual expression,
\begin{equation}
    \vecv_\mathrm{SB} = \frac{\langle\vecF_\mathrm{SB}\rangle\btimes\vecB}{qB^2}c = \frac{mc}{qB^2}
    \begin{bmatrix}
    (s-2)\om \langle u_x\rangle B_z \\ -2\om \langle u_y\rangle B_z \\ 2\om \langle u_y\rangle B_y - (s-2)\om \langle u_x\rangle B_x
    \end{bmatrix},
\end{equation}
where $\langle ...\rangle$ indicates averaging over the particle gyration. Since the sign of the drift depends on the particle charge, in general particles of equal mass but opposite charge will experience opposite drifts in all spatial directions.

We now show that the $\vecv_\mathrm{SB}$ drift is directly related to the dynamics leading to the measured differential energization. First, we note that the initial conditions in our MRI simulations, $\vecB=(0,0,B_{z,0})$, lead to a drift motion exclusively in the $xy$-plane. Such motion is precisely the well-known epicyclic motion of particles in orbit around massive objects (e.g.\ \citealt{balbushawley1998}), which is here modified by the presence of a vertical magnetic field. In Figure \ref{fig:diffdrift} (left panel), we show a portion of the trajectory of one electron and one positron (with $m=|q|=1$) moving according to equation \eqref{eq:riquelmeu} in the presence of a weak magnetic field $B_{z,0}$ such that $\omrat=1$, with initial velocity $\vecv=(0,0.1,0)$. We observe that the usual particle gyromotion is here modified by a drift of the guiding center along $\pm y$ depending on the particle charge $q=\mp1$. Moreover, the gyromotion is in itself different for the two particles, with the electron attaining higher gyrofrequency and smaller gyroradius (and vice versa for the positron). Although the guiding-center drift is purely in the toroidal direction in this case, this differential gyromotion ought to manifest in our 2D ($x$-$z$) simulations as well, given that the particle oscillation along $x$ occurs differently for opposite charges. This difference in particle trajectories implies that charge neutrality cannot be maintained at all times, and a net current must arise.

While the differential gyromotion is present, at least in the beginning, in all our simulations, it is arguably of minor importance in the overall system evolution. Indeed, in section \ref{sec:ene2D} we have shown that significant differential heating primarily manifests during and after the linear stage, when the magnetic-field geometry differs substantially from the initial (pre-instability) phase. We find that relevant differential heating arises due to the channel-flow configuration of $B_x$ and $B_y$ during the later stages: $B_z$ here is heavily subdominant (see section \ref{sec:evol_phases}), and the main drift component of $\vecv_\mathrm{SB}$ is along $z$. In Figure \ref{fig:diffdrift} (right panels, top row) we plot the spatial distribution of $B_x$ and $B_y$ (normalized over the initial field $B_0$) for a 2D, small-box ($2\times2\lmri^2$) simulation with parameters $\omrat=120$, $\beta_0=624$ at $t=3.3P_0$, when channel flows are present and current sheets have formed at the channel interfaces. In the same Figure (right panels, bottom row), we also plot the spatial distribution of $v_{\mathrm{SB,z}}$ for electrons and positrons taken at the same moment in time. We observe that, in the vicinity of current sheets, both species are drifting along $z$; however, the particular magnetic-field configuration causes this drift to be directed \emph{toward} current sheets in the case of electrons, and \emph{away} from them in the case of positrons. In a standard current-sheet (e.g.\ Harris-like) magnetic configuration, particles drift towards current sheets due to the usual $\vecE\btimes\vecB$ motion irrespective of their charge; this global behavior is still verified in our case, but the addition of the $\vecv_\mathrm{SB}$ drift now implies that charges of opposite sign drift toward current sheets at different rates.

Since particles experience most of the energy gain via reconnection in current sheets, this differential drift motion can explain the large difference in particle energy observed between the two species. Electrons, which flow toward current sheets faster, are capable of gaining more energy over shorter times; this precisely corresponds to the faster electron energization rates reported in section \ref{sec:ene2D}. Eventually, all available magnetic energy is distributed, via reconnection, unequally among the pair-plasma species, with electrons experiencing a larger energy gain.

The differential heating we report here for the first time (to the best of our knowledge) is a genuine physical effect which, however, is arguably of very minimal importance in realistic scenarios. Indeed, it can immediately be observed that the drift speed $v_\mathrm{SB}$ is inversely proportional to $\omrat$ (and indeed, for the $\omrat=120$ case shown in Figure \ref{fig:diffdrift}, $v_\mathrm{SB}/c\ll1$). In realistic environments where $\omrat\sim 10^7$, this drift would be completely negligible and therefore no differential heating would be observed. In our simulations (as well as in any other presented in literature), however, we are bound to employ much smaller values of $\omrat$ due to otherwise excessive computational costs. Therefore, when measuring particle heating in simulations, this effect must be taken into account; particularly in the context of future electron-ion numerical experiments, where analyzing particle heating is of main interest for observations of accreting compact objects. In such simulations, it will be of major importance to quantify the ``spurious'' differential heating described here, in order to correctly evaluate the energy partition between particle species.

\begin{figure*}
\centering
\includegraphics[width=1\textwidth, trim={0mm 80mm 0mm 0mm}, clip]{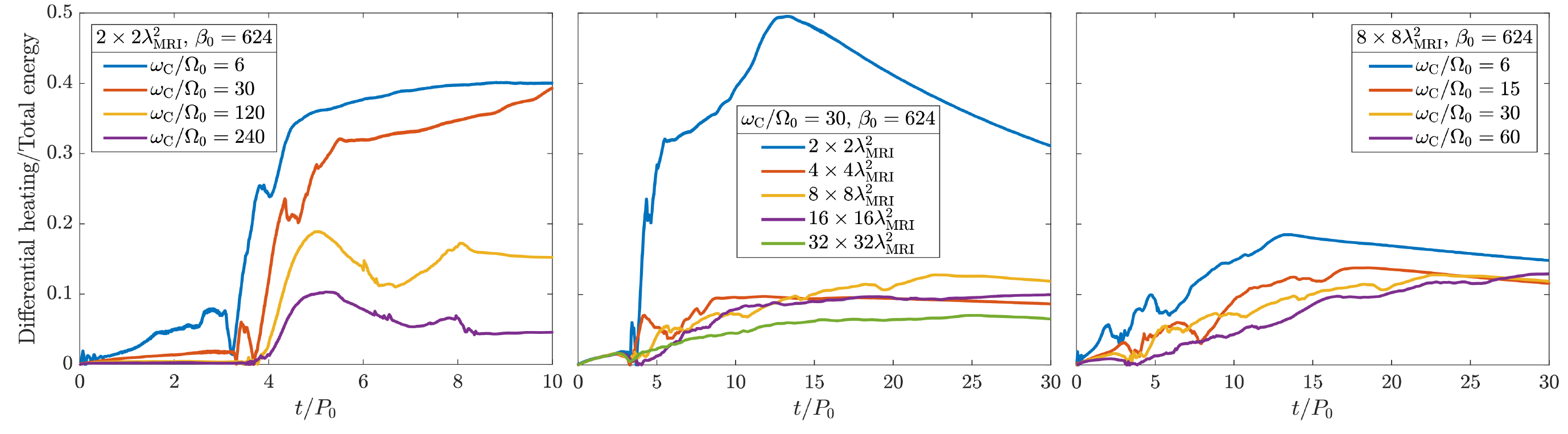}
\caption{Differential heating in 2D pair-plasma MRI simulations, measured as the difference in kinetic energy between electrons and positrons normalized over the instantaneous total energy of the system. Simulations with small box size and variable $\omrat$ (left), variable box size and fixed $\omrat$ (center), and large box size and variable $\omrat$ (right) are compared.}
\label{fig:diffene}
\end{figure*}

We conclude by reporting the differential heating measured in our 2D simulation campaign presented in section \ref{sec:2D}. This is shown in Figure \ref{fig:diffene}, where the difference in electron and positron total energy $mc^2(\gamma-1)$ is normalized over the instantaneous total energy of the system, in order to measure the relative importance of differential heating with respect to the global energetics. We observe that larger $\omrat$ values and larger boxes can ameliorate this issue substantially, keeping the differential heating below 10\%. While it is clear that a larger $\omrat$ implies smaller drift speeds and therefore smaller differences in the rate at which particles flow towards current sheets, justifying the positive effect of larger box sizes is less straightforward. Qualitatively, we can consider that smaller boxes tend to create persistent large-scale channel flows that are less likely to be destroyed by unstable modes and result in developed turbulence (as illustrated in section \ref{sec:ene2D}); this large-scale, ordered distribution of magnetic fields is the most favorable for particles to drift undisturbed toward currents sheets at different rates. In large boxes where turbulence develops, particles are less likely to experience such long-lived, undisturbed motion, being instead continuously scattered across the turbulent structures. Differential heating can still accumulate over long times, reaching however much smaller levels.

\section{Energy conservation in the shearing box}
\label{app:energycons}

\begin{figure*}
\centering
\includegraphics[width=1\textwidth, trim={0mm 45mm 0mm 0mm}, clip]{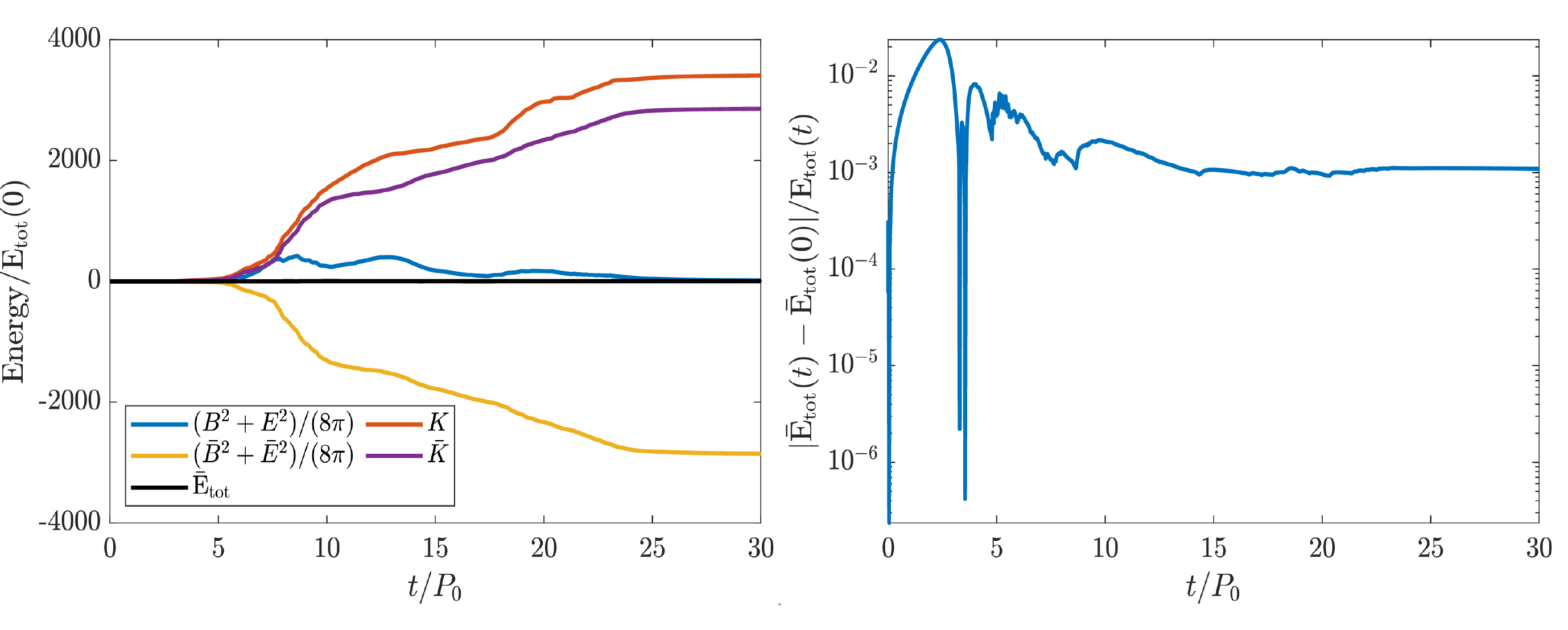}
\caption{Energy conservation in a 2D pair-plasma MRI simulation with parameters $\omrat=30$, $\beta_0=624$, and box size $8\times8\lmri^2$. Left: evolution of the total electromagnetic and kinetic energy compared with the modified electromagnetic and kinetic energy, where shearing-box source terms have been subtracted. The modified total energy $\bar{\mathrm{E}}_\mathrm{tot}$ is a conserved quantity in the absence of numerical errors. Right: evolution during the run of the error on the conserved $\bar{\mathrm{E}}_\mathrm{tot}$ relative to the instantaneous total energy $\mathrm{E}_\mathrm{tot}$ (the latter includes source terms and is not conserved).}
\label{fig:energy_cons}
\end{figure*}

Considering for simplicity the 2D shearing box in shearing coordinates, equations \KSBSC~manifestly include source terms affecting the field and particle evolution. These source terms translate into energy sources that are not present in standard (non-shearing-box) PIC simulations. Evaluating energy conservation in the presence of these sources is not straightforward.

To obtain a measure of energy conservation, we compute the energy sources directly at runtime. For the fields, the energy source terms are obtained by dotting Faraday's and Amp\`ere's laws with $\vecB$ and $\vecE$, respectively:
\begin{equation}
    \vecB\bcdot\frac{\pd\vecB}{\pd t} = \frac{1}{2}\frac{\pd B^2}{\pd t} = \mbox{(standard energy terms)} - s\om B_xB_y,
\end{equation}
\begin{equation}
    \vecE\bcdot\frac{\pd\vecE}{\pd t} = \frac{1}{2}\frac{\pd E^2}{\pd t} = \mbox{(standard energy terms)} - s\om E_xE_y,
\end{equation}
where in addition to standard energy-exchange terms we now have extra source terms stemming from the shearing-box paradigm. For the particles, we dot the momentum equation with $\vecu/\gamma$,
\begin{equation}
    \frac{\vecu}{\gamma}\bcdot\frac{\rmd\vecu}{\rmd t} = \frac{\rmd \gamma}{\rmd t} = \mbox{(standard energy terms)} + s\om\frac{u_xu_y}{\gamma},
\end{equation}
which produces one extra energy source. All terms deriving from the shearing-box framework are thus proportional to $\om$.

To quantify the importance of the shearing-box sources, we compute each source term in the whole simulation box during a representative 2D pair-plasma MRI simulation with parameters $\omrat=30$, $\beta_0=624$, and box size $8\times8\lmri^2$. The source terms are then integrated in time, to obtain the net energy variation associated with these terms. Subtracting the net contribution of each term from the electromagnetic and particle energy produces a new measure of energy that only accounts for standard energy-exchange channels (i.e.\ without external energy sources related to the shearing box). In simulations, the total energies including the source terms are given by
\begin{equation}
    B^2(t) = \sum_g B^2_g(t)\Delta V_g,
\end{equation}
\begin{equation}
    E^2(t) = \sum_g E^2_g(t)\Delta V_g,
\end{equation}
for the electromagnetic energy at each grid cell $g$ (with cell volume $\Delta V_g$), and
\begin{equation}
    K(t) = \sum_p m_pc^2(\gamma_p(t)-1),
\end{equation}
for the kinetic energy of all particles $p$. By subtracting the shearing-box terms, the \emph{modified} field energies (i.e.\ without shearing-box sources) are
\begin{equation}
    \frac{1}{2}\bar{B}^2(t) \equiv \frac{1}{2}B^2(t) - \int_0^t dt \sum_g \Delta V_g \left(-s\om B_{x,g}(t)B_{y,g}(t)\right),
\end{equation}
\begin{equation}
    \frac{1}{2}\bar{E}^2(t) \equiv \frac{1}{2}E^2(t) - \int_0^t dt \sum_g \Delta V_g \left(-s\om E_{x,g}(t)E_{y,g}(t)\right),
\end{equation}
and the modified particle energy is
\begin{equation}
    \bar{K}(t) \equiv K(t) - \int_0^t dt \sum_p s\om\frac{u_{x,p}(t)u_{y,p}(t)}{\gamma_p(t)}.
\end{equation}
With these definitions, the modified energies are such that
\begin{equation}
    \pd_t\bar{\mathrm{E}}_\mathrm{tot} =\pd_t\left(\frac{\bar{B}^2+\bar{E}^2}{8\pi} + \bar{K}\right) = 0,
\end{equation}
i.e.\ the sum of the modified electromagnetic and kinetic energy integrated over the simulation box remains constant in time (as in standard PIC simulations), in the absence of numerical errors. Note that conversely, the total energy including sources $\mathrm{E}_\mathrm{tot} \equiv (B^2+E^2)/(8\pi) + K$ is not constant: the sources continuously inject energy into the system.

Figure \ref{fig:energy_cons} (left panel) shows the evolution in time of the modified electromagnetic and kinetic energy (integrated over the box) as well as of the sum of the two. The plot shows that the modified electromagnetic energy becomes \emph{negative}: this is because the external energy input for the electromagnetic fields is much larger than the initial energy, and subtracting this energy source results in observing exclusively the standard process of magnetic energy dissipation, e.g.\ via reconnection. Because of the external source, more magnetic energy than initially available can be dissipated, and the modified electromagnetic energy can fall below 0. The decrease in electromagnetic energy corresponds almost exactly to an increase in particle energy via dissipation processes. The total (electromagnetic plus kinetic) modified energy is much smaller than its individual components, owing to the fact that the modified electromagnetic energy becomes negative.

The right panel of Figure \ref{fig:energy_cons} shows a measure of energy conservation calculated as $|\bar{\mathrm{E}}_\mathrm{tot}(t)-\bar{\mathrm{E}}_\mathrm{tot}(0)|/\mathrm{E}_\mathrm{tot}(t)$. This quantity is such that the numerator expresses the deviation from exact energy conservation in terms of the absolute error on the conserved quantity $\bar{\mathrm{E}}_\mathrm{tot}(t)$; normalizing over the instantaneous total energy (including sources) serves to provide a measure of how important the numerical errors are compared to the actual amount of energy in the system at a certain time. We see that the error is at most of order $10^{-2}$, i.e.\ numerical energy dissipation is negligible throughout a typical run.



\bibliographystyle{aasjournal}
\end{document}